\documentclass{jaa}
\usepackage{natbib}
\usepackage{graphicx}
\begin{document}\sloppy

\title{
An Automated Pipeline for Ultra-Violet Imaging Telescope (UVIT) 
}

\author{S. K. Ghosh\textsuperscript{1,4,*}, S. N. Tandon\textsuperscript{2,5},
S. K. Singh\textsuperscript{3}, D. S. Shelat\textsuperscript{3,4},
P. Tahlani\textsuperscript{3}, A. K. Singh\textsuperscript{3},
T. P. Srinivasan\textsuperscript{3},P. Joseph\textsuperscript{5}, A. Devaraj\textsuperscript{5},
K. George\textsuperscript{5,6}, R. Mohan\textsuperscript{5}, 
J. Postma\textsuperscript{7}, C. S. Stalin\textsuperscript{5}}

\affilOne{\textsuperscript{1}Tata Institute of Fundamental Research, Mumbai 400005, India\\}
\affilTwo{\textsuperscript{2}Inter-University Centre for Astronomy \& Astrophysics, Pune 411007, India\\}
\affilThree{\textsuperscript{3}Space Application Centre (ISRO), Ahmedabad 380015, India\\}
\affilFour{\textsuperscript{4}National Centre for Radio-Astrophysics (NCRA-TIFR), Pune 411007, India\\}
\affilFive{\textsuperscript{5}Indian Institute of Astrophysics, Bangalore 560034, India\\}
\affilSix{\textsuperscript{6}Ludwig-Maximilians-Universität, Munich, Germany \\}
\affilSeven{\textsuperscript{7}University of Calgary, Calgary, Canada \\}

\twocolumn[{

\maketitle

\corres{swarna@tifr.res.in}

\date{Received: date / Accepted: date}

\begin{abstract}

We describe a versatile pipeline for processing the data collected by the
 Ultra-Violet Imaging Telescope (UVIT) on board Indian Multi-wavelength
 astronomical satellite ASTROSAT.The UVIT instrument carries out simultaneous
 astronomical imaging through selected filters / gratings in Far-Ultra-Violet
 (FUV), Near-Ultra-Violet \& visible (VIS) bands of the targeted circular
 sky field ($\sim$ 0.5 deg dia). This pipeline converts the data (Level-1)
 emanating from UVIT in their raw primitive format supplemented by inputs
 from the spacecraft sub-systems into UV sky images (\& slitless grating
 spectra) and associated products readily usable by astronomers  (Level-2).
 The primary products include maps of Intensity (rate of photon arrival),
 error on Intensity and effective Exposure. The pipeline is open source,
 extensively user configurable with  many selectable parameters and its
 execution is fully automated. The key ingredients of the pipeline includes –
 extraction of drift in pointing of the spacecraft, and disturbances in pointing due to
 internal movements; application of various corrections to measured position
 in the detector for each photon - e.g.
  differential pointing with respect to a reference frame for shift and add operation,
 systematic
 effects and artifacts in the optics of the telescopes and detectors,
 exposure tracking on the sky, alignment of sky products from multi-episode
 exposures to generate a consolidated set and astrometry. Detailed logs
 of operations and intermediate products for every processing stage are
 accessible via user selectable options. While large number of selectable
 parameters are available for the user, a well characterized ``standard default"
 set is used for executing this pipeline at the Payload Operation Centre
 (POC) for UVIT and selected products are archived and disseminated by
 the Indian Space Research Organization (ISRO) through its ISSDC portal.

\end{abstract}

\keywords{telescopes: UVIT --- instrumentation: pipeline}

}]
\msinfo{04 October 2021}

\section{Introduction}

   The ASTROSAT is the first Indian astronomical space mission which
 covers a wide range of the electromagnetic  spectrum (Agrawal, 2006;
 Singh et al., 2014). This satellite
 carried five major scientific instruments to provide
 a platform for multi-wavelength
 astronomy from a low earth orbit with a small inclination to the equator.
 Four instruments covered
 different energies of soft to hard X-ray astronomy while the fifth one,
 Ultra-Violet Imaging Telescope (UVIT), enabled astronomical measurements
 over wide instantaneous field of view with high angular resolution,
 in the ultraviolet region. A summary of characteristics for each of
 these 5 instruments on board ASTROSAT are presented in Table 1. The
 primary aim of UVIT is to simultaneously image the desired region of
 the sky in Far-Ultra-Violet (FUV; 1300 – 1800 \AA) and Near-Ultra-Violet
 (NUV; 2000 – 3000 \AA) bands for a field of $\sim$ 28${}^{\prime}$
 diameter with a resolution $\sim$ 1.5${}^{\prime\prime}$ (Tandon et al
 2017a \& 2017b). The telescope for NUV band also provided for imaging
 in optical / visible band (VIS; 3200 – 5500 \AA), whose primary aim
 was to provide improved aspect of the telescope bore site every second.
 All three bands simultaneously view almost identical region of the sky.
 More details of UVIT and its in-orbit performance are presented in
 Tandon et al. (2017c) and references therein.
 The NUV band went non-functional on March 30, 2018 and remained so
 despite  multiple attempts to revive it.


\begin{table*}[th]
\footnotesize
\caption{Scientific experiments on board Indian multi-wavelength astronomical satellite ASTROSAT}

\begin{tabular}{|l|l|l|l|l|l|}
\hline
\textbf{Instrument}&\textbf{Ultra-Violet}&\textbf{Soft Xray}&\textbf{Large Area}
 &\textbf{Cadmium-Zinc-}&\textbf{Scanning Sky} \\
 &\textbf{Imaging}&\textbf{Telescope}&\textbf{Xenon}
 &\textbf{Telluride}&\textbf{Monitor} \\
 &\textbf{Telescope}&\textbf{(SXT)}&\textbf{Proportional}
 &\textbf{Imaging}&\textbf{(SSM)} \\
 &\textbf{UVIT}& &\textbf{Counter}
 &\textbf{Telescope}& \\
 & & &\textbf{(LAXPC)}
 &\textbf{CZTI)}& \\
\hline
Waveband &Far UV: 130-180 nm &Soft X-ray: &Soft \& Hard &Hard X-ray: &Soft \& Hard \\ 
 &Near UV: 200-300 nm &0.3 - 8 keV &X-ray: &25 - 150 keV &X-ray : \\ 
 &Visible: 320-550 nm & &3 - 100 keV & &2.5 - 10 keV \\ 
\hline
Field of View &$\sim$28 arc-min &$\sim$40 arc-min &0.9 deg$\times$0.9 deg &4.6 deg$\times$4.6 deg &22-27 deg$\times$\\
 & & & & &100 deg\\
\hline
Optics &Twin Ritchie Chretian &Conical foil &Collimator &2-D coded &1-D coded\\
 &2-mirror system &(Wolter-I) & &mask &mask\\
 & & Mirrors & & &\\
\hline
Detector &Photon &X-ray CCD &Proportional &CdZnTe &Position\\
 &Counting &at the &Counter &Arrays &sensitive\\
 &(Intensified) &focal plane & & &Proportional\\
 &CMOS imagers & & & &Counter\\
\hline
Angular &1.2-1.5 arc-sec &2 arc-min &$\sim$1-5 arc-min &8 arc-min &$\sim$10-13 arc-min\\
Resolution &(in Far \& &&(in scan &&along coding\\
&Near UV) &&mode) &&axis \& 2.5\\
&&&&&deg across\\
\hline
Energy &10-50 nm &90 eV at 1.5 keV &12-15\% &6\% at &25\%\\
Resolution &as per filter &136 eV at 5.9 keV &in 22-60 keV &100 keV &at 6 keV\\
 &selection & & & &\\
\hline
Geometrical &1250 &250 &10,800 (total &976 &173\\
Collection &for each of the& &for 3 & &(total for\\
Area (cm$^{2}$) &3 wavebands & &identical & &3 units)\\
 & & &units) & &\\
\hline
Effective &$\sim$10 for Far &90 at 1.5 keV &6000 &415 &11\\
Area (cm$^{2}$) &UV band; &&at 5-20 keV &photometric; &at 2.5 keV;\\
 &$\sim$10 for Near &&(total for 3 &335 &51\\
 &UV band; &&identical &spectroscopic &at 5 keV;\\
 &$\sim$40 for Visible &&units) &&(total for\\
 &band; && &&3 units)\\
\hline
Time &2 milli-sec &$\sim$2.4 sec for &10 micro-sec &20 micro-sec &0.1 milli-sec\\
Resolution &&full field; &&& \\
&&$\sim$0.28 sec for &&&\\ 
&&central area &&&\\ 
\hline
Sensitivity &20 mag (5$\sigma$) &$\sim$10$^{-13}$ erg cm$^{-2}$ &1 milli-Crab &20 milli-Crab &$\sim$28-40 milli-\\
&in 160 sec for &sec$^{-1}$ (5$\sigma$)&(3$\sigma$) in &(5$\sigma$) in &Crab (3$\sigma$)\\
&for Far UV &in 20,000 sec& 1,000 sec &10,000 sec &in 600 sec\\
\hline
Total Mass (kg) & 230 & 70 & 419 & 56 & 65.5 \\
\hline
%
%
%
\end{tabular}
\end{table*}



    The subject matter of this paper is the automated pipeline for
 processing of the raw data received from the instrument and the spacecraft
 to produce images for FUV and NUV in standard units. The processing
 involves two separate steps. The first step, called L1 (Level-1), processes the
 telemetery data to provide a bundle of data for each target of observation.
 The L1 data bundle for UVIT consists of Science (Imaging) data from all
 3 Detectors and Auxilliary (Aux) data containing inputs regarding UVIT
 Filters and various information from Spacecraft Systems, e.g. time
 calibration, attitude of satellite reference axes, position in orbit,
 gyro signals, and house keeping information. The second process,
 called L2 (Level-2), uses the bundle provided by L1 to make images for FUV and NUV
 in standard units for use by astronomers. The process L1 is implemented
 at a centre of 
 Indian Space Research Organization (ISRO), while the process L2, which is theme of this paper,
 is implemented at UVIT-Payload Operation Centre (UVIT-POC) at Indian
 Institute of Astrophysics, Bangalore. 
 It is ISRO's exclusive right \& responsibility to operate, maintain and upgrade the archives
 for disseminating all scientific data products from UVIT
 (which includes both L1 \& L2).

  The motivation for this paper is as follows.
 It provides an authentic
 description of processes involved in the translation from the Level-1
 to the Level-2, which uses this pipeline. 
 The ISRO’s archive hosts 
 the L2 products generated using a ``default" set of selectable
 parameters which are optimal for catering to most general users.
 The present work not only explains the implications of these
 selections, but also provides details about various additional options 
 some of which could be useful to researchers with specific science aim.
 In addition, for future refinements of the pipeline, this work 
 provides the basic reference. 

 Structure of this paper is as follows.
 In Section 2, we present a brief overview of imaging operations with
 UVIT and the complete end-to-end flow of data beginning from their
 collection to the final products that are ready for scientific analysis
 and interpretation by astronomers. In Section 3, detailed description
 of all the processing steps involved in generating Level-2 starting
 from Level-1 data is presented (hereafter referred to as UVIT Level-2
 Pipeline, UL2P).
 This section first describes the higher level structure of the software,
 and details of the  lower level structure follow subsequently.
 Section 4 contains 
 experience during commissioning the pipeline, 
selected sample results,
 success rates and yields of UL2P and their general archiving details.
 Section 5 presents a summary along with comments on possible further
 improvements for future.
\section{Overview}

\subsection{Imaging with UVIT}

 The imaging components of UVIT include optics, detectors and filter wheels. 
 The optics consists of twin identical Ritchey-Chretian telescopes
 as shown in Fig. 1. (more details can be found in Kumar et al 2012 a).

\begin{figure*} [th]
\centering
\includegraphics[scale=0.40]{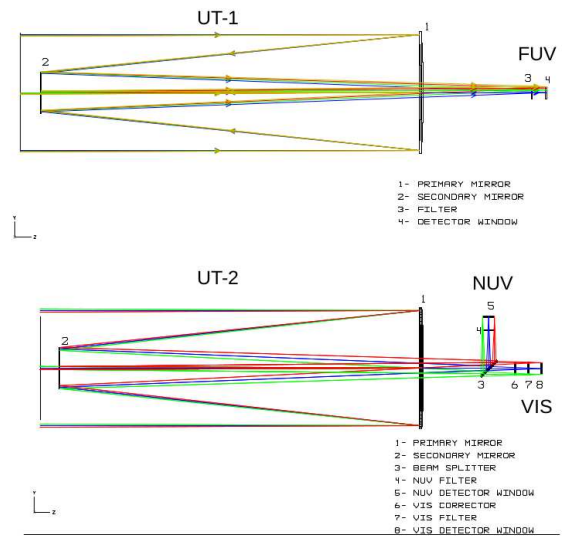}
\caption{
Optical layout of the two UVIT Telescopes, UT-1 and UT-
2, each employing a pair of Primary \& Secondary mirrors. The
UT-1 is dedicated for the Far-UV (FUV) band. The UT-2 feeds
the Near-UV (NUV) band via reflection of a beam-splitter and
the Optical (VIS) band through it.
}
\end{figure*}

 The detector system for each of the 3 bands consist of intensified C-MOS imager
 (Star-250; 512$\times$512 pixels),
 whose gain can be configured through its high voltage settings.
 While a high gain deploys
 Photon Counting mode
 (PC; $\sim$ 29 frames/s for the full field) used for the two UV bands,
 a low gain effects Integration mode (INT; $\sim$ 1 frame/s) used 
 for the VIS band.  Details of these modes are available in the Appendix-1 (Sec. A1.1). 

 A single clock (among the 3 bands) is selected for the operational ease
 of inter-band time alignment.

 The Filter Wheel Drive system of each band positions the selected filter
 (or grating) and is parked at a light blocking position when not in use.

 The electronic controller for each band configures and operates 
 the detector read out system. The key selections include : 
  (a) size of the sky field -
 full field (512$\times$512 pixels) or a smaller square window
 (100$\times$100 / 150$\times$150 / 200$\times$200 / 300$\times$300 / 350$\times$350)
 for obtaining higher frame-rates, roughly  in inverse proportion the number of pixels read
 ;
  \& (b) the  imaging mode (PC / INT).

  The imaging sessions are scheduled only during dark part
 of any orbit to avoid scattered solar radiation. Every planned sky field
 undergoes detailed checks for its safety
  against overexposure due to any unacceptably bright object in the field or 
moon within a cone of  defined half angle
from the  axis. 
 Additionally, onboard safety 
 measures with autonomous swift action ensure detector safety against accidental exposure 
 to bright field.

  The spacecraft operations usually translate observations of a specific
 astronomical target, with the filter and window size, of a typical science
 proposal into a sequence of smaller segments of imaging sessions, i.e.
 {\it Episodes}, spread over multiple orbits, as a single orbit can support
 imaging for a maximum of $\sim$ 2000 seconds, primarily due to the requirement
 of dark side of the orbit, and other factors like angle constraint to avoid
 earth occultation, South Atlantic Anomaly etc can further curtail this
 duration.

\subsection{End to end data flow}

 The Fig. 2 shows the scheme of onboard interfaces between the ASTROSAT spacecraft
 and UVIT for control of operations using commands as well as collection of
 data to be communicated to the ground station. In addition to planned 
 operations, the spacecraft systems also process emergency situations
 and guides UVIT to a safe condition.
 The data collected correspond to sky images, status
 \& settings of various parameters of subsystems and health monitoring.

\begin{figure*} [th]
\centering
\includegraphics[scale=0.40]{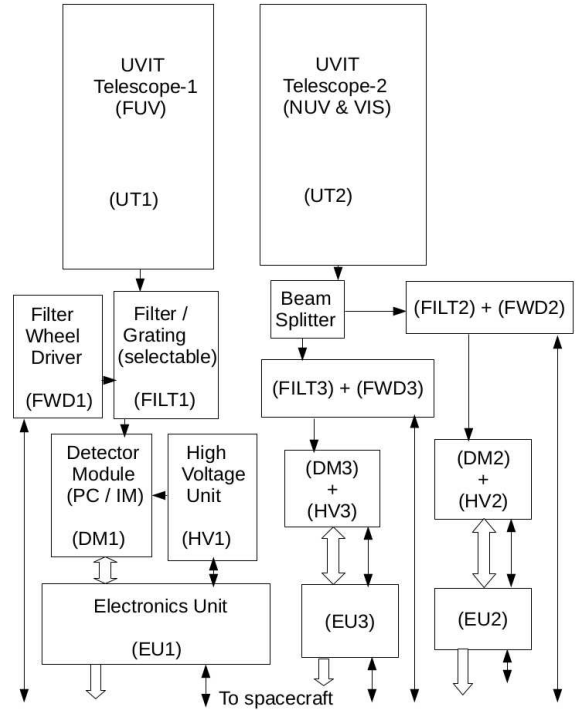}
\caption{
Schematic of the UVIT Payload.
The telescope UT1 (UT2 \& a Beam Splitter) feeds all sub-systems for the FUV (NUV \& VIS) band/(s).  
The sub-systems represent functionalities including the selection of desired Filter
 (or Grating), 
configuration of Detector read out \& settings of its High Voltages, and
two way data communications with the spacecraft.
}
\end{figure*}

  A block diagram displaying data flow and interconnections on ground
 is presented in Fig. 3. After the required processing at the ISRO’s
 ground segment systems, the data for UVIT are provided as FITS files
 called Level-1 data. These includes Science Data from the three detectors,
 and all the additional information on UVIT and the SC (named Aux Data)
 necessary for the next processing step. All the parameters of the
 detectors are made available in the Science Data, and the Aux Data
 provides status \& positions of the three filter wheels, time calibration
 table, orbital position, velocity, attitude information of the reference
 axes of the satellite, data from Gyros, and housekeeping information, at
 appropriate sampling rates.

\begin{figure*} [th]
\centering
\includegraphics[scale=0.40]{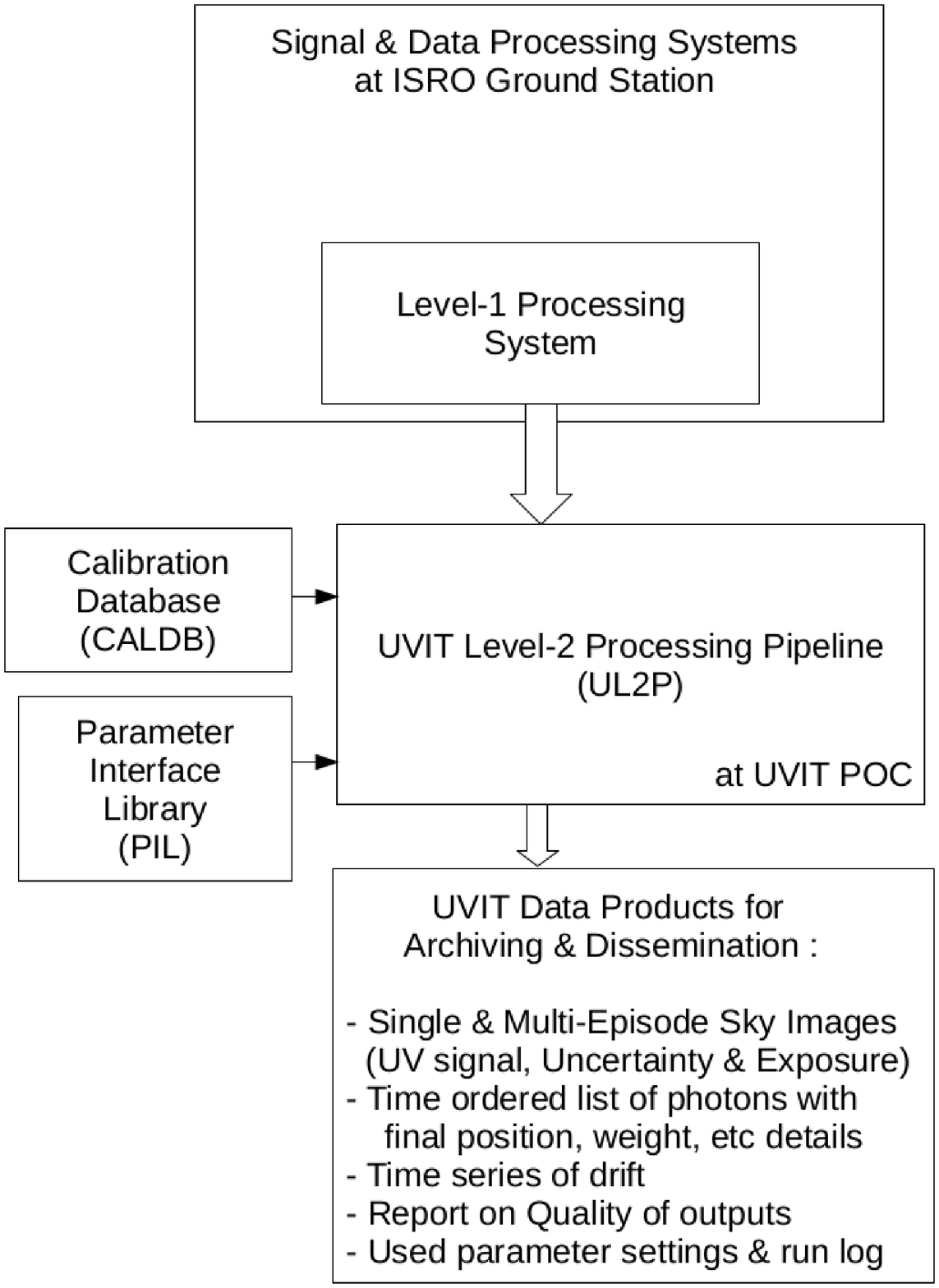}
\caption{
UVIT / ASTROSAT Ground Segment Systems.
The UVIT Payload Operation Centre (POC) executes the Level-2 Pipeline
using various inputs and generates the astronomer ready Sky Image Products
\& other outputs for archiving and dissemination.
}
\end{figure*}

  Each observing session (Episode, for a specific band, filter \& window setting),
 is no longer than
 $\sim$ 2000 s, i.e. dark part of an orbit. All the Level-1 data products
 (Science Data \& Aux Data) for all the Episodes for a target are
 organized into a single bundle named as ``merged Level-1". In the
 merged bundle Science Data for each band are assembled in individual
 directories and all the Aux Data are assembled in another directory.
 The Level-2 pipelines uses the directories of Science Data and Aux
 Data along with the calibration data base for generating the final
 images in standard units and coordinates. The data on raw images are
 provided by Science Data, data on the filters and orientation of the
 satellite are provided by Aux Data. Data on variation of sensitivity
 over the field and geometrical distortions are taken from the
 calibration data base. Details of operations executed in the pipeline
 are described in Section 3.

\section{Level-2 Pipeline}

   This section gives a full description of the pipeline. The section is
 divided in two parts: in the first part functionality of the pipeline is
 described fully, and in the second part the actual implementation is
 described. A list of acronyms used in this document as well as frequently
 used names for a few key user selectable parameters is presented in Table 2.
 Additionally, a complete list of all parameters are available at the very
 end (Appendix-11).
 For most readers the functionality of the pipeline
 would suffice without going into the detailed description, which is full
 of many intricate technical details, unless they want to use / run the pipeline themselves
 for generation of the images.


\begin{table*}[th]
\tiny
\centering
\caption{List of frequently used acronyms, names and selectable parameters referred to in this document.}
\begin{tabular}{|l|l|l|}
\hline
\textbf{Serial}&\textbf{Acronym / Name }&\textbf{Expanded form / }\\
\textbf{No.}&\textbf{/ Selectable parameter}&\textbf{Description}\\
\hline
 & {\bf Acronym :} &  \\
\hline
1 & CAL\_DB & Calibration Database  \\
\hline
2 & CR & Cosmic Rays  \\
\hline
3 & CRC & Cyclic Redundancy Check   \\
\hline
4 & ENT & Effective Number of Photon   \\
\hline
5 & FITS & Flexible Image Transport System   \\
\hline
6 & FUV       & Far-UltraViolet band of UVIT                             \\
\hline
7 & GTI      & Good Time Interval                             \\
\hline
8  & ICRS   & International Celestial Reference System                             \\
\hline
9  & INT      & INTegration Mode of imaging                              \\
\hline
10  & ISRO    & Indian Space Research Organization                              \\
\hline
11  & ISSDC  & Indian Space Science Data Centre                             \\
\hline
12  & L1       & Level-1 data (input for the pipeline)                              \\
\hline
13  & L2      & Level-2 data  (output of the pipeline)                              \\
\hline
14  & L2\_PC      & Level-2 Sky Imaging chain for data \\ 
  &       & collected in Photon Counting (PC) mode  \\
\hline
15  & L2\_INT      & Level-2 Sky Imaging chain for data \\ 
  &       & collected in Integration (INT) mode \\ 
\hline
16  &  MCP     &  Micro-Channel Plates (used as                            \\
  &       & Intensifier in the Detector Module)                          \\
\hline
17 & MJD & Modified Julian Date \\
\hline
18 & mL1 &Merged Level-1 data bundle (input for the  \\
 &  &pipeline; provided by ISRO / ISSDC) \\
\hline
19 & NUV & Near-UltraViolet band of UVIT \\
\hline
20 & OE &Other Episode/s (other than the Reference Episode \\
 &  &among the ``group" of Episodes to be combined, \\
 &  &with identical Band, Filter \& Window configuration) \\
\hline
21 & PC  & Photon Counting Mode of imaging \\
\hline
22 & PIL & Parameter Interface Library \\
\hline
23 & POC  &Payload Operation Centre (located at the Indian \\
 &  &Institute of Astrophysics, Bangalore) \\
\hline
24 & PSF & Point Spread Function \\
\hline
25 & RA\_INT & Level-2 Relative Aspect chain for data \\
 &  &collected in Integration (INT) mode \\
\hline
26 & RA\_PC &Level-2 Relative Aspect chain for data \\
 &  &collected in Photon Counting (PC) mode \\
\hline
27 & RAS &Relative Aspect Series (time series of 3 variables \\
 &  &quantifying drift; Roll-Pitch-Yaw / X-Y-$\theta$) \\
\hline
28 & RE &Reference Episode (with longest exposure time) \\
 &  &among the ``group" of Episodes to be combined, \\
 &  &each with identical Band, Filter \& Window \\
 &  &configuration \\
\hline
29 & RF &Reference Frame; defines the instance of time for \\
 &  &an Episode, with respect to which drift corrections \\
 &  &are measured \& applied to frames at future times  \\
\hline
30 & SAA & South Atlantic Anomaly \\
\hline
31 & SC & Spacecraft \\
\hline
32 & SSM & Scanning  X-ray Sky Monitor \\
\hline
33 & UL2P & UVIT Level-2 Pipeline \\
\hline
34 & USNO  &United States Naval Observatory (catalogue of \\
 & &stars) \\
\hline
35 & UTC & Universal Time Clock \\
\hline
36 & UV & Ultra-Violet \\
\hline
37 & UVIT & Ultra-Violet Imaging Telescope \\
\hline
38 & VIS & Visible (VIS) band of UVIT \\
\hline
&{\bf Name / Selectable parameter :} & \\  
\hline
1 & Aux & Auxiliary data (from spacecraft sub-systems) \\
\hline
2 & Episode &Smallest unit of imaging session as per mission \\
 & &operations; corresponding data consist of time \\
 & &series of frames read out at a planned rate \\
\hline
3 & EXP\_TIME &Total effective Exposure Time (seconds) based on \\
 & &no. of frames contributing to the product \\
\hline
4 & Exposure  &2-D Array, its elements hold no. of frames \\
 &  &exposed with valid pixels of the detector \\
\hline
5 & N\_acc &No. of successive VIS frames to be stacked for \\
 &  &generating individual accumulated images, \\
 &  &`Acc\_frame' (from which star are identified for \\
 & &drift tracking; used in RA\_INT chain) \\
\hline
6 & N\_avg &Centroids of detected stars from N\_avg successive \\
 &  &accumulated images (Acc\_frames) are combined \\
 & &to select a wider time bin for drift computation \\
 & (used in RA\_INT chain) \\
\hline
7 & N\_combine &(1) No. of successive NUV /FUV frames to be \\
 & &accumulated for generating time series of images, \\
 & &`Comb\_frame', from which drift is extracted \\
 & &(used in RA\_PC chain); \\
 & &(2)Division of Episode into multiple pseudo- \\
 & &Episodes, each consisting of N\_combine \\
 & &consecutive frames (when ``pseudo-Episode" \\
 & &option is used in L2\_PC chain); \\
\hline
8 & N\_p &No. of packets containing image data in a frame \\
\hline
9 & ObsID &Observation ID identifies an unique sky pointing \\
 & &towards a target \\
\hline
10 & pseudo-Episode &Part of an Episode when divided into multiple \\
 & &data chunks (each with N\_combine frames)   \\
\hline
11 & Sig &2-D array, with elements holding accumulated \\
 &  &ENPs from drift corrected frames \\
\hline
12 & Signal & 2-D array, sky image in ``counts/sec" unit  \\
\hline
13 & Uncertainty &2-D array,  with elements holding statistical \\
 & &uncertainty from no. of photon of events \\
 & &contributing to the pixel (in ``counts/sec" unit) \\
\hline
14 & utcFlag &Switch to select either Universal Time Clock or \\
 &  &UVIT’s Master Clock for timing \\
\hline
\end{tabular}
\end{table*}


\subsection{Functionality}

 This sub-section provides a qualitative description of
 the operations carried out by the pipeline
 for generating the astronomer ready output products.
 This includes the primary aim, various inputs and the key
 processing {\it steps}. 
 The terminology used here are as follows. 
 The term ``Step" has been introduced at the very highest level of description
 for ease of comprehension.
 A ``Chain" refers to a completely 
 stand alone sequence of smaller processing ``Block"-s.
 The functionality of several Blocks used  
 for different imaging modes and/ or Chains are similar, hence they have been 
 implemented as ``Modules" callable with appropriate selection of
 switches. The codes for remaining Blocks are embodied within
 respective Chains. 
 Individual Chains are described in later sections 3.2.1, 3.2.2 \& 3.2.3. 
 While a few brief generic comments about Modules appear in section 3.2.7,
 individual Modules are described in Appendix 1 (Sec A1.1, to A1.20).
 The functionality of Blocks that are not Modules are given within the 
 description of Chains.
 
\subsubsection{Primary Aim of Level-2 Pipeline}

 The UVIT Level-2 Pipeline (UL2P) is designed to translate
 ISSDC /ISRO provided Level-1 data (mL1)
 into science ready products in astronomer friendly format, 
 utilizing instrument calibration database  
 (CAL\_DB) as well as user settings of selectable parameters \& switches (PIL)
 (see Fig 3).
 The primary products include  the two UV sky image arrays as detected photon count rates
 in standard coordinate system (Right Ascension – Declination J2000)
 as well as primitive detector coordinate system (X-Y axes of the sensor)
 for every imaging Episode.
 The corresponding Uncertainty Arrays and sky Exposure Arrays are also 
 generated. The subsidiary products include
 time series of spacecraft drift, photon centroid list for the 
 image with flat field weight, etc.
 In addition to the above products for every Episode of observation,
  similar products are generated,
 by combining data for all the Episodes, for the total exposure (in all the orbits)
 with identical Filter-Window combination.
 The above is achieved through an one time run of a fully automated
 Driver Scheme, using a single configuration file for user selections.

\subsubsection{Input Data}

  There are three types of inputs required for running the pipeline, viz.,
 the bundled Level-1
 data (merged Level-1; mL1), the calibration database (CAL\_DB) and the user settings
 of all selectable parameters and switches (Parameter Interface Library; PIL).

  The mL1 represents collection of all data pertaining to one specific pointing
 of the spacecraft to the astronomical target (referred to by an unique Observation ID, 
 {\it ObsID}).
 It consists of science data from
 all sky exposures executed by UVIT during that pointing as well as auxiliary
 data from the spacecraft systems (e.g. orbit \& attitude information).
 The science data are packed as a sequence of ``frame"-s, one
 for each individual short exposure. Their content depend on the mode of
 operation of the detector, viz., Integration (INT; used for VIS band; $\sim$ 1 frame per sec)
 or Photon Counting (PC; used for NUV \& FUV bands; $\sim$ 29 fps).
 While the raw pixel values of the entire field (512x512) are preserved for INT mode, 
 centroids (computed onboard) of individual detected photons constitute the frame 
 for the PC mode. 
 More details about exact contents of these data are available elsewhere (Appendix-1, Sec. A1.1).

 An {\it ``Episode"} of exposure in any band is
 defined as one continuous imaging session with a specific selection of filter and
 window size.
 Thus, an Episode can either
 last the whole usable dark-side of an orbit or a fraction of it in case
 the settings are changed. 
 There is an option in the pipeline to divide the imaging duration of one Episode
 into multiple equal segments for processing independently. These segments
 are referred to as  ``Pseudo-Episode" individually.
 The nomenclature of grouping of science data
 used here is displayed in Fig. 4.
 The CAL\_DB contains organized datasets representing calibration arrays
 for each band (FUV, NUV \& VIS) quantifying flat field correction across the detector,
 distortions, bad pixels, etc. The complete details of CAL\_DB are presented in
 Appendix-5.

  The PIL is a single configuration file containing numerical values of all
 selectable parameters and ON/OFF settings of various switches. 
 While complete details are available in Appendix-11, a few key parameters
 are also listed in Table-2.

\begin{figure*} [th]
\centering
\includegraphics[scale=0.40]{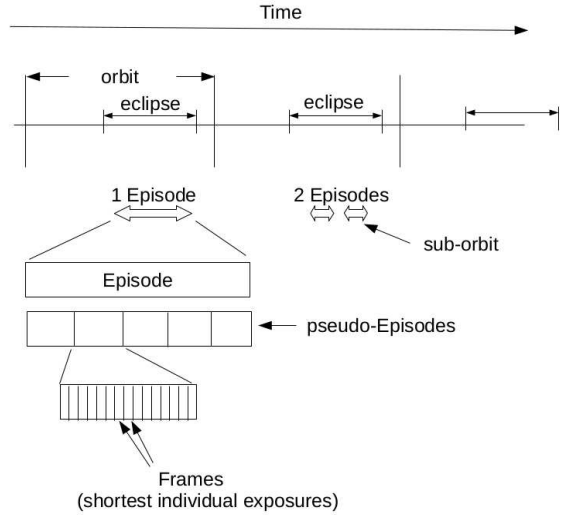}
\caption{
Nomenclature for data groupings.
Each ``orbit" corresponds $\sim$ 100 minutes - the time taken by the 
spacecraft to go around the Earth once. The term ``eclipse" refers
to that part of every orbit when the spacecraft is in the Earth's shadow
blocking the Sun. UVIT observations are mandatorily scheduled during
eclipses for instrument safety. The entire eclipse duration could either be
used for sky exposure with a fixed settings for data collection leading to a single ``Episode"
or the time can be sub-divided into 2 or more exposures (with different
settings) as per scientific needs. The pipeline has an option
to sub-divide the data collected during one Episode into multiple
pseudo-Episodes. The imaging data collection during an Episode involves repeated 
read outs of the detector -  data corresponding to an individual read out
is called a ``Frame". 
}
\end{figure*}

\subsubsection{The Processing  Steps}

 The generation of UV images from the inputs described above,
 involve the following main processing steps which are identical
 for NUV and FUV bands.
 1) At first the drift of spacecraft pointing
 is estimated using stars detected in images captured in VIS frames
 (sampled $\sim$ 1 frame-per-sec, fps). There is an alternate provision for finding drift
 from NUV / FUV band data also in absence of VIS band data.  
 2) Next, individual UV frames 
 are added to generate a cumulative image in the detector coordinates,
 after applying drift corrections and various instrumental
 corrections (which varies across the field of view) to the detected photons
 of each individual frame.
 Appropriate transformation is applied to generate the image in RA-Dec coordinates also.  
 The corresponding Uncertainty to the UV image (from photon statistics) and the
 Exposure on the sky are also generated, which complete the set of image products.
 The above description is applicable for each single imaging session
 (Episode) lasting $\sim$ 2000 sec or less (see Fig. 5). 
 3) Image products from successive imaging sessions with identical combination
 of filter and window size (for the same UV band), are combined to generate the
 products for the total exposure 
 by applying
 corrections for the relative shift and rotation determined using brigher
 UV stars detected in individual Episodes (described in a later sub-section 
  named Driver Scheme, Sec 3.2.4; see Fig. 9).

\begin{figure*} [th]
\centering
\includegraphics[scale=0.30]{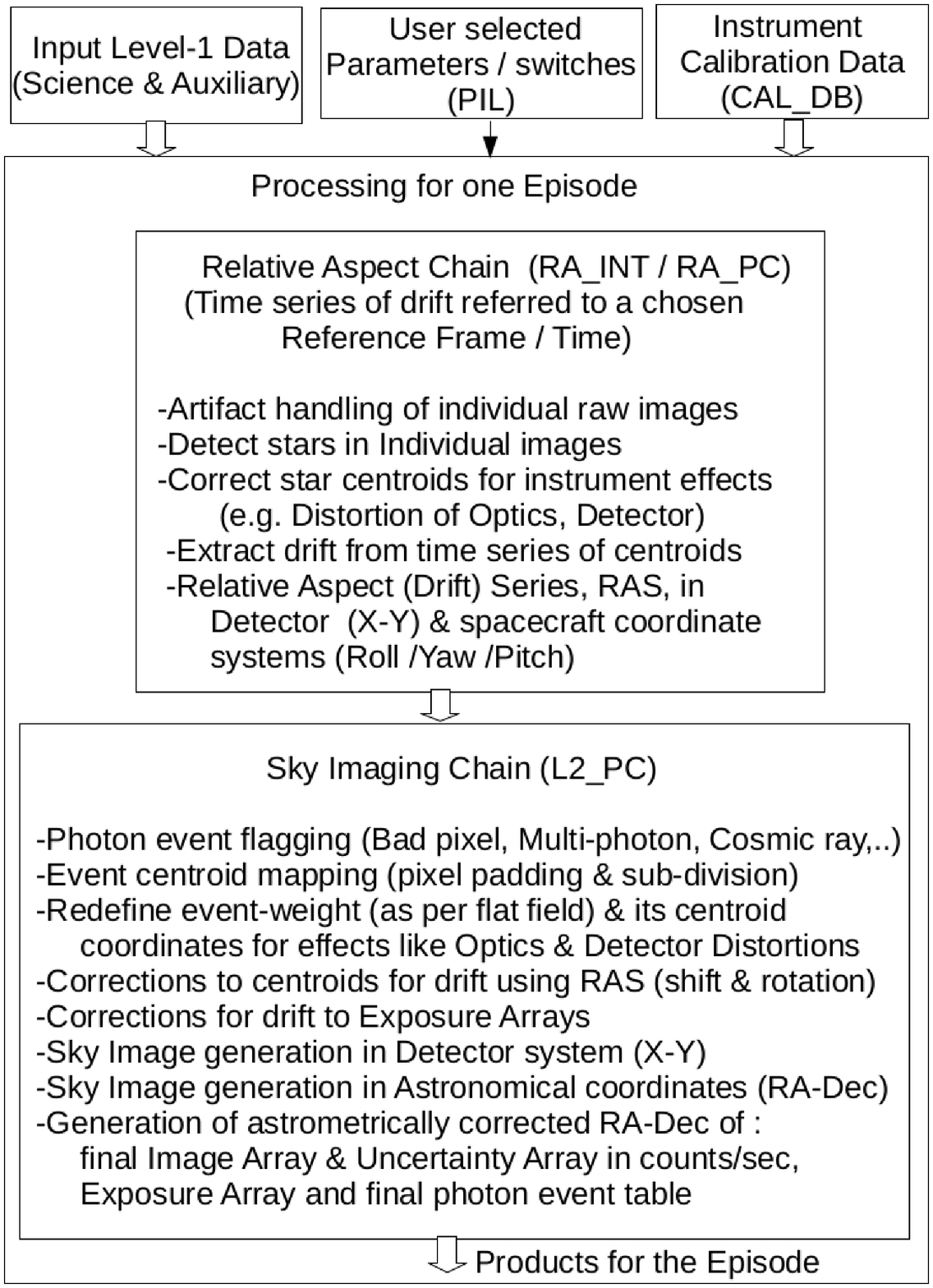}
\caption{
Schematic diagram showing generation of UV sky image corresponding to 
a single uninterrupted observing session called {\it Episode}.
Starting with Level-1 data for the Episode along with UVIT calibration data (CAL\_DB)
\& user  selected options (PIL), two stand alone processing chains
 (Relative Aspect chain, RA\_INT or RA\_PC;
 \& L2\_PC) generate the output products.
 Actions in the first block (Relative Aspect chain to extract drift of the spacecraft)
 include : correcting for artifacts \& detection of stars in raw VIS images
  for RA\_INT (\& similarly from UV images generated by stacking photon centroids,
 in case of RA\_PC)
 applying corrections to star centroids for instrument effects (distortions),
 determine drift (shift \& rotation) as a function of time (from centroids)
 in Detector coordinates and finally convert to spacecraft coordinates.  
 The second block (Sky Imaging chain to generate UV sky image \& other 
 products) begins with input UV data (list of centroids of photon events),
 flags affected events for rejection, computes event weight to include
 flat field, applies corrections to centroids for various systematic effects
 of the instrument \& drift (applicable for corresponding time instance)
 determined in the first block, to obtain true sky arrival directions of individual
 UV photons. The UV sky image is generated by accumulating individual photons
 in both the Detector and Astronomical coordinates directly. Corresponding 
 Uncertainty (from photon statistics) as well as sky Exposure maps are also
 generated.  
}
\end{figure*}

 The first two  steps, \#1 \& \#2  above are stand alone processing
 chains named - (i) the  Relative Aspect chain which generates
 a time series of drift relative to position at some reference time
 as defined by the corresponding Reference Frame;    
 (ii) the  Sky Image chain which
 generates astronomical images \& associated products.
 As the exposures can be taken either in
 photon counting (PC) mode or in integration (INT) mode, each of the
 chains has two versions (accordingly the chains are designated as :
 RA\_INT / RA\_PC \& L2\_PC / L2\_INT).
 The last  step (\#3) is called Driver Scheme which combines images \& associated
 products (Exposure \& Uncertainty Arrays) obtained from all the Episodes,
 corresponding to the same combination of filter and window, after
 aligning them to correct for relative shift and rotation.

 The key
 functionalities of these chains and the Driver Scheme are summarized
  along with their implementation 
 in
 the  next section (Sec. 3.2).
 The chains in turn call various processing
 Modules, whose technical descriptions appear in Appendix-1.

 \subsection{Implementation}

  This subsection provides various details of implementation of the
 pipeline. 
 First the processing chains RA\_INT, RA\_PC \& L2\_PC and
 the Driver Scheme are summarized,
 followed by the strategy for addressing the situation of data loss
 in the input bundle as well as handling of special situations. 
 Finally general concepts behind the design of modules are presented
 (details of individual modules can be found in Appendix-1).  

  The figures displaying the processing blocks of various chains follow
 a convention to distinguish Modules from rest by shading the latter
 (Fig.s 6, 7 \& 8). In addition the blocks which have rarely been
 turned on are identified in the respective captions for the figure.

\medskip
 \subsubsection{Relative Aspect chain for Integration mode {\it (RA\_INT) :}}

  The purpose of this chain is to obtain time series of instantaneous
 shift and rotation of the field, relative to the position at any chosen
 time or set of frames, every second or so. This chain operates on VIS
 band data which are collected in INT mode. The typical accuracy obtained
 is $\sim$ 0.1${}^{\prime\prime}$ for the full field. The chain operates
 on data of one Episode at a time. Various steps of the processing are
 shown in Fig. 6 (RA\_INT) and are described below in the sequence in
 which they are executed and the numbers in parenthesis refer to the
 block numbers in that figure.

\begin{figure*} [th]
\centering
\includegraphics[scale=0.40]{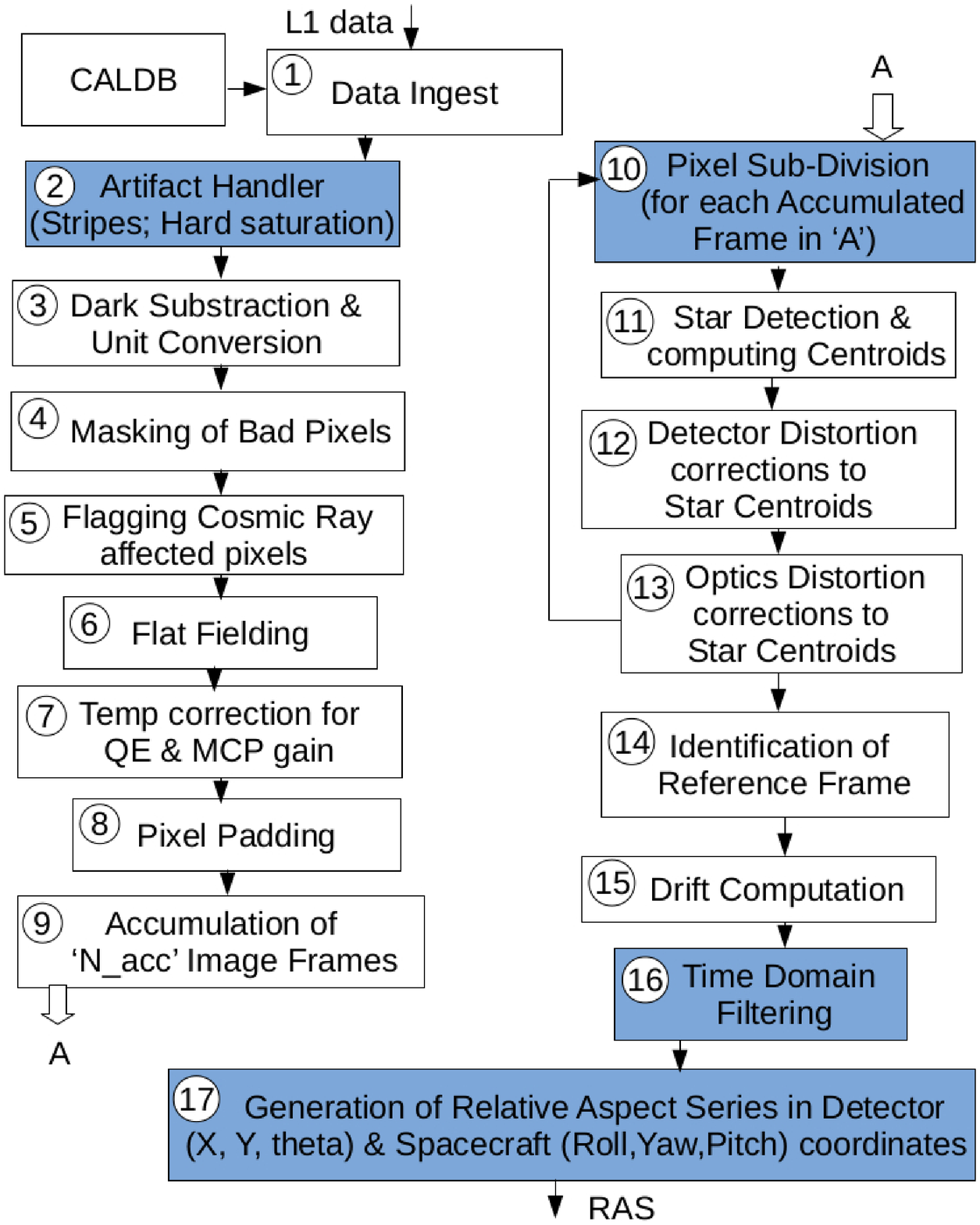}
\caption{
Schematic block diagram of the Relative Aspect
chain for Integration mode (RA\_INT). The block numbers 7 and 10 have
 never been turned on (except for testing \& validation).
 The manual mode of star selection is not shown.
 The shaded blocks do not belong to any Module.
}
\end{figure*}

   {\it Data Ingest (1)} : Import of all the user selected switch settings
 and values from a parameter file. Relevant data on calibration and on
 the exposure are taken from CAL\_DB and mL1 respectively. These
 data are: a) from calibration: the tables /arrays of bad pixels,
 distortions, flat-field, dark \& exposure template; b) from mL1:
 i) science data originating from the instrument: frames of INT mode
 images in selected band, and ii) auxiliary data from the S/C: aspect
 of S/C, time calibration, house keeping information on instrument
 health, good-time /bad-time flags. Handling of discontinuities and
 abnormal data. Removal of engineering grade science data collected
 during gradual ramping up of the highvoltages at the beginning of
 the Episode, as per safety protocol. Extraction of individual image
 frames.

  {\it Artifact Handler (2)} : The unexpected artifacts in the image
 and saturated pixels are flagged here enabling further processing to
 ignore them.

 {\it Dark Subtraction \& Unit Conversion (3)} : Subtraction
 of `dark' frame from each individual image frame followed by conversion
 of the signal values from ``count per exposure" to ``count per second"
 unit.

 {\it Masking of Bad Pixels (4)} : Flagging of the locations corresponding
 to `Bad Pixel'-s in every image frame.

 {\it Flagging Cosmic Ray affected pixels (5)} : Identification of
 Cosmic Ray affected pixels (in every image frame) and their flagging.

 {\it Flat Fielding (6)} : Multiplicative correction for non-uniform
 response across the field applied to all image frames.

 {\it Pixel Padding (8)} : Expansion of all images from 512$\times$512
 to 600$\times$600 size by symmetrically adding pixels along all four
 sides populated with a flag. This enlargement allows accommodation of
 movement of the instantaneous sky field due to spacecraft drift over
 the duration of the Episode. While this functionality is not explicitly
 required in this Chain, the corresponding Module (\& also the following
 processing steps) being common with the Sky Imaging Chain, this step
 is retained to keep the array dimensions compatible.

 {\it Image Accumulation (9)} : Selection of time step for drift
 computation by accumulating and averaging selected number, N\_acc,
 of successive image frames generating a series of accumulated frames,
 Acc\_frame.

 {\it Detection of stars \& finding Centroids (11)} : Analysis of
 individual image Acc\_frames to identify stars using an algorithm which
 uses dynamic threshold, distribution of light around local maxima and
 neighbourhood criteria. Computation of X \& Y centroids for every
 detected star and intensity ordered tabulation.

 {\it Correction for Detector Distortion (12)} : Application of additive
 corrections to the centroids of stars for the detector distortion.

 {\it Correction for Optics Distortion (13)} : Application of additive
 corrections to the centroids of stars for the distortion due to telescope
 optics.

 {\it Identification of Reference Frame \& Binning of time step (14)} :
 Selection of timing reference for drift computation process, by identifying
 a particular Acc\_frame to be the Reference Frame, RF (by skipping the
 initial N\_skip number of Acc\_frame-s). Beginning with the RF and till the end of the
 Episode, an additional level of binning of time step is carried out
 optionally, by averaging star centroids exytracted from N\_avg number of
 successive Acc\_frame-s.
 (For default usage, N\_avg = 1).

 {\it Drift Computation (15)} : Computation of time series of relative
 drifts (shifts along X \& Y axes and a rotation about the central pixel)
 from the differences in centroids of matching stars tabulated for
 successive time samples.
 The centroids for multiple stars are used in a least square fitting
 scheme having a choice of either equal or intensity based weighting.
 A manual mode of inter active selection of stars is also available to
 mitigate rare situations arising due to residual artifacts in the
 image data unaddressed by the handler block (\#2).
 Subsequent translation of these drifts for every
 time sample to the integral drifts, with respect to the timing reference
 corresponding to the Reference Frame.

 {\it Time Domain Filtering (16)} : The time series of drift in three
 variables ($x_{shift}$, $y_{shift}$ \& $\theta$) are low-pass filtered
 in time domain, with user selectable parameters for smoothening them.

 {\it Generation of Relative Aspect Series (17)} : The time series of
 drifts are tabulated as Relative Aspect Series, RAS, in Detector ($X$, $Y$,
 $\theta$) as well as Spacecraft (Roll,Yaw,Pitch) coordinates systems
 as finalproduct from this RA\_INT chain.

 The Appendix-8 presents the directory structure of final products from
 the RA\_INT chain.

 \subsubsection{ Relative Aspect chain for Photon Counting mode {\it (RA\_PC) :}}

   The aim of the RA\_PC chain is identical to that for the RA\_INT chain,
 but its functional details are different since it operates on the data
 collected in Photon Counting mode (e.g. when NUV or FUV data is used to
 obtain the time series of drift, Relative Aspect Series). This chain also
 operates on data of one Episode at a time. Various steps of the
 processing are shown in Fig. 7 (RA\_PC) and are described below in the
 sequence in which they are executed.

\begin{figure*} [th]
\centering
\includegraphics[scale=0.40]{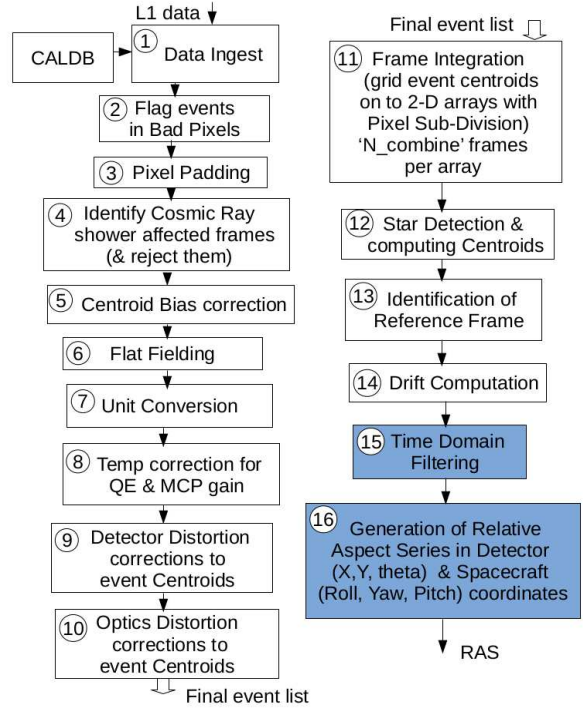}
\caption{
Schematic block diagram of the Relative Aspect
chain for Photon Counting mode (RA\_PC)
The block number 8 has
 never been turned on (except for testing \& validation).
 The shaded blocks do not belong to any Module.
}
\end{figure*}

{\it Data Ingest (1)} : Similar to the case of RA\_INT (see above), except
 that the science data originating from selected band of the instrument
 operated in PC mode consist of a table with details of individual photon
 events detected in successive exposed frames in the Episode. These details
 (e.g. centroids) are generated from the raw frames, through on board
 processing. In this step, a master table of photon events along with
 relevant data is created.

 {\it Flagging of events in Bad Pixels (2)} : A column is appended to the
 master table which holds the Bad Pixel flag. The events with their X \& Y
 centroid coordinates overlapping with any of the bad pixels are flagged.
 Later processing steps ignores the flagged events.

 {\it Pixel Padding (3)} : The centroid coordinates are modified to translate
 their range from (1 to 512) to (1 to 600) along each axis thereby effectively
 incorporating symmetric padding along all four sides. This allows
 accommodation of movement of the instantaneous sky field due to spacecraft
 drift over the duration of the Episode. Here again, this step is retained
 for compatibility of the shared Modules between the Relative Aspect \&
 Sky Imaging Chains.

 {\it Rejection of Cosmic Ray affected frames (4)} : Identification of
 individual frames affected by showers due to Cosmic Ray events and discarding
 all events in them. The process uses statistical distribution of number
 of events in all frames of the Episode and user selected parameters to
 arrive at the threshold value for flagging. This rejection is useful for
 dark fields where contribution of showers to the background of photon events
 can be a large fraction of the total. If the rejection is not desired, the
 parameters can be set accordingly.

 {\it Centroid Bias correction (5)} : Application of corrections for
 systematic bias, due to use of a simple algorithm on board, to the centroid
 values.

 {\it Flat Fielding (6)} : The `Effective Number of Photon' (ENP) for every
 event is changed from its initial value of unity by application of
 multiplicative correction for non-uniform response across the field.

 {\it Unit Conversion (7)} : The entries for all events are modified from
 `ENP per frame' to `ENP per second' unit.

 {\it Correction for Detector Distortion (9)} : Application of additive
 corrections to the centroids of photon events for the detector distortion.

 {\it Correction for Optics Distortion (10)} : Application of additive
 corrections to the centroids of photon events for the distortion due to
 telescope optics.

{\it Frame Integration (11)} : Transformation of centroid coordinates for
 all photons from (600, 600) range to (4800, 4800) to implement pixel
 sub-division. Generation of a time series of two dimensional sky image
 arrays, Comb\_frame (4800$\times$4800), by accumulating photon events
 from selected number of consecutive frames (N\_combine). The functionality
 of this selectable parameter N\_combine for this RA\_PC chain is
 equivalent to the parameter N\_acc for RA\_INT chain. The distinct
 nomenclature is chosen to emphasize the large difference in their typical
 values (N\_acc $\sim$ 1; N\_combine $\sim$ 30).

  At this stage of processing of the RA\_PC chain, 2-D sky images are
 available starting from PC modedata. Accordingly, the subsequent steps
 leading to extraction of drift are very similar to the corresponding
 steps of RA\_INT.

 {\it Detection of stars \& finding Centroids (12)} : Similar to that for
 RA\_INT chain (step \#11) above.

 {\it Identification of Reference Frame \& Binning of time step (13)} : Similar
 to that for RA\_INT chain (step \#14) above.

 {\it Drift Computation (14)} : Similar to that for RA\_INT chain
 (step \#15) above.

 {\it Time Domain Filtering (15)} : Similar to that for RA\_INT chain
 (step \#16) above.

 {\it Generation of Relative Aspect Series (16)} : Similar to that for
 RA\_INT chain (step \#17) above.

\medskip
 \subsubsection{ Sky Image chain for Photon Counting mode {\it (L2\_PC) :}}

 The Sky Image chains L2\_PC and L2\_INT generate astronomical images
 for observations carried out in Photon Counting and Integration
 mode respectively .
 Primary aim of UVIT is to image in NUV and FUV bands, both
 of which are observed in the PC mode. Hence, the functionality of
 only the L2\_PC
 chain is presented here, which is the more
 significant among the two versions.

 The L2\_PC chain processes data from one Episode
 and one band (NUV / FUV) at each instance. The Fig. 8 displays the
 schematic block diagram for this chain. Its steps are described below
 in the sequence of their execution. Certain steps here are similar to
 corresponding steps in the RA\_PC chain described earlier.

\begin{figure*} [th]
\centering
\includegraphics[scale=0.40]{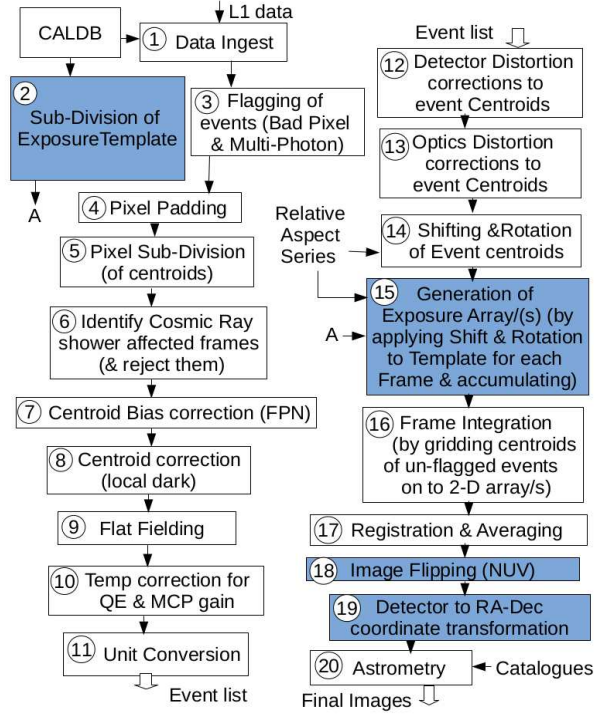}
\caption{
Schematic block diagram of the Sky Imaging
chain for Photon Counting mode (L2\_PC)
The block numbers 8 and 10 have
 never been turned on (except for testing \& validation).
 The shaded blocks do not belong to any Module.
}
\end{figure*}

{\it Data Ingest (1)} : Importing selected switch settings and parameters,
 calibration and exposure data, spacecraft aspect, time calibration, etc.
 as well as science data (see step \#1 of RA\_INT \& RA\_PC for details).
 Generation a master table of all photons along with their details for
 the entire Episode.

 {\it Creation of Exposure Template (2)} :
    The master templates of sky exposure are created for every combination of band, mode
 and window size by carrying out pixel sub-division of corresponding arrays
 available from the CAL\_DB. The resulting sky Exposure Template arrays are of
 4800$\times$4800 size.
In order to
 track true exposure of the sky on the detector, the relevant master array will be
 offset according to drift corrections applicable for the time instances and
 stacked together in a later processing block.

 {\it Flagging of events in Bad Pixels; \& with Multi-Photon signature (3)} :
 Flagging of events with centroids located within a Bad Pixel (similar to
 step \#2 of RA\_PC). Identification of potential multi-photon cases from
 event details along with user selected parameters and flagging them- this
 selection has never been used.

 {\it Pixel Padding (4)} : Modification of the centroid coordinates of
 photon events by translating their range from (1 to 512) to (1 to 600)
 to accommodate spacecraft drift (similar to step \#3 of RA\_PC).

 {\it Pixel Sub-Division of centroids (5)} : Transformation of centroid
 coordinates for all photons from (600, 600) range to (4800, 4800) by
 pixel sub-division.

 {\it Rejection of Cosmic Ray affected frames (6)} : Identification and
 rejection of frames affected by Cosmic Ray showers (similar to step \#4
 of RA\_PC).

 {\it Centroid Bias correction (7)} : Application of corrections for
 systematic bias to the centroid values (similar to step \#5 of RA\_PC).

 {\it Flat Fielding (9)} : Application of multiplicative correction for
 non-uniform response across the field (similar to step \#6 of RA\_PC).

 {\it Unit Conversion (11)} : The entries for all events are modified from
 `ENP per frame' to `ENP per second' unit (similar to step \#7 of RA\_PC).

 {\it Correction for Detector Distortion (12)} : Application of additive
 corrections to the centroids of photonevents for the detector distortion
 (similar to step \#9 of RA\_PC).

 {\it Correction for Optics Distortion (13)} : Application of additive
 corrections to the centroids of photon events for the distortion due to
 telescope optics (similar to step \#10 of RA\_PC).

 {\it Corrections for drift (14)} : Extraction of corrections for spacecraft
 drift applicable to individual frames by time interpolation of the
 corresponding Relative Aspect Series, RAS. The RAS should be available
 from a prior run of the Relative Aspect chain (RA\_INT or RA\_PC) covering
 the time duration of the Episode. Application of drift correction (involving
 2 shifts \& a rotation) to centroids of the individual photons.

{\it Generation of drift corrected Exposure Array/(s) (15)} : Application
 of drift corrections to a time series of Exposure Template (generated in
 step \#2 earlier) arrays, one each corresponding to every frame of
 the Episode or pseudo-Episode, and their accumulation leading to
 `Exposure' array/(s).

 {\it Frame Integration (16)} : Consideration of the entire Episode or
 division into multiple pseudo-Episodes each consisting of N\_combine
 consecutive frames, Comb\_frame (after discarding initial N\_discard
 frames). Generation of 2-D UV sky image/(s) by gridding every un-flagged
 photon of the Episode / (pseudo-Episode) onto 4800$\times$4800 `Signal'
 array/(s) as per its (X, Y) centroids in the detector coordinate system.
 Generation of corresponding statistical `Uncertainty' array/(s),
{ which ideally should be based on counting of photon events in each pixel
 but here it has been computed from the `Signal' \& corresponding
 `Exposure' arrays. This approximation leads to a systematic error in the
  `Uncertainty' due to variation of the Detector's response across
  the field, which is $<$10\% for central 24$^{\prime}$ diameter circle of the
   field  (see Tandon et al 2017c, 2020).
}

 {\it Registration \& Averaging (17)} : Combination of multiple `Signal',
 `Exposure' \& `Uncertainty' arrays from pseudo-Episodes after aligning
 them for any relative shifts and rotation determined using brightest
 point sources detected in `Signal' arrays. Conversion to physical units
 for entries in the final combined set of `Signal' (count/second),
 `Exposure' (second) \& `Uncertainty' (count/ second) arrays. All these
 arrays are of 4800$\times$4800 size in the detector (X, Y) coordinate
 system.

 It is noteworthy that this block is effective only if `pseudo-Episode'
 option has been selected, in which the data from an Episode is divided
 into parts of selected size (default configuration of the pipeline does
 not exercise this option). The process of combining images involves
 transformation of individual sub-pixels (at 9600$\times$9600, followed
 by 2x2 binning), unlike the use of individual photon centroids while
 combining multiple Episodes in the Driver Scheme described later (Sec. 3.2.4).

 {\it Image Flipping (NUV) (18)} : Flipping the set of image arrays -
 `Signal', `Exposure' \& `Uncertainty', for the NUV band only, about X-axis
 to undo the effect of the folding plane mirror (see Fig. 1).

 {\it Transformation to astronomical RA-Dec coordinates (19)} : Conversion
 from detector to astronomical coordinate system Right Ascension, Declination
 (ICRS J2000) using attitude information of the spacecraft, for all 3 image
 arrays. The UV image (`Signal') is re-generated by coordinate transformation
 of individual photon centroids to minimize any loss of angular resolution
 (in case pseudo-Episode option has not been exercised; else image
 sub-pixels are transformed).
{ At this stage, the outermost regions of the nearly circular field
is truncated based on sky exposure less than 10\% of the peak exposure.
The process of this truncation for all the three image arrays, viz.,
 `Signal', `Exposure' \& `Uncertainty' are carried out consistently.}

 {\it Astrometry using optical catalogue (20)} : Correlation of bright stars
 detected in the UV image with USNO A2 optical star catalogue to determine
 astrometric corrections (shifts and rotation). Application of these
 corrections to all the 3 arrays, viz., `Signal', `Exposure' \& `Uncertainty',
 with improved absolute aspect. The probability of success for the astrometry
 step in this chain (for an individual Episode) is lower than that for the
 similar operation on the multi-Episode combined image in Driver
 Scheme (Section 3.2.4) with higher total sky exposure.

 The Appendix-9 presents the directory structure of final products from the
 L2\_PC chain.

\medskip
 \subsubsection{ Driver Scheme}

  The Driver Scheme operates at the highest level of hierarchy of the data
 processing in the pipeline. It handles the entire merged Level-1 dataset (`mL1')
 at one go and carries out all necessary processing to finally generate the
 full set of final deliverable products for all 3 bands, for different pairs
 of filters and window sizes used in each band. Often to achieve a long
 exposure under identical configuration, the observations are carried out
 as a series of Episodes. The observations with same combination of filter
 \& window size need to be segregated together to generate a single set of
 products consolidating the total integration time planned.
 The method employed to combine Episodes requires that individual Episodes are
 considered as single data chunks (\& not sub-divided into pseudo-Episodes).
 This requires
 a few processing steps in addition to those handling individual Episodes,
 which are also executed in a systematic manner by the Driver Scheme. The Driver
 Scheme internally calls the Relative Aspect and Sky Imaging chains (described
 in Sec. 3.2.1 / 3.2.2 \& 3.2.3) multiple times as required for any specific
 mL1 dataset. The Fig. 9 displays the functional details of operations of
 the Driver Scheme. The top level directories for the final products from this
 Driver Scheme along with additional information regarding next level
 contents corresponding to the multi-Episode ``group"-s are presented in
 Appendix-10. Given a merged dataset (`mL1'), the Driver Scheme identifies
 science data segments for the 3 bands which correspond to the same Episode
 (from time stamps). It is pertinent to note the role Reference Frame
 (described earlier in step \#14 of RA\_INT; Sec. 3.2.1) plays in
 connecting data from the band used to extract the drift, usually VIS,
 but NUV in absence of VIS, and the band in which sky images are to be
 generated (NUV / FUV). It provides the critical timing reference for
 calculating drifts for each Frame of the UV bands from the time series
 of drifts generated by Relative Aspect chains.

\begin{figure*} [th]
\centering
\includegraphics[scale=0.35]{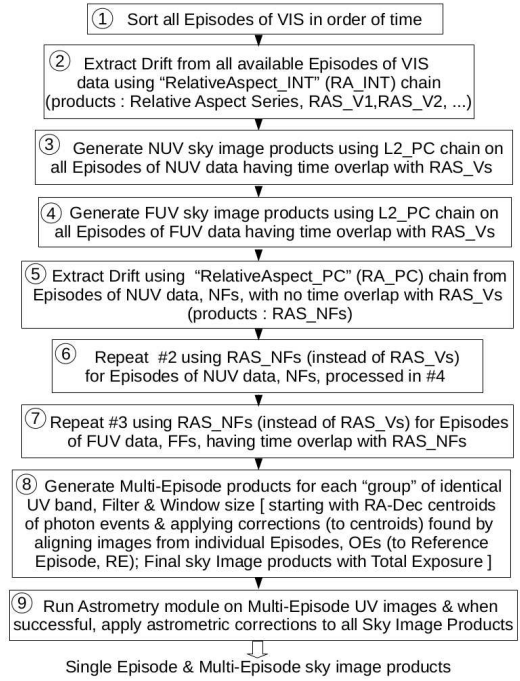}
\caption{
Schematic diagram showing sequence of steps of Driver Scheme (Default case)
which operates autonomously on the entire dataset for a specific ObsID (mL1;
consists of many Episodes)
and generates all requisite products in one go.
An {\it Episode} represents the dataset from a single imaging session
whereas collection of all imaging sessions during that particular pointing
of the spacecraft to the target constitutes an ObsID.
The two processing chains (each operating for a single Episode at a time),
viz., Relative Aspect (RA\_INT /RA\_PC) and
Sky Imaging (L2\_PC) are used repeatatively. The former derives the spacecraft
drift relative to a chosen reference direction. The latter generates
UV images from detected photons after applying corrections for drift
and instrument systematics. The Relative Aspect chain, RA\_INT, uses VIS data
and in its absence, the RA\_PC version for NUV/ FUV data.
Sky Image Products (UV, Uncertainty \& Exposure arrays)
corresponding to multiple Episodes with identical filter
 \& window size are generated by combining data from individual Episodes.
}
\end{figure*}

  {\it Default configuration :}

  1) Identification of time ordered sequence of available Episodes of
 VIS band data (say, V1, V2, ... Vn).

  2) Running of Relative Aspect chain
 for Integration mode, RA\_INT, using VIS band for tracking on each of
 these individual Episodes to generate corresponding drift series (RAS\_V,
 Relative Aspect Series from VIS). Label these drift series as RAS\_V1,
 RAS\_V2, ... RAS\_Vn and the corresponding time ranges to be TR1, TR2,
 .... TRn. All these RAS\_V products are available in individual
 sub-directories corresponding to each Episode, inside the directory ``/uvit"
 (see Appendix-10).

 3) Identification of data sets for individual Episodes of NUV band and
 any combination of Filter \& Window size (N1, N2, .... Np), which have
 time overlap with any of the time ranges (TRi : i=1,2, ...n) for which
 RAS\_Vs are available from the previous step. Running of Sky Imaging
 chain L2\_PC on each of these `p' Episodes of NUV data. Flagging of the
 remaining Episodes of NUV band data (if any) which do not have time
 overlap with any of the time ranges (TRi : i=1,2, ...n), and label
 them as NF1, NF2, ... NFm. Products for individual Episodes are stored
 under unique directories (``/\_NUV\_1" etc; Appendix-10).

 4) Similar to previous step \#2 but for FUV band data. Identification
 of individual Episodes of FUV band and any combination of Filter \& Window
 size (F1, F2, .... Fq), which have time overlap with any of the time
 ranges (TRi : i=1,2, ...n) for which RAS\_Vs are available from the
 step \#1. Running of Sky Imaging chain L2\_PC on these `q' Episodes of
 FUV data. Flagging of the remaining Episodes of FUV band data (if any)
 which do not have time overlap with any of the time ranges (TRi : i=1,2,
 ...n), and label them as FF1, FF2, ... FFk. Products for individual
 Episodes are stored under unique directories (``/\_FUV\_1" etc; Appendix-10).

 5) Execution of Relative Aspect chain for Photon Counting mode, RA\_PC,
 using NUV band for tracking on each of the flagged Episodes of NUV band
 data NF1, NF2, ... NFm (identified in step \#2) to generate corresponding
 drift series RAS\_NF1, RAS\_NF2, ... RAS\_NFm. Label the corresponding
 time ranges of these RAS\_NFs to be TR\_NF1, TR\_NF2, .... TR\_NFm.
 All these RAS\_NF products are available in individual sub-directories
 corresponding to each Episode, inside the directory ``/\_RAPC" (see Appendix-10).

6) Execution of Sky Imaging chain L2\_PC on each of the flagged Episodes
 (total `m') of NUV band data NF1, NF2, ... using the corresponding drift
 series RAS\_NF1, RAS\_NF2, ... (generated in step \#4). Products from this
 step are stored in additional directories in continuation with those generated at
 the step \#2 above.

 7) Execution of Sky Imaging chain L2\_PC on those flagged Episodes of FUV
 band data among FF1, FF2, ... FFk, which have time overlap with any of
 the time ranges TR\_NF1, TR\_NF2, .... TR\_NFm using the drift series
 generated using NUV data in step \#4, viz., RAS\_NF1, RAS\_NF2, ... RAS\_NFm.
 Products from this step are stored in additional directories in continuation
 with those generated at the step \#3 above.

 8) After the executions of the Sky Imaging chain has been completed for
 all Episodes (completing all the 7 steps described above), their products
 are segregated for identical combinations of the 3 key parameters, viz.,
 UV band, filter \& window size, into ``group"-s. A process attempts to
 combine all images in RA-Dec (ICRS) system from individual Episodes
 (with no subdivision into pseudo-Episodes)
 belonging to a particular group using the following steps : (i) Identify
 the Reference Episode, RE, with largest sky exposure, EXP\_TIME; (ii)
 Order the remaining `Other Episode'-s (OE\_1, OE\_2, ... OE\_n) in
 descending order of EXP\_TIME; (iii) Attempt to match brightest stars
 (avoiding outer annular region with sky exposure $<$ 20\% of peak)
 from RE with brightest stars from other orbits (OE\_1, OE\_2, ...) one
 by one. Let `k' cases be successful (k $\le$ n); (iv) Centroid coordinates
 of individual bright stars are used to determine relative Shifts \&
 Rotation between each pair of (RE \& OE\_j; j : 1, 2, ... k); (v) Application
 of identical shifts \& rotation corrections to the centroids of to all
 photon events of the particular individual Episode (``OE\_j"); (vi)
 Application of shifts \& rotation corrections to the corresponding
 Exposure arrays, which involve transformation of individual sub-pixels
 at 9600$\times$9600 level and re-gridding. Generation of a combined
 Exposure array by stacking all `k' components; (vii) Populating photon
 events into `aligned\_to\_RE' arrays and convertion to combined Signal
 array (``count / second") by pixel by pixel arithmetic operations
 including corresponding aligned and combined Exposure array;
 { One important caveat deserves mention here : while the Exposure
arrays from individual Episodes carry the 10\% cut, the table of
photons which populate `aligned\_to\_RE' arrays do not. This leads
to significantly erroneous values for Signal at the outer regions.
For accurate photometry of stars /targets in such affected outer regions
 the three arrays, i.e.
 Signal array, Exposure array and Uncertainty array,
 from each Episode should be added with the shifts and
 rotations as applicable.}
 (viii)
 Generation of the combined Uncertainty array. All products resulting
 from combining of multiple Episodes record in their headers a log
 identifying the Episode selected as RE, complete list of all the OE-s
 as well as the Episodes which contributed to these multi-Episode products.
 It is noteworthy that the final combined multi-Episode UV image (Signal
 array) is generated directly from transformed centroids of individual
 photons, just like for individual Episodes, thereby retaining their
 angular precision despite undergoing alignment operations involving
 rotation. However, the generation of Exposure and Uncertainty arrays
 do face some loss of precision, even though their impact is significantly
 reduced by employing pixel sub-division as described above. The process
 of combining observations from multiple Episodes includes a check on the
 difference in the Roll angle values (which is a slowly varying function
 of time) between the Reference Episode \& each of the Other Episodes
 in order to mitigate their occasional incorrect entries for attitude noticed in
 the L1 data. This difference is required to be less than 2 degrees for
 the Episode to be considered for the combining step.

 Products generated by this step, corresponding to individual ``group"
 of multi-Episodes are stored under a directory name uniquely identifying
 that ``group" (e.g. ``/NUV\_Final\_F1\_W511"; Appendix-10).

 9) Execution of Astrometry Module to determine finer corrections \& apply
 them, to the sky coordinates of the multi-Episode combined image products
 (Signal, Exposure \& Uncertainty arrays) for each ``group" generated in
 the previous step. Astrometric corrections are determined by correlating stars
 identified from the Signal array with standard astronomical catalogues
 of stars. If successful, identical corrections (involving shifts and
 rotation of arrays through pixel sub-division, transformation \& binning
 scheme) are applied to each of the set of 3 arrays. For more technical
 details about this Astrometry Module see Appendix-1 (Sec. A1.20).
 Products generated by this final step for astrometry, for every ``group"
 of multi-Episodes are stored under an unique directory (e.g.
 ``/NUV\_FullFrameAst\_F1\_W511"; Appendix-10).

 {\it Forced NUV tracking configuration :}

 There is an additional option in the Driver Scheme to completely ignore
 the VIS datasets \& use only NUV data for generating the drift series
 (RAS) \& then make UV images of the sky (both NUV \& FUV). In absence
 of NUV data, the FUV itself can also be used to generate RAS followed by making
 FUV images of the sky. The Fig. 10 shows possible cases arising due to
 different kinds of time overlaps between VIS \& UV band Level-1 data.
 Normal situation as per observation planning is case `A', which is ideal
 and handled using VIS data for tracking. However, sometimes other cases
 with partial (`B' / `C') or no overlap (`D') are also noticed which occur
 due to details of data flow up to the generation of Level-1 data. While the
 `default' setting of the Driver Scheme (tracking by VIS when any amount
 of time overlap exists, \& force UV tracking when none exists) handles
 the cases A \& D optimally, the remaining cases (B \& C) lead to sub-optimal
 usage of data (limited the overlap part only). This is mitigated by forcing
 drift tracking with UV data on all the data (ignoring VIS completely).
 Of course, success with UV tracking critically depends on existence of at
 least one UV bright star within the field of view. Accordingly, the strategy
 followed at the UVIT-POC is to execute the Driver Scheme twice, once each
 with the two options (default \& forced NUV tracking) and provide two
 sets of Level-2 products generated by them, allowing users the choice.

\begin{figure*} [th]
\centering
\includegraphics[scale=0.40]{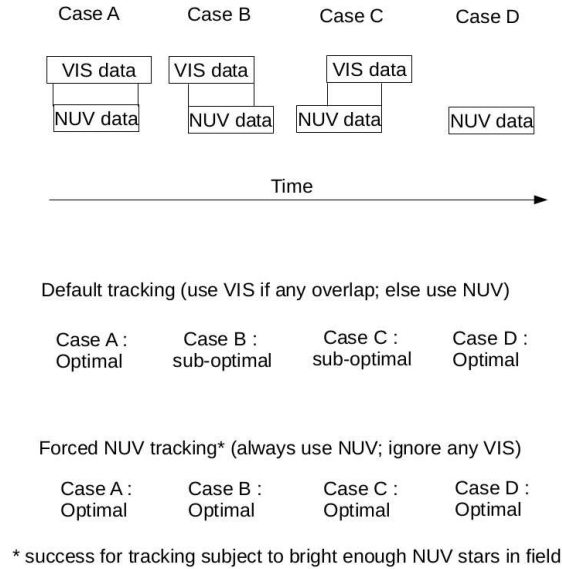}
\caption{
The extraction of spacecraft Drift using VIS \& NUV data under different
situations of time overlap between sky exposures of VIS and NUV bands.
The logic used under ``Default" settings of the Driver Scheme, viz., use VIS when any
time overlap with UV imaging exists, leads to sub-optimal results
for Cases `B' \& `C'. This is avoided with ``forced" NUV tracking.
Accordingly, pipeline products generated using both these strategies
 are provided for the users (when NUV data are available).
}
\end{figure*}

\subsubsection{Strategy for missing data}

 In the design of the pipeline, special emphasis has been given to incorporate
 graceful degradation in view of occasional loss of data at various stages
 of communication between UVIT and ground station as well as errors encountered
 in the input products from Level-1. For example – (a) in case of incomplete
 extraction of drift (e.g. due to large chunk of missing data), images are
 generated using products for whatever duration they could be salvaged);
 (b) in complete absence of VIS data for any Episode, the drift extraction
 is automatically attempted using the NUV data itself; (c) the scheme for
 combining multiple Episodes, employs a logic which leads to best possible
 outcomes in the event of sub-optimal signal to noise situation; (d) crash
 situations of individual processes / chains do not hamper the pipeline from
 full exploration of the entire data set. The pipeline attempts to maximize the sky
 exposure translating to scientifically useful products in spite of the
 limitations in the input data and artifacts therein. The pipeline has been
 implemented primarily in C++ with total $\sim$ 30,000 lines of code.

 \subsubsection{Handling special situation}

 An Episode of observation has been defined earlier as an uninterrupted
 period of data collection. The input Level-1 science data for an Episode
 consisting of time sequence of image frames however often show multiple
 types of anomalies – e.g. missing part / full frames, discontinuities in
 frame number, frame time \& other artifacts. Most such issues have been
 addressed by gradually improved algorithms incorporated in the Data
 Ingest block of the pipeline during early in orbit Performance Verification
 phase of ASTROSAT \& UVIT (see Appendix-1, Sec. A1.1 for details).
 The term pseudo-Episode has also been introduced as the user chooses to
 divide the data from a normal Episode into multiple parts, Comb\_frames
 (N\_combine frames each), to technically address possible misalignments
 at shorter time scales. For example, when the temporal evolution of
 spacecraft drift in a particular Episode includes discrete large and
 fast drifts components, division into pseudo-Episodes is useful.
 Another example of forced use of pseudo-Episode due to a different reason
 follows. Under some specific situation arising out of design of UVIT
 hardware as well as software preceding the pipeline, even a single
 uninterrupted data collection needs to be sub-divided into smaller
 pseudo-Episodes (for example : 16-bit Frame Counter overflow in $\sim$
 100 s, during PC mode imaging with the smallest window size 100$\times$100),
 each holding consecutive 2$^{16}$ frames. The last pseudo-Episode would in
 general contain a smaller number of frames. The pipeline treats each
 pseudo-Episode at par with a normal Episode.

 \subsubsection{Modules}

 The pipeline has been implemented in a modular fashion keeping in mind
 several blocks of processing are common to different chains and their
 versions based on imaging mode. Accordingly these are implemented as
 stand alone Modules for flexibility of being called by different chains.
 Whenever relevant, these Modules internally support both the imaging modes,
 viz., Integration (INT) \& Photon Counting (PC). Many of them can be
 turned ON or OFF depending on the user need and / or enabling testing,
 diagnosing and evaluation. The resulting products from each Module follow
 a naming convention with name suffix embedded in them. A master list of
 all the Modules, their functionality and linkages to different chains
 are presented in Table 3. The large number of selectable switches and
 variable parameters used by these Modules are provided in Appendix-11.
 Some generic comments on Modules follow prior to their description.
 Every Module has several primitive switches regarding overwriting of
 existing file structure (`clobber'), inclusion of complete log of
 operations on product headers (`history'), choice for timing reference -
 using the Universal Time Clock or UVIT’s Master Clock (`utcFlag'),
 storing outputs from individual block (`Write\_to disk' flags), etc.
 The headers of every product from individual Modules record settings
 of all selectable switches and parameters. In addition, values of
 various corrections factors (e.g. events rejected due to parity error,
 CRC, etc), frames affected by Cosmic Ray showers \& discarded, frames
 affected by artifacts in data, initial frames affected by mandatory
 safety checks which are discarded, etc. are logged in the headers.
 Every failure to achieve convergence or successful completion of any
 block is recorded in a master log with a message (including the text
 string ``CRASH") which helps to track it and relevant details for its
 later investigation and diagnosis. Individual Modules are described with
 full technical details in Appendix-1. They are presented in an order
 following their sequence of appearance in the chains for drift extraction
 – Relative Aspect (RA\_INT \& RA\_PC; section 3.1.4.1) followed by
 the Modules exclusively for the chains for Sky Imaging (L2\_PC \& L2\_INT;
 section 3.1.4.2). In general, within every Chain, almost all Modules
 process all the frames of an Episode before transferring control to
 the next Module (only exceptions being the sequence of blocks 10, 11,
 12 \& 13 in RA\_INT chain which form a looping segment; see Fig. 6).

 There is also a library of utilities that are commonly used by several
 Modules \& processing blocks.


%
\begin{table*}[th]
\scriptsize
\centering
\caption{List of Modules used in the Pipeline and their linkages to Chains}
\begin{tabular}{|l|l|l|l|l|}
\hline
\textbf{Serial} & \textbf{Functionality of the Module} & \textbf{Name of the} &
  \textbf{References to} &  \textbf{Usage} \\
\textbf{No.} & & \textbf{Module} & \textbf{Module description} & \textbf{of the Module} \\
 &  &  & \textbf{ (Appendix 1)} & \textbf{(Chains)} \\
\hline
1 &Ingest the observational data &DataIngest &A1.1; &RA\_INT, RA\_PC, \\
 &as well as calibration databases &  &Fig.14 &L2\_PC, L2\_INT  \\
\hline
2 &Unit conversion from `event' to &uvtUnitConversion&A1.2 &RA\_INT, RA\_PC, \\
 &`event-per-second' &&  &L2\_PC, L2\_INT \\
\hline
3 &Flagging of pixels / events &uvtMaskBadPix &A1.3 &RA\_INT, RA\_PC, \\
 &corresponding to unusable regions of & & &L2\_PC, L2\_INT \\
 &the detector & & &  \\
\hline
4 &Identifying frames affected by &uvtCosmicRayCorr &A1.4 &RA\_INT, RA\_PC, \\
 &Cosmic Ray showers& & &L2\_PC, L2\_INT \\
\hline
5&Applying correction for response &uvtFlatFieldCorr &A1.5 &RA\_INT, RA\_PC, \\
&variation over the detector area & & &L2\_PC, L2\_INT \\
\hline
6 &Correction for temperature &uvtQEMCPCorr &A1.6 &RA\_INT, RA\_PC, \\
 &dependence of : Quantum Efficiency & &  &L2\_PC, L2\_INT \\
 &(QE) of detector gain of Micro- & && \\
 &Channel-Plate (MCP) & &  & \\
\hline
7 &Symmetric expansion of 2-D arrays to &uvtPixPadding &A1.7 &RA\_INT, RA\_PC, \\
 &accommodate spacecraft drifts & & &L2\_PC, L2\_INT \\
\hline
8 &Frame Accumulation by stackings &uvtAccEveryTsec &A1.8 &RA\_INT \\
 &of successive frames & & & \\
\hline
9 &Generating finer grid of 2-D arrays to &uvtSubDivision &A1.9 &RA\_INT, RA\_PC, \\
 &achieve / preserve higher resolution & & &L2\_PC, L2\_INT \\
\hline
10 &Star Detection - identify brighter &uvtDetectStar &A1.10; &RA\_INT, \\
 &sources in the sky image & &Fig.15 &RA\_PC \\
\hline
11 &Correction for Detector Distortion - &uvtDetectDistCorr &A1.11 &RA\_INT, RA\_PC, \\
 &photon event / star centroids corrected & & &L2\_PC \\
 &for local defects inherent in the & & & \\
 &Detector & & & \\
\hline
12 &Correction for Distortion in the &uvtOpticAssDistCorr &A1.12 &RA\_INT, \\
 &optical assembly -  photon event / star & & &RA\_PC, L2\_PC \\
 &centroids corrected for local & & & \\
 &distortions in the Telescope optics & & & \\
\hline
13 &Identification of a Reference Frame &uvtRefFrameCal &A1.13 &RA\_INT, \\
 &with respect to which all `relative' & & &RA\_PC \\
 &shifts and / or rotations etc are & & & \\
 &computed & & & \\
\hline
14 &Computation of drift (for temporal &uvtComputeDrift &A1.14; &RA\_INT, \\
 &scale of several seconds) from VIS / & &Fig.16 &RA\_PC \\
 &NUV images to generate  Relative & & & \\
 &Aspect Series (RAS) & & & \\
\hline
15 &Computation of jitter (for sub-second/ &uvtComputeJitter &&RA\_INT, \\
 &second time scale) from Gyro signals & &&RA\_PC \\
\hline
16 &Computation of inter-band alignment &uvtComputeThermal &&RA\_INT, \\
 &due to thermal effects & &&RA\_PC \\
\hline
17 &Correction to the photon event &uvtCentroidCorr &A1.15 &L2\_PC \\
 &centroids for error in on board & & & \\
 &estimation of dark counts of the & & & \\
 &CMOS sensor & & & \\
\hline
18 &Correction for the  Fixed Pattern &uvtCentroidBias &A1.16 &RA\_PC, \\
 &Noise to the photon event centroids & &  &L2\_PC \\
\hline
19 &Correction for drift by applying Shift &uvtShiftRot &A1.17 &L2\_PC, \\
 &\& Rotation & & &L2\_INT \\
\hline
20 &Frame Integration - dividing the full &uvtFrameIntegration &A1.18 &RA\_PC, \\
 &dataset into smaller sub-sets, as an option & & &L2\_PC \\
\hline
21 &Generation of Exposure Weighted &uvtFindWtdMean &&L2\_INT \\
 &Mean Images \& Exposure Arrays & && \\
\hline
22 &Registration \& Averaging - aligning &uvtRegAvg &A1.19 &L2\_PC, \\
 &images from sub-sets of an episode & & &L2\_INT \\
 &and combining into a single image & & & \\
\hline
23 &Establishing absolute aspect of the &uvtFullFrameAst &A1.20 &L2\_PC, \\
 &image by matching catalogued stars & & &L2\_INT, \\
 & & & &Driver Scheme \\
\hline
%
\end{tabular}
\end{table*}


%
%

  \section{Commissioning experience, operation and performance of the pipeline}

 Extensive testing of each Module / algorithm as well as the four 
 chains have been carried out over an extended period. This began
 with data collected in the laboratory in the year 2015, well before the
 launch of UVIT/ ASTROSAT. The spacecraft drifts were simulated
 with jigs created to generate relative motion between a UV source
 and the detector assembly. Most debugging could be completed
 only after launch with in-orbit data from Performance Verification
 phase. Some tweaking of strategies and choices of algorithms were
 also needed. Many of the parameters related to instrument
 characteristics were refined based on the data from in-orbit calibration,
 whose starting values were from lab tests. Two noteworthy parameters
 which needed most careful and laborious analyses are : (i) Relative
 shifts between time stamps of the frames in the three bands – these 
 are required for translating time-series of the drift derived from
 one band, e.g. VIS, to the other bands, e.g. NUV and FUV, \& (ii)
 relative angles between the coordinate system (X-Y axes) among the 
 three detectors as well as with respect to the spacecraft coordinates
 (Yaw-Pitch). A common Master Clock (from the selected band) is used
 by all the bands for time stamping individual frames. The time offsets
 between the NUV (or FUV) band (irrespective of window size
 selection) with respective to the VIS band (as well as between NUV
 \& FUV bands, which is needed in case of drift tracking using NUV)
 have been calibrated. These are presented in Appendix-2. The
 details of relative angles between coordinate systems are presented
 in Appendix-3.

  The UL2P described above have been in regular use at the UVIT Payload
 Operation Centre (POC) for generating standard bundle of products
 for archiving and dissemination to the community of astronomers by
 the ISSDC. The UVIT Driver Module is executed twice on each merged Level-1 data
 set – once using VIS band for drift extraction (RA\_INT chain) and the
 other time forcing the use of NUV band for extracting drift (RA\_PC
 chain). A nominal set of default parameters are used for both the runs.
 The standard bundle of products for dissemination is selected by the 
 POC \& ISSDC, which includes the most important items required by
 common end users (astronomers). The Table 4 list all contents of
 this bundle. It includes products for each combination of UV Band,
 Filter \& Window size from individual episode as well as combined
 over multiple episodes (for details see Appendices 2, 3 \& 10). For the
 single episode case, constituents are : Drift series (RAS; final output
 of RA\_INT/ RA\_PC chain; from `uvtComputeDrift' directory; Event list
 (including final centroids in both Detector \& Astronomical coordinates;
 from `uvtShiftRot' directory; one final post-Astrometry UV sky map
 (Signal; RA-Dec; ICRS; J2000; from `uvtFullFrameAst' directory), and 
 three pre-Astrometry sky maps of UV (Signal), Exposure \& Uncertainty
 (all in Detector X-Y coordinates; from `uvtFlippedRegImage' directory). 
 For the combined multi-episode case the constituents are : one final
 post-Astrometry UV sky map (Signal; RA-Dec; ICRS; from the directory 
 like - `XUV\_FullFrameAst\_F1\_W511', where X=F or N for FUV and NUV
 respectively), and three pre-Astrometry sky maps of UV (Signal),
 Exposure \& Uncertainty (all in Detector X-Y coordinates; from the
 directory like `XUV\_Final\_F1\_W511'). There are two sets of these
 above constituents – one set each for drift extraction using VIS \& NUV.
 The above is further repeated for each combination of UV Band, Filter
 \& Window size. The complete details of all products (including the
 standard bundle) from the pipeline are presented in Appendices 2, 3 \& 10.
 The current version (v6.3) of UL2P described here has been in use at
 the POC since mid 2018. Prior to that, an earlier version (v5.7) 
 was in use.

               The UL2P is completely open source and its distribution 
 has been simplified by creating an ``Installer" compatible with most
 Unix operating systems. Relevant instructions for installation and
 other related information are publicly available for download from
 multiple sites (https://uvit.iiap.res.in/Downloads; \\
 http://astrosat-ssc.iucaa.in/?q=uvitData; \\
 https://www.tifr.res.in/{\char`\~}uvit/).
{ Two caveats must be noted by users of this pipeline.
1) As has been mentioned earlier (Sections 3.2.3 \& 3.2.4)
the precision of photometry in multi-Episode UV images for
the outmost regions, are compromised
by the fact that Exposure arrays from individual Episodes undergo
a cut ($<$ 10\%) but not the corresponding photons while 
combining Episodes;
2) The scheme of applying large angle rotation to the Exposure
array (e.g. from Detector X-Y system to astronomical RA-Dec
coordinate system) leads to small but detectable artifacts of
repeated patterns with amplitude $\sim$ few percent.
Both these issues have been discovered rather recently and they
will be addressed in the next upgrade of the pipeline.
The corrective measure for \#1 will include applying the exposure based
    cut ($<$ 10\% of peak) on the Signal (unit : count /second) array only (i.e.
    leaving Exposure \& Uncertainty arrays without any cut)
  at the individual Episode processing level.
  This will preserve photometric precision of the multi-Episode arrays
   generated by the Driver Scheme.
    For the \#2, an improved logic for mapping pixels from pre- to
post-rotation Exposure arrays will be employed.
}

%

\begin{table*}[th]
\scriptsize
\centering
\caption{Standard set of pipeline data products bundled by UVIT POC for dissemination and archiving by the ISRO /ISSDC}
\begin{tabular}{|l|l|l|l|l|l|l|}
\hline
\textbf{Product} &\textbf{Description}&\textbf{Drift}&\textbf{Drift}&\textbf{Drift}&\textbf{Drift}& \textbf{Total} \\
\textbf{type} & &\textbf{extraction}&\textbf{extraction}&\textbf{extraction}&\textbf{extraction}& \textbf{No.} \\
& &\textbf{using VIS}&\textbf{using VIS}&\textbf{using NUV}&\textbf{using NUV}&\textbf{of} \\
& & \textbf{(RAS\_VIS)}&\textbf{(RAS\_VIS)}&\textbf{(RAS\_NUV)*}&\textbf{(RAS\_NUV)*}&\textbf{files}  \\
& & & & & & \\
& & \textbf{Sky Image }&\textbf{Sky Image }&\textbf{Sky Image }&\textbf{Sky Image }&  \\
& &\textbf{in NUV band}&\textbf{in FUV band}&\textbf{in NUV band}&\textbf{in FUV band}&  \\
& & & & & & \\
& &&&\it{*for obsn. till }&\it{*for obsn. till}&  \\
& &&&\it{March 30, 2018}&\it{March 30, 2018}&  \\
\hline
{\it Products for each } &&&&&& \\
{\it Individual Episode :} &&&&&& \\
\hline
%
Sky Image &FITS image &1 &1 &1 &1 &4 \\
(Detector coordinates; X-Y) &(4800x4800 pixels) & & & & & \\
\hline
Sky Image, post Astrometry &FITS image &1 &1 &1 &1 &4 \\
(Astronomical coordinates; &(4800x4800 pixels) & & & & & \\
RA-Dec, ICRS) & & & & & & \\
\hline
Exposure Map &FITS image &1 &1 &1 &1 &4 \\
(Detector coordinates; X-Y) &(4800x4800 pixels) & & & & & \\
\hline
Uncertainty Map &FITS image &1 &1 &1 &1 &4 \\
(Detector coordinates; X-Y) &(4800x4800 pixels) & & & & & \\
\hline
 Photon Events List &FITS binary table &1 &1 &1 &1 &4 \\
 (Photon centroids in both &&&&&& \\
 Detector \& Astronomical &&&&&& \\
 coordinates) &&&&&& \\
\hline
 Drift Series (RAS file) &FITS binary table &1 &1 &1 &1 &2 \\
 (In Detector \& Spacecraft &&&&&& \\
 coordinates) &&&&&& \\
\hline
{\it Products for each ``Group"} &&&&&& \\
{\it of Multi-Episodes :} &&&&&& \\
\hline
Sky Image, post Astrometry &FITS image &1 &1 &1 &1 &4 \\
(Astronomical coordinates; &(4800x4800 pixels) & & & & & \\
RA-Dec, ICRS) & & & & & & \\
\hline
Sky Image, pre Astrometry &FITS image &1 &1 &1 &1 &4 \\
(Astronomical coordinates; &(4800x4800 pixels) & & & & & \\
RA-Dec, ICRS) & & & & & & \\
\hline
Exposure Map, post &FITS image &1 &1 &1 &1 &4 \\
Astrometry (Astronomical &(4800x4800 pixels) & & & & & \\
coordinates; RA-Dec, ICRS) & & & & & & \\
\hline
Uncertainty Map, post &FITS image &1 &1 &1 &1 &4 \\
 Astrometry (Astronomical &(4800x4800 pixels) & & & & & \\
coordinates; RA-Dec, ICRS) & & & & & & \\
\hline
{\it Common Products :} & & & & & &\\
\hline
Data Quality Report &XML & & & & & 1 \\
\hline
README &TXT & & & & & 1 \\
\hline
ChangeLog &TXT & & & & & 1 \\
\hline
DISCLAIMER &TXT & & & & & 1 \\
\hline
Pipeline Parameter Files &TXT & & & & & 2 \\
\hline
\end{tabular}
\end{table*}


%

 Next, some typical examples of results are
 presented.
 Figure 11 displays two typical examples of drifts extracted (along
 Detector X \& Y axes) by the Relative Aspect chain RA\_INT using
 VIS band data. It spans $\sim$ 1200 / 1500 seconds of time. The peak
 to peak variation of drift is around $\sim$ 160 arc seconds. The sharp
 spikes (positive in one \& negative in another) observed in these
 plots appear at regular intervals corresponding to systematic jerks imparted
 by planned mechanical movement of the payload Scanning X-ray Sky
 Monitor (SSM). The drift extraction process has been able to
 successfully mitigate any major adverse effect of such rather harsh
 disturbances on the PSF.

\begin{figure*} [th]
\centering
\includegraphics[scale=0.40]{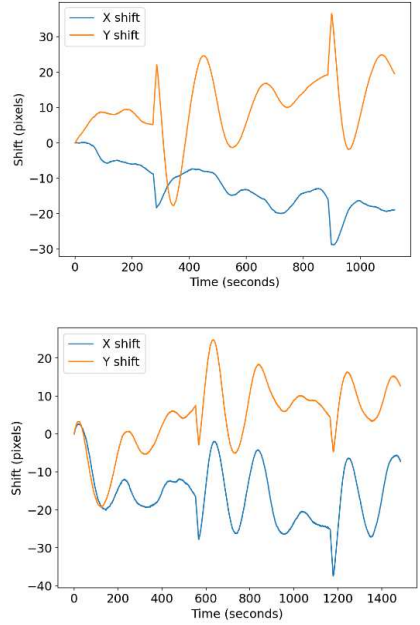}
\caption{
Two examples of drifts extracted by the Relative Aspect
chain RA\_INT from VIS data spanning 1200 / 1500 sec. The
peak to peak movements are $\sim$ 160${}^{\prime\prime}$ 
(1 pixel $\sim$ 3.3${}^{\prime\prime}$).
}
\end{figure*}

  Next, as an example we present UV images of an astronomical source, viz.,
 the nearby spiral galaxy NGC 300 at a distance of $\sim$ 1.9 Mpc,
 processed by the pipeline to demonstrate its efficacy. The Far-
 UV image is displayed in Fig. 12, which uses 14,300 s exposure with 
 in the filter F148W ($\lambda_{mean}$ = 148.1 nm, $\Delta\lambda$ = 50 nm).
 The Fig. 13 shows a colour composite Near-UV image of NGC 300 using
 observations with exposures of 20,000 s in filter N242W (blue;
 $\lambda_{mean}$ = 241.8 nm, $\Delta\lambda$ = 78.5 nm) \&
 2,500 s in filter N263M (green; $\lambda_{mean}$ = 263.2 nm,
 $\Delta\lambda$ = 27.5 nm).

   While these sample images show the utility and 
 success of the pipeline in extracting astronomically significant high
 quality images qualitatively, the POC carries out systematic analyses
 of the products to quantify their quality objectively.
   The performance of this Level-2 pipeline has been reported
  by Ghosh et al. (2021), which had used UVIT observations
 carried out till mid-2020. Here we have updated it including the latest status.

  The FUV and
 NUV images generated by the pipeline are analyzed for Point Spread
 Function, PSF, for a minimum of three stars (point sources) spread
 over the entirefield of view. The extensive in-orbit calibrations
 have characterized the PSFs for both the FUV and NUV bands by a
 compact central core and an extended pedestal (Tandon et al 2017c, 2020).
 Based on these anticipated structure of the PSFs a quality score
 has been defined. The criteria followed to assign the quality score are : 
 the best score of `10' is assigned if the pedestal is $<$ 20\% and
 the full width at half maxima, FWHM, of the PSF is $<$ 1.6 arc-sec; 
 similarly `9' \& `8' correspond to FWHM being $<$ 1.8 arc-sec \& $<$
  2.0 arc-sec, respectively. For every increase in the pedestal by
 5\% (beyond 20\%), the quality score is reduces by 2. This score is
 estimated for both the bands (FUV and NUV) and the mean of these
 values is quoted in the quality report which always accompanies
 the UL2P product bundle.
                          The distributions of quality score in FUV
 \& NUV bands 
for all observations carried out till recently (09-Dec-2021) 
 are shown in Table 5. 
                                      As can
 be seen, the pipeline products for NUV band achieve a quality factor
 of 9 or above for $\sim$  90\% of observations. For the FUV band,
 this fraction is relatively lower at $\sim$ 54\%, which increases
 to $\sim$ 66\% if the quality factor of 8 or above are considered.
 The relative superiority of the quality score of images in
 NUV over FUV is related to the fact that the UVIT instrument achieved
 better angular resolution in the NUV band. Since the NUV band shares
 the same telescope with VIS band and FUV uses a separate
 dedicated telescope (see Fig. 2), a possible cause could be temporal
 changes in the relative angle between the two telescope structures
 within the time scale of an orbit. Activating the `pseudo-Episode'
 option for L2\_PC chain for the FUV band may improve its quality score.

\begin{figure*} [th]
\centering
\includegraphics[scale=0.40]{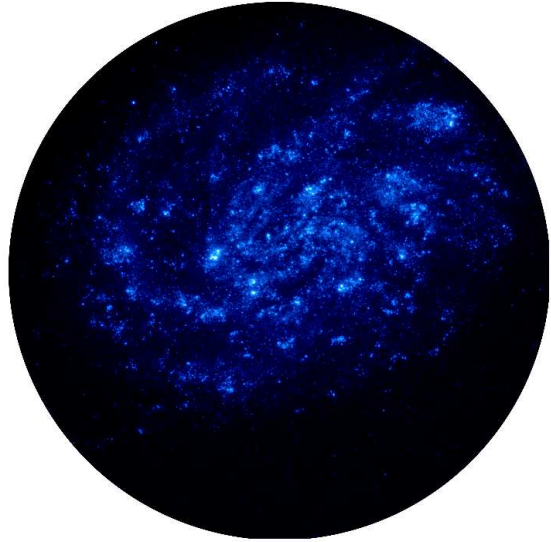}
\caption{
Far-UV image of the Spiral galaxy NGC 300 generated
 by the pipeline using exposure of 14,300 s with the
 filter F148W (${\lambda}_{mean}$ = 148.1 nm, ${\Delta\lambda}$ = 50 nm).
}
\end{figure*}

\begin{figure*} [th]
\centering
\includegraphics[scale=0.40]{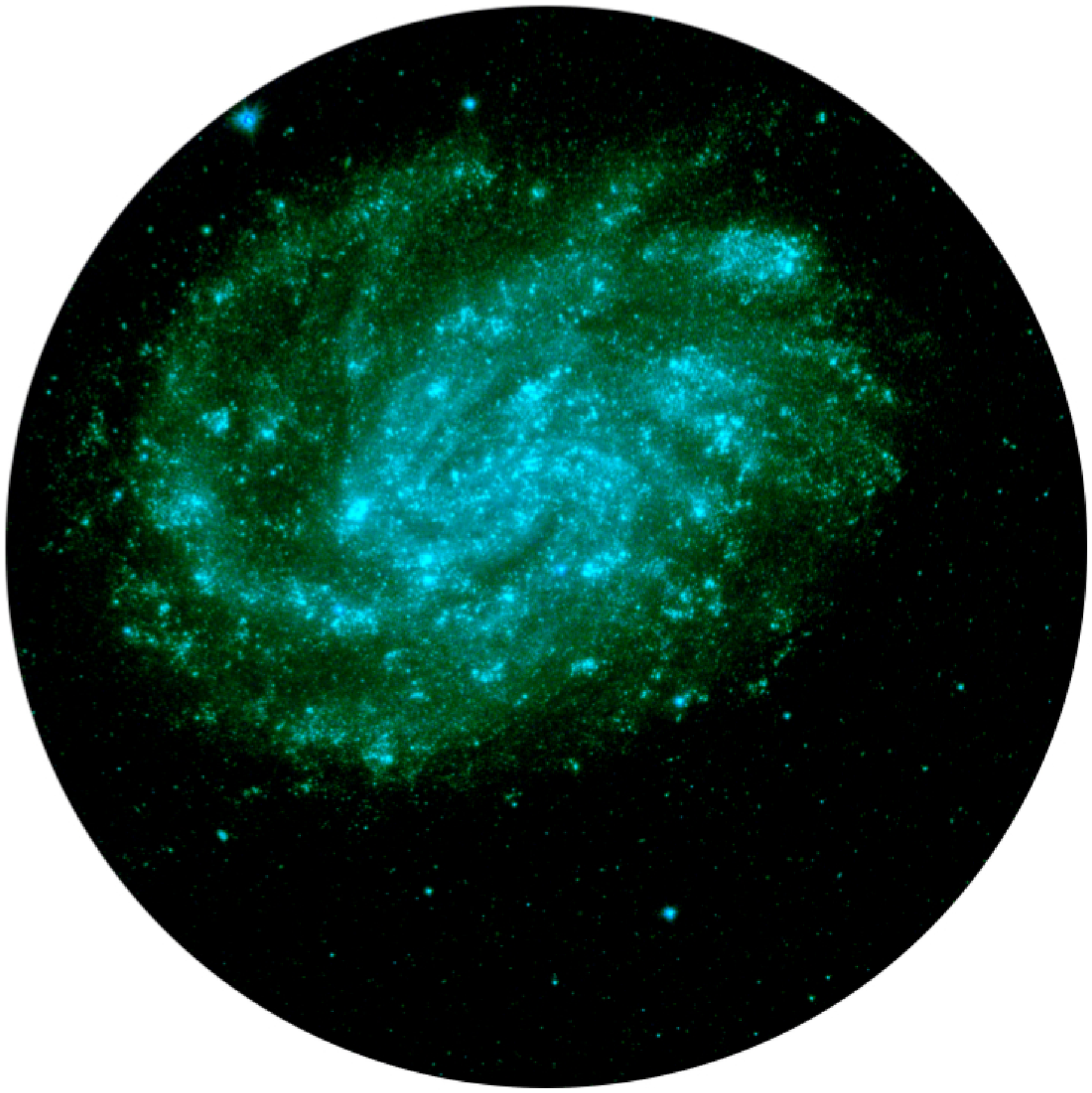}
\caption{
Colour composite NUV image of the Spiral galaxy NGC 300 generated
 by the pipeline using exposures
 of 20,000 s in filter N242W (blue; ${\lambda}_{mean}$ = 241.8 nm,
 ${\Delta\lambda}$ = 78.5 nm) and 2,500 s in filter N263M
 (green; ${\lambda}_{mean}$ = 263.2 nm, ${\Delta\lambda}$ = 27.5 nm).
}
\end{figure*}

   One of the key parameters quantifying eventual success rate of the
 UL2P is the fraction of total on sky exposure finally contributing
 to the UV image product. The ASTROSAT operations group schedules sky
 exposures based on approved time allocated (say T\_MCAP), for a 
 particular target from a scientific Proposal. Let the Level-1 data
 successfully generated at the ground station of ISRO be for
 duration T\_L1. This forms the input for the UL2P. Let the final
 UV image product generated by UL2P be T\_L2 (somewhat like EXP\_TIME
 keyword in header). Then yield of UL2P (for that particular
 episode of observation) can be defined by the ratio - (T\_L2 / T\_L1).
 The distribution of this yield for both NUV \& FUV observations carried
 out till recently (09-Dec-2021) are presented in Table 6.
                                         The fraction of all 
 pointings in this sample achieving yield above 0.9 (0.8) is $\sim$ 72\%
 (86\%) for the NUV band. The corresponding fraction for the FUV band is 
 $\sim$ 75\% (88\%). These values indicate that there is scope for
 improvements in the pipeline.


\begin{table*}[th]
\centering
\caption{Distribution of the quality values (from achieved PSF
size \& amount of pedestal) for FUV and NUV band images.
}
\begin{tabular}{|l|l|l|l|l|l|l|l|l|l|}
\hline
{Quality bin (range)} &(1-2)&(2-3)&(3-4)&(4-5)&(5-6)&(6-7)&(7-8)&(8-9)&(9-10) \\
\hline
{NUV band (No. of ObsIDs)} & 0 & 0 & 0 & 3 & 6 & 5 & 7 & 18 & 335 \\
\hline
{FUV band (No. of ObsIDs)} & 15 & 22 & 30 & 41 & 58 & 84 & 84 & 118 & 527 \\
\hline

\end{tabular}
\end{table*}





\begin{table*}[th]
\centering
\caption{Distribution of the yield values (percentage 
of input Level-1 data
translated to Level-2 image products, for FUV \& NUV bands)
achieved by the pipeline.
}
\begin{tabular}{|l|l|l|l|l|l|l|l|l|l|l|}
\hline
{Yield bin (\%)} & 10 & 20 & 30 & 40 & 50 & 60 & 70 & 80 & 90 & 100 \\
\hline
{NUV band (No. of ObsIDs)} & 3 & 0 & 1 & 0 & 7 & 4 & 11 & 27 & 54 & 275 \\
\hline
{FUV band (No. of ObsIDs)} & 9 & 5 & 3 & 5 & 9 & 24 & 28 & 72 & 149 & 797 \\
\hline

\end{tabular}
\end{table*}



\medskip
\medskip
 \section{Summary and possible future improvements}

 A versatile data processing pipeline (UL2P) has been developed for
 the UVIT payload on board ASTROSAT satellite. This pipeline runs
 in a fully automated mode and is ideally suited for the Payload Operation
 Centre, where huge amounts of data collected from UVIT need to be processed
 swiftly without human intervention. The pipeline generates astronomer
 ready sky images in UV. Its main components are : handling of occasional
 artifacts and anomalies in Level-1 data, extraction of spacecraft
 drifts, and disturbances due to internal movements; application of 
 various corrections to measured direction of arrival for each photon
 – e.g. spacecraft motion (with respect to planned inertial pointing
 towards target direction), systematic effects and artifacts in the
 optics of the telescopes and detectors, exposure tracking on the sky,
 alignment of sky products from multi-episode exposures to generate a
 consolidated set and astrometry. Detailed logs of operations are
 maintained on headers of products and intermediate products for every
 processing stage are accessible via user selectable options.
 This pipeline has been tested extensively before releasing for regular
 use at the Payload Operation Centre (POC) located at the Indian Institute
 of Astrophysics (IIA), Bangalore. Every UVIT dataset from ISRO/ISSDC
 is received by the POC which after processing sends the corresponding
 standard products back to ISRO/ISSDC for archiving and dissemination.
 The current version of UL2P described here has been in
 use since February 2018
 and has continued to 
 successfully generate relevant products for the astronomer community
 leading to scientific results world wide.
                                           While large number of
 selectable parameters are available for the user, a well characterized
 ``standard default" set is used for executing this pipeline at the
 Payload Operation Centre (POC) for UVIT and selected products are
 archived and disseminated by the Indian Space Research Organization
 (ISRO) through its ISSDC portal. Users preferring to use selections
 of parameters different from the ``standard default", can download,
 install and run UL2P themselves following instructions provided online.
 There are a few areas where improvements to this pipeline would help
 the end users. These include - (1) improvement in efficiency of combining
 multi-orbit images, (2) improvement in success rate for astrometry
 correction and (3) improvement in the precision of astrometry. Each
 of these improvements are linked to use of optical stars (from VIS
 images) instead of the current scheme of using UV stars detected in
 UVIT images with significant signal to noise ratio. The scientifically
 important research areas involving extremely faint fields will benefit
 most from these.

%

\medskip

{\bf Appendix-1 : Modules used in the UVIT Level-2 Pipeline} \\

This appendix provides complete technical details of the processing blocks
 used by the UVIT Level-2 pipeline. Given the two different available modes
 of imaging operations the processing chains have distinct versions for
 each of these modes. There are several computational blocks with commonality in
 their function among the chains / versions. Hence, such processes are
 captured as Modules with selectable switches and parameters to meet the
 requirements of all similar applications (see Table 3 for a list of all
 Modules and their linkages to different chains). A complete list of
 switches and parameters for these Modules is available in Appendix-11.
 The Modules are called by the chains and the resulting products from each
 Module follow a naming convention with name suffix embedded in them for ease
 of tracking as well as higher level of scripting for autonomous operations.
 Individual Modules are described next, in a natural order of their sequence
 of appearance in the chains for drift extraction : Relative Aspect (RA\_INT
 \& RA\_PC) followed by those exclusively for the Sky Imaging (L2\_PC \&
 L2\_INT). There are 20 Modules presented below in sub-sections numbered :
 A1.1, A1.2, ... to A1.20. Every Module consists of several blocks operating
 in a sequence, and in general, each block operates on all the frames of an
 Episode before transferring control to the next block (with only exception of a
 few in the RA\_INT chain).

\medskip
{\bf A1.1 DataIngest} \\

 The Data Ingest is the very first block for any of the processing chains and
 it prepares all inputs data for processing in the subsequent Modules. Its key
 functionality is to import input data, check its integrity, sort them into
 various groups / tables and carry out certain computations. It directly interfaces
 with the ``merged Level-1" (mL1) data bundle which is the primary input to the
 pipeline, the Calibration Database, CAL\_DB, as well as the user selected
 parameter set (Parameter Interface Library; PIL).

  First a brief description of the contents of input data bundle mL1 is presented.
 The mL1 contains all the Level-1 data products {Science Data \&
  Aux Data} for the same 
 Observation ID (ObsID; it includes Proposal ID, Target serial number \&
  Pointing sequence number) in an organized manner.
 Generally the observations corresponding to a specific ObsID lasts 
 multiple successive orbits.
 The directory
 structure of a representative data set (mL1) is shown in Appendix-4.
The Appendix-5 shows the directory structure of the 
Calibration Database, CAL\_DB.
 The notation followed to represent multi-level directory structure in 
 Appendix-4 as well as similar other Appendices (5 to 10) is as follows :
 a normal dash symbol represents files in current directory and  
 a longer horizontal shift or dash represents the next lower sub-directory
 (followed by the sub-directory name) \& the lines immediately following it
 give contents of that sub-directory.
 A few files with very long names are broken into two lines with the second
 line beginning with a large number of dots. 
  
 The Science Data are segregated into the three
band wise directories. These directories in turn
host multiple sub-directories, each of which correspond to one specific ``Episode" of 
data collection. Each sub-directory is populated by 3 files (FITS tables),
the imaging data and two files (GTI \& BTI) containing flags for a large
 number of parameters pertaining to the spacecraft \& UVIT. 

 The imaging data are packed as a sequence of ``frame"-s, one
 for each individual exposure, preserving their time order. Each frame
 comprises of an integral number of fixed size ``packets" (2048 bytes) whose contents
 and format depend on the imaging mode (INT/PC) configured.
 Each packet begins with the Synchronization Mark, Primary Header and a
 Secondary Header (includes an Image Frame Counter \& various other Counters
 with timing information) followed by 2016 bytes of
 image data \& ending with a parity-code (CRC).
 The very first packet of every frame always holds the data
 for all settings for the Detector Module and the data for system status
 and health monitoring only. The second packet onwards contains image data.
 The total number of packets in any frame, $N_p$ depends on the imaging
 mode. The last packet for the frame usually needs padding to complete
 the stipulated fixed size (2048 bytes).
 In case of Integration mode (INT), the raw pixel ADUs (2-bytes per pixel)
 are packed in the usual sequence of 2-D array covering the entire selected
 window region. Hence, the value of $N_p$ is always predictable for INT mode,
 e.g. maximum value of 262 corresponds to full 512$\times$512 window setting.
 On the other hand, in Photon Counting mode (PC), the raw pixel signals
 are processed on board to identify individual photon event and compute its
 centroid along two detector based axes (X, Y). After this processing, the
 data from the frame comprises of all extracted photon events packed as a
 sequence, with each photon event contributing 6 contiguous bytes. Accordingly,
 each packet can accommodate up to 336 photons. As a result, the value of
 $N_p$ in PC mode may vary from frame to frame, since it depends on the
 total number of photon events identified.
 The contents of the 6 bytes for every detected photon event are : centroid
 along X (2 bytes), centroid along Y (2 bytes), measures of asymmetry in
 the raw event foot-print \& local background (2 bytes).

  The Auxiliary (Aux) data correspond to the entire 
 time duration covering the full merged data set.
 They are FITS tables containing uninterrupted continuous 
 time series of various measured / derived parameters mostly
 originating in spacecraft sub-systems.  
 Examples of Aux data that are used directly by the pipeline follow.  
 The time correlation between the
  spacecraft’s clock (raw \& calibrated) and internal clocks of UVIT (`*.tct').
 The attitude of the spacecraft :
  tabulated as quaternions representing rotation angles with respect to
 reference coordinate system - Roll, Yaw \& Pitch (`*.att'). The signals
  from the angular rate sensors (Gyros) for the 3 axes (Roll, Yaw \&
  Pitch) which are tabulated separately (`*.gyr').

  The Fig. 14 provides the sequence of operations carried out by Data Ingest.
 DataIngest initially carries out the following functions - (i) import mL1 data,
 (ii) set up structured paths for Science data and AUX data, (iii) establish
 links to Calibration Database, (iv) import and interpret settings of various
 user selectable switches and values of parameters, from Parameter Interface
 Library, PIL, queries. 

   As a part of check on data integrity at a primitive level, the user selected
 action is carried out on individual data packets which failed the CRC test.
 Next, the headers of individual science data blocks corresponding to distinct
 Episodes are interpreted for various settings : respective observing Band,
 imaging Mode, Filter, the band selected for Master Clock. For each Episode,
 a time correlation between UVIT’s Master Clock and the absolute time
 (Universal Time Clock, UTC, stamped by the Level-1 process) is constructed.
 Occasionally, the UTC shows erratic behaviour. Even when user selects to use
 the UVIT clock (i.e. with `utcFlag' switch is OFF), a robust error tolerant
 scheme has been employed to establish a scheme capable of predicting Modified
 Julian Date (MJD) from UVIT clock, albeit with limited precision ($\sim$ 1 s).

\begin{figure*} [th]
\centering
\includegraphics[scale=0.40]{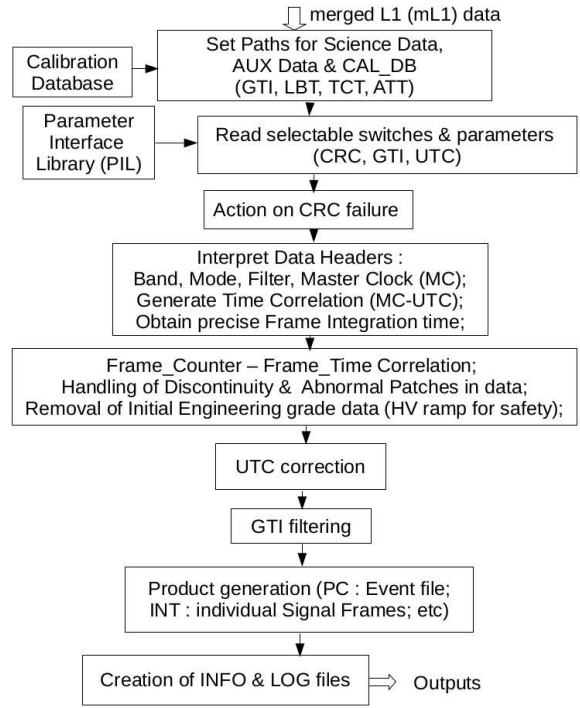}
\caption{
Block Diagram for DataIngest Module.
This module first collects all relevant input data from observations,
calibration and user selections of configurable parameters.
It checks integrity and quality of observational data and separates them
bandwise in suitable formats (as per imaging mode : INT / PC) for processing by the later stages.
 It carries out time correlation and 
certain other functions  as described in the text.
}
\end{figure*}

 Next step deals with identification and treatment of corrupt and unusable data.
 The frame number and its corresponding time stamp must always increment in a
 regular fashion as per the selected parameters related to the frame read-out
 rate (with the exception of natural return to zero on overflow of the
 respective counters). This provides a good handle for identifying certain
 kinds of data corruption. The nature of unusual data patch includes the
 following occurrences noticed in most Level-1 data sets (in PC mode) :
 mismatch between Frame Number and Frame Time (one of them being in the correct
 sequence correct and the other showing a JUMP `spike', which can be positive
 or negative); completely discontinuous Frame Number \& Frame Time inserted
 within normal good data sequence; data from a particular frame repeated
 elsewhere in the same data set (i.e. for the same observing Episode); abrupt
 discontinuity in Frame Number which violates monotonicity, etc. These issues
 have come to light progressively during the mission and appropriate logic
 incorporated within this DataIngest Module. As a result this Module has
 evolved over longer time. The identified `outlier' frames are discarded from
 further consideration. The fraction of input Level-1 data successfully
 translated by DataIngest Module to reorganized output usable for the
 subsequent Modules depends on frequency of occurrences of unusual data
 patches. Some statistical information based on past $\sim$ 4 years’
 operation of UVIT are provided in Section 4.

 Initial part of every Episode of observation includes data collected during
 ramping up of high voltages for detector safety, which is not scientifically
 useful since the gain is changing with time. These time stretched are
 identified and discarded. The primary time variable to be used extensively by
 the subsequent steps of the pipeline is generated by applying UTC correction
 to UVIT’s Master Clock (only if `utcFlag' is selected to be ON). In case
 the Good Time Interval (GTI) filtering is turned ON, the time stretches of
 data not meeting the criteria selected by the user are discarded. The precise
 value for exposure time per frame is computed and recorded. In actual
 processing so far, GTI filtering has never been used.

  The next phase of processing in DataIngest depends on the imaging Mode.
 For INT, individual frames containing raw pixel counts are extracted and
 stored in separate 2-D FITS image files in a directory named `SignalFrames'
 with nomenclature of each file having Frame number and Time stamp
 embedded in it. This directory also stores a list of extracted images.

  For PC, sequence of individual frames with centroid coordinates in detector
 X-Y coordinate system and other details like Frame number, UVIT Master clock
 time, UTC time, diagnostic measures of event’s raw 5x5 pixel footprint -
 `Max-Min' \& `Min' values of the 4 corners for every detected photon
 are all packed into a single `events' file. It holds the data from the
 specific imaging Episode as one master list of photon events in chronological
 order as FITS table.

  For both the Modes, the following products are included : UVIT-UTC time
 correlation table, log of data loss and an `info' file for passing various
 computed \& PIL parameters to the subsequent Modules. 

  Some long term effects of exposure to radiation (SEE) on the on board
 electronic components were noticed midway through the mission life. They
 were recoverable by hardware RESET to the system at periodic intervals.
 However, some fraction of the data collected were affected beyond the designed
 immunity in the pipeline. As a result additional software patches were
 developed and applied at a later stage to mitigate these effects. While 
 nature of these fixes belongs to the DataIngest module, they were accommodated
 in the envelope code for the chain (RA\_INT) and described in Section 3.2.1.

  A few general remarks follows, on the convention followed throughout the
 pipeline for flagging of unusal data / locations. First about flagging pixels 
 /elements of a 2-D image / array, which need not be considered for processing 
 due to either considered as ``bad" due to any defect or representing a region
 outside the active area of the detector. An unique value (-9999) is assigned
 to such pixels / elements for their easy recognition as an ``invalid" pixel
 / element by all processing blocks. No arithmetic operation is carried out
 on such flagged pixels / invalid elements \& their values are left unaltered.
 On the other hand, the flagging of individual photons in the master event list
 are carried out by appending an additional column with an entry with a value
 `1' (= good) / `0' (=bad) indicating their inclusion / exclusion for
 consideration in further processing (e.g. event located outside the detector’s
 active area; asymmetry in footprint signifying multi-photon event).

 Directory structures of the products of DataIngest Module for INT and PC modes
 are provided in Appendix-6 \& Appendix-7 respectively.

\medskip
{\bf A1.2 uvtUnitConversion} \\

 Aim of this Module, (called by all the versions of the chains) is to translate the image data into
physical unit, making it independent of the frame integration time, selected for the observation.
For INT mode data, first a dark frame, D(ix,jy), taken from CAL\_DB is subtracted from the raw frame,
RI(ix, jy), and then divided by the frame integration time, $T_i$, (earlier computed by DataIngest
Module) making the pixel values to be `signal/s'. (The code is designed to derive this dark frame by
linear interpolation in time between dark frames taken before the sky-exposure and after the sky-
exposure. But, in practice this dark frame is taken from the CAL\_DB of ground calibration.)
This process is repeated for every image frame.

\begin{verbatim}
I_uc(ix,jy)= [RI(ix,jy)-D(ix,jy)]/(T_i); 
\end{verbatim}

for all pixels (ix, jy) of the image,
\& repeated for all images;
For PC mode data, each photon event is assigned a value of ‘event/sec’ (1 divided by the frame
integration time) and recorded in an additional column of the master table of events. This column is
referred to as Effective Number of Photons per second (ENP).

\begin{verbatim}
 ENP_uc(k) = [ 1 /(T_i) ]; 
\end{verbatim}
for all events in master table `k' = 0 to `k\_max'

Note the suffix `uc' which identifies the outputs with the Module. Similar naming convention is
followed for all other Modules also.

\medskip
{\bf A1.3 uvtMaskBadPix} \\

This Module is called by all versions of the processing chains. The functionality of this Module
includes identifying and flagging of : (1) bad pixels in INT mode image arrays, \& (2) events in PC
mode master table with centroids located in bad pixels as well as those satisfying criteria for being
classified as multi-photon event.
BAD\_PIXELS files, for each of the three bands, are arranged as per the mode of observations and the
chosen window size, in the CAL\_DB. All the inactive pixels, either because of being outside the
detector’s circular field of view or beacause of selection of one of the smaller window options, are
represented as `bad pixels' by populating `0' entries (rest with `1' entries), in the corresponding
BAD\_PIXELS file BP(ix, jy).
For INT mode the flagging is effected on all bad pixels of each image frame.

\begin{verbatim}
 if BP(ix, jy) = 0, then 
 I_bp(ix, jy) = INVALID_PIX_VALUE; 
 else I_bp(ix, jy) = I_uc(ix, jy); 
\end{verbatim}

As stated earlier (Sec A1.1), a value of `-9999' is assigned for the `INVALID\_PIX\_VALUE' flag.
For PC mode, every photon event is assigned a value for the variable `BAD\_FLAG' depending on the
integral parts of its centroid coordinates. The centroid coordinates, (x, y), are real numbers represented
by an integer part (in the range : 0 - 511) and a fractional part. Consider the integer parts of the
centroids (x, y) to be (ix, jy). A new column for `BAD\_FLAG' is appended to the master table of
events, which is populated with :

\begin{verbatim}
BAD_FLAG(k) = BP(ix, jy),
\end{verbatim}

 where centroid coordinates for the event `k' are (x, y);for all events in master table `k' = 0 to `k\_max'
This Module carries out another function for the PC mode, viz., flagging for Multi-photon events.
The option for discarding events having two neighbouring photons was introduced with the aim of
achieving higher angular resolution. On board processing in PC, mode for detecting individual photon
events from raw frames, also extract diagnostic information as stated earlier. The identification of a
imulti-photon event is based on this diagnostic data, ( `Max-Min' \& `Min' among the raw counts from 4
corners of the 5x5 photon event footprint, ) and the selected threshold value, `mult\_photon\_thrsld'. For
events meeting the above condition are flagged for potential rejection (= `0') in the
`MULTI\_PHOTON' column appended in the master table of events.

\begin{verbatim}
if [Max-Min](k) > mult_photon_thrsld, 
 then MULTI_PHOTON(k) = 0;
else MULTI_PHOTON(k) = 1; 
\end{verbatim}

for all events in master list `k' = 0 to `k\_max'
Subsequent processing has the option to discard every frame with even one event with multi-photon
flag set for rejection from further consideration. However, this option has never been used.

\medskip
{\bf A1.4 uvtCosmicRayCorr} \\

This Module handles the effects of Cosmic Rays and is called by all versions of the processing chains.
There is a provision for by passing this Module.

This Module attempts to mitigate the effects of individual primary Cosmic Ray (CR) hits as well as
shower of secondary charged particles generated by interaction of CR within and near the detector.
Some of the charged particles would pass through the CMOS imager leading to a large signal in
individual pixels, while other charged particles and photons would generate light pulses some of which
look like photon events in the PC mode. In the INT mode, the former can be detected as unusually
large signals in individual pixels the latter cannot be isolated as these would be similar to genuine
photon events. In the PC mode, some of the hits would generate a large shower of events in a frame
which can be distinguished statistically from the normal frames. For INT data, pixels with values
above a selected threshold, CR\_thrshld, are flagged and the process is applied on all the frames.

\begin{verbatim}
 if I_bp(ix, jy) > CR_thrshld, then 
   I_cr(ix, jy) = INVALID_PIX_VALUE; 
 else I_cr(ix, jy) = I_bp(ix, jy); 
\end{verbatim}

 for all pixels (ix, jy) of the frame.

For PC mode data, CR affected frames are first identified based on their event count being larger than
a dynamically determined `event\_count\_threshold' and then flagged. All events from these flagged
frames are ignored for further processing. The value of `event\_count\_threshold' is computed using two
user selected parameters (p \& q) as follows :

$$ event\_count\_threshold = $$
$$ AVG + p*(AVG)^{0.5} + q/((AVG)^{0.5}) $$

where, AVG is an estimate for the average number of true photon events per frame determined from
the entire data set for that observation episode. The statistical scheme employed for finding AVG
iteratively removes the CR affected frames from the sample as ``outliers". The second term in the
equation addresses fluctuations in case of very low values of event counts encountered for fields very
dark in the UV. At first the CR affected frames are identified using their unique frame numbers and
then all photon events from those CR affected frames are removed from the master list
 of events (\& separately recorded in the log of CR affected frames). As a result, the total number of useful frames is
reduced. The loss would depend on the orbit and distribution of matter in the spacecraft, and for UVIT
it is $\sim$ 5 frames/s. Therefore, for optimal results this correction is only useful for dark fields where
contribution of the Cosmic Rays to the photon counts is not much less than the contribution of the
background.

\medskip
{\bf A1.5 uvtFlatFieldCorr}  \\

This Module, called by all the versions of the chains, applies a multiplicative correction for
variations of response of the detector and transmission of the selected filter across the field. The
correction factor arrays are available from the CAL\_DB (FLAT\_FIELDS\_FILTER). The array
appropriate for the band, mode and filter is selected to apply this correction. For INT mode this 2-D
correction array, FF(ix, jy), is multiplied to every image frame, leaving flagged pixels (flagging
process assigned an unique `invalid' value for its easy identification) unaltered.

\begin{verbatim}
 if I_cr(ix,jy)=INVALID_PIX_VALUE, then 
  I_ff(ix,jy) = INVALID_PIX_VALUE; 
 else I_ff(ix,jy)=I_cr(ix,jy)*FF(ix,jy);
\end{verbatim}

 for all pixels (ix, jy) of the frame.

In case of PC mode data, the value in ENP column for each event is replaced after multiplying it with
the correction value from the array element corresponding to that event’s centroid coordinates.

\begin{verbatim}
 ENP_ff(k :ix,jy)= ENP_uc(k)*FF(ix,jy); 
\end{verbatim}

 where integer part of the centroid coordinates of
the event `k' are (ix, jy); for all events in master list `k' = 0 to `l\_max'

\medskip
{\bf A1.6 uvtQEMCPCorr} \\

This Module corrects for the thermal effects on the detector’s response - through variation of
Quantum Efficiency and gain of the Micro\_Channel\_Plate assembly. This Module is called by all the
versions of the chains and it can be turned off. It uses Aux (LBT) data for reading the instantaneous
temperature of the detector Module (Camera Proximity Unit, CPU) and estimates corresponding
multiplicative correction factor by interpolating the relevant table from CAL\_DB (QE\_TEMP). The in
orbit thermal stability of the CPUs have been so good that so far this correction was never required.

\medskip
{\bf A1.7 uvtPixPadding} \\

This Module is called by all versions of the processing chains.
In order to accommodate movements of the UVIT field on sky due to spacecraft drift over the
duration of Episode, it is necessary to enlarge the size of the 2-D arrays representing the sky images.
For INT mode all image frames are expanded from 512$\times$512 to 600$\times$600 size,
 I\_pp(ix, jy), by
introducing a 44 pixel wide boundary on all four sides and assigning each of these added pixels with
the value of `Bad Pixel' flag. For the PC mode data, centroid coordinates (cx, cy) for all photon events
are incremented by (+44, +44) along both axes :

\begin{verbatim}
cx_pp= cx + 44; & cy_pp= cy + 44; 
\end{verbatim}

 for all events in master list `k' = 0 to `l\_max'. It may be noted that the size of the padding chosen leads to a natural limitation of inclusion of the
sky exposure data corresponding to spacecraft drifts at one go. However, a data set with unusually large
total drift can still be handled by sub-dividing it into multiple (non-overlapping) time ranges chosen
intelligently based on the drift profiles along the two axes (X, Y). Its implementation involves
selections of parameters for the Module uvtRefFrameCal described later on (see Sec. A1.13). The
pipeline will then need to be executed as many times as the number of sub-divided datasets.

\medskip
{\bf A1.8 uvtAccEveryTsec}  \\

This Module is called only by the Relative Aspect chain for Integration mode data (RA\_INT).
Functionality of this Module is to accumulate selected number, `N\_acc', of successive INT mode
frames to enhance the signal for detection of faint stars with a good S/N ratio. A pixel by pixel
averaging is carried out resulting in accumulated frames, Acc\_frame-s. This Module does not handle
PC mode (instead this functionality is embedded within the RA\_PC chain directly).

\medskip
{\bf A1.9 uvtSubDivision}  \\

This Module is called by all versions of the processing chains, with the functionality of dividing the
image pixels or their coordinates in to sub-pixels.

The target angular resolution for UV bands of UVIT was $<$ 1.8${}^{\prime\prime}$, while the pixel scale for the final
512$\times$512 CMOS sensor is $\sim$ 3.3${}^{\prime\prime}$. The detector system and on board processing for PC mode ensures
precision of determining event centroid coordinates (X, Y) to 1/32 of each pixel ($\sim$ 0.1${}^{\prime\prime}$). As a result,
there is a need for sub-dividing various arrays. A subdivision by 8x8 has been used to get more than 4
effective pixels within the required resolution (i.e. mapping 600$\times$600 to 4800$\times$4800). For PC mode,
photon event centroids :

\begin{verbatim}
cx_sd= cx_pp*8; & cy_sd= cy_pp*8; 
\end{verbatim}

for all events in master list `k' = 0 to `l\_max'.

For INT mode processing, an user selectable factor, sub\_div\_factor, m (= 2 / 4 / 8) has been
introduced for sub-division. This Module implements this sub-division where, original contents of the
individual pixels are equally divided among the corresponding sub-divided pixels depending on value
of sub\_div\_factor, m. The sub-pixels corresponding to flagged pixels are retained as flagged.

\begin{verbatim}
if I_acc(ix, jy) = INVALID_PIX_VALUE, then 
   I_sd(i_m_x, j_m_y) = INVALID_PIX_VALUE; 
else I_sd(i_m_x,j_m_y)=I_acc(ix,jy)/(m^2);
\end{verbatim}

 over (m x m) sub-pixels; 

for all pixels (ix, jy) of the image,
\& repeated for all the images;

\medskip
{\bf A1.10 uvtDetectStar}  \\ 

This Module is for identifying stars in an image and quantifying coordinates of their geometric
centroids. It is called by all versions of the chains.

This Module processes individual sky images to identify candidate stars from local maxima in light
distribution, confirm them to be stars from refined analyses of neighbourhood pixels and compute the
centroid coordinates. Finally an intensity ordered table of stars is generated. This Module supports both
INT and PC mode data. The functionality of this Module is very basic \& general purpose, accordinglyit is called explicitly by the two chains extracting spacecraft aspect (RA\_INT \& RA\_PC) and implicitly
by the remaining two chains for their larger Modules : uvtRegAvg (Sec. A1.19), uvtFullFrameAst (Sec.
A1.20), which are described later. Fig. 15 presents a schematic flowchart for this Module. In brief, the
implementation is as follows. The values of all valid pixels in the input sky image frame are used to
determine the `mean (MN)' \& `standard deviation (SD)' for the background, after removing `outliers'
(due to presence of stars) by using an iterative scheme. An initial starting `threshold' for detecting a star
is defined as (MN + p * SD), where `p' is user selectable and is initially kept at a conservatively high
value ($\sim$ 10 - 50). At first all pixels with values above this threshold are considered to be associated
with potential stars. If their number is lower than the user selected value for minimum number of stars,
`min\_stars', then the value of `p' is reduced gradually and the process repeated, till the requirement is
met.

\begin{figure*} [th]
\centering
\includegraphics[scale=0.40]{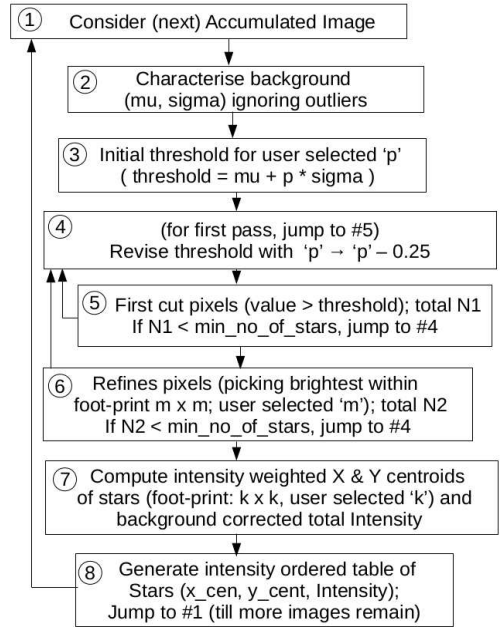}
\caption{
 Flow chart for star detection (uvtDetectStar module).
The algorithm for detecting stars is based on identifying local peaks above
a threshold which is sum of average background and the standard deviation multiplied
by a factor, `p'. The value of `p' is tweaked iteratively (initially starting with a high value)
till the desired number of stars are identified. Centroid positions and intensity
for the stars are tabulated.  
}
\end{figure*}

Multiple neighbouring pixels at this stage could correspond to the same star. In the next step, a
neighbourhood criterion using a square box foot print for the star image is used to identify the brightest
pixel in each box to get a refined list of pixels (rp\_ix, rp\_jy). Size of the box is user selectable but the
optimal size is found to be 15x15. Whether the candidate mimics signatures of a genuine star is tested
by an elaborate logic using the values of the neighbouring pixels, thus providing immunity against
noise spikes. If the number in this refined list does not meet the `min\_stars' condition, then the value
of `p' is reduced further and both the steps described above are repeated till the condition is met (or `p'
reaches a value $<$ 5 \& the process aborts with appropriate error log).

In the final step, intensity weighted centroid and background corrected total intensity in the box are
computed. They are computed as :

\begin{verbatim}
c_l_x=[sum{sig(ix,jy)*ix}/sum{sig(ix,jy)}] 
c_l_y=[sum{sig(ix,jy)*jy}/sum{sig(ix,jy)}]
  int_l = [sum{sig(ix,jy)}] 
\end{verbatim}

where sum is evaluated for `ix' \& `jy' running through [rp\_ix – (k+1)/2] to [rp\_ix + (k+1)/2] \& [rp\_jy
– (k+1)/2] to [rp\_jy + (k+1)/2], respectively , where `k+2' is size of the box and `sig' are background
corrected values of signal.

Finally, an intensity ordered (descending) table of intensities (int\_l) and centroids (c\_l\_x, c\_l\_y) is
generated.

\medskip
{\bf A1.11 uvtDetectDistCorr}  \\

This Module is called by both INT \& PC versions of the Relative Aspect chain and the PC version
of the Sky Imaging chain.

This Module applies corrections for the distortion inherent to the detector Module (primarily due to
the fibre optic taper). This distortion modifies the expected linear relation between location of
electronic detection of a photon event and its original location of arrival on the photo-cathode. Hence,
in order to determine the true position for each photon event, a correction needs to be applied. This
correction has been quantified by extensive calibrations in the laboratory and confirmed in orbit using
well studied astronomical targets. The corrections are identical for all the filters of the band and are
available in the CAL\_DB (DISTORTION/ DETECTOR/ band) as 512$\times$512 arrays. For PC mode, the
correction is applied to centroids of every individual photon event \& for INT mode to the centroids of
each detected star tabulated by the uvtDetectStar Module (Sec. A1.10). The correction values
applicable to the nearest pixel coordinates for each centroid (c\_x, c\_y) are applied :

\begin{verbatim}
c_x_ddc = c_x - 
det_dist_corr_x{nint(c_x-44),nint(c_y-44)};
c_y_ddc = c_y - 
det_dist_corr_y{nint(c_x-44),nint(c_y-44)}
\end{verbatim}

where the arrays `det\_dist\_corr\_x' \& `det\_dist\_corr\_y' are from CAL\_DB (for appropriate band
FUV / NUV / VIS). An offset in indexing (by 44; see Sec A1.7) is needed since the arrays in CAL\_DB
are in 512$\times$512 system (\& all coordinates subsequent to uvtPixPadding Module have them translated to
600x600 system to accommodate spacecraft drift).

\medskip
{\bf A1.12 uvtOpticAssDistCorr}  \\

This Module is called by both INT \& PC versions of the Relative Aspect chain and the PC version of
the Sky Imaging chain.

This Module corrects for distortion due to the Telescope Optical Assembly and is very similar to that
discussed above (uvtDetectDistCorr in Sec. A1.11). This distortion has been estimated from design of
the telescope and is very small. The corrections to be applied are available in the CAL\_DB
(DISTORTION /OPTICS /band/filter). It may be noted that since this particular correction originates
from design of the telescope optics alone, the CAL\_DB arrays are identical for all filter-band
combinations. For PC mode, the correction is applied to centroids of every individual photon event \&
for INT mode to the centroids of each detected star tabulated by the uvtDetectStar Module (Sec.
A1.10). The correction values applicable to the nearest pixel coordinates for each centroid (from the
earlier Module, say, c\_x\_ddc, c\_y\_ddc) are applied :

\begin{verbatim}
c_x_oadc = c_x_ddc - 
 optic_assm_dist_corr_x{nint(c_x_ddc-44),
                      nint(c_y_ddc-44)};
c_y_oadc = c_y_ddc - 
 optic_assm_dist_corr_y{nint(c_x_ddc-44), 
                      nint(c_y_ddc-44)}
\end{verbatim}

where the arrays `optic\_assm\_dist\_corr\_x' \& \\
 `optic\_assm\_dist\_corr\_y' are from CAL\_DB (for
appropriate band FUV / NUV / VIS \& filter F1/ F2 / ...) in 512$\times$512 system.

\medskip
{\bf A1.13 uvtRefFrameCal}  \\

This Module selects timing reference and step size for subsequent extraction of drift series. It is
called by both INT \& PC versions of the Relative Aspect chain.
This Module is responsible for two functionalities : (1) generate the Reference Frame defined by an
ensemble of centroids and intensities of the detected stars with respect to which all relative aspects
for future instances of time will be computed, \& (2) to prepare tables, similar to the preceding
``(1)", of
detected stars for all subsequent time samples. These are then used to find the drifts in a later
processing Module (uvtComputeDrift in Sec. A1.14).
It begins with the tabulated lists of distortion corrected centroids of stars detected. Each table
represents one time sample. The sequence tables as a time series have been generated earlier from
individual Acc\_frame-s (by uvtDetectStar in Sec. A1.10). There are two user selected parameters for
this Module, viz., number of initial time samples to be ignored, N\_skip, \& the number of successive
time samples to be combined as a block, N\_avg, to effectively widen the time bin for computing the
average positions of the stars. The resulting time series from the sequence of blocks will subsequently
be used for computation of the drift.
The very first block is used to generate the Reference Frame by associating stars using neighbourhood
criteria on the centroids from N\_avg number of individual Acc\_frame-s. This is followed by recording
the average values for centroid along X, centroid along Y, Intensity \& Time. This process is repeated
for the subsequent blocks. The selectable parameter N\_skip provides a scheme to handle data set withunusually large drift, by breaking down the full time range into multiple parts.

\medskip
{\bf A1.14 uvtComputeDrift }  \\

This Module is called by both INT \& PC versions of the Relative Aspect chain.
It supports both INT \& PC mode input data and carries out the following tasks : (1) compute the
drift which includes shifts and rotation of the field of selected UVIT band as a function of time in
Detector coordinate system to generate Relative Aspect Series (RAS); (2) time domain digital low pass
filtering of the RAS using selectable parameters; \& (3) to translate the filtered RAS to the spacecraft
coordinate system, viz., Roll, Yaw \& Pitch using mechanical and optical mounting details of the bands
on the spacecraft. The Fig. 16 presents the schematic flow of processing in this uvtComputeDrift
Module.

The drift computation involves at first reading in all the tables of centroids \& intensities for stars,
from products of the earlier Module uvtRefFrameCal (Sec. A1.13). In the next step, within a loop, two
consecutive tables are considered which are referred to using subscripts `p' \& `c' for `previous' and
`current' respectively. The stars common to both the tables are identified and associated as pairs using
their centroid coordinates and an user selected neighbourhood criterion representing maximum
expected relative drift.

The generation of the RAS involves computation of best estimates the three variables
 `$\delta$x\_shift',
`$\delta$y\_shift' \& `$\delta\theta$' in detector coordinate system, representing the best fit for transformation of centroids
from `table\_p' to `table\_c'. This best estimation is carried out using the centroids and a selectable
algorithm (3 choices available). The most commonly used algorithm finds solution for least sum of
squares of deviations, while giving equal weight to all the stars to avoid bias due to some bright stars in
few locations. The algorithm is linearised by use of the approximation
 ``sin $\delta\theta$ = $\delta\theta$ and cos $\delta\theta$ =1"
which is valid as the relative rotations within any Episode are $<$ 0.1 degree.

\begin{figure*} [th]
\centering
\includegraphics[scale=0.40]{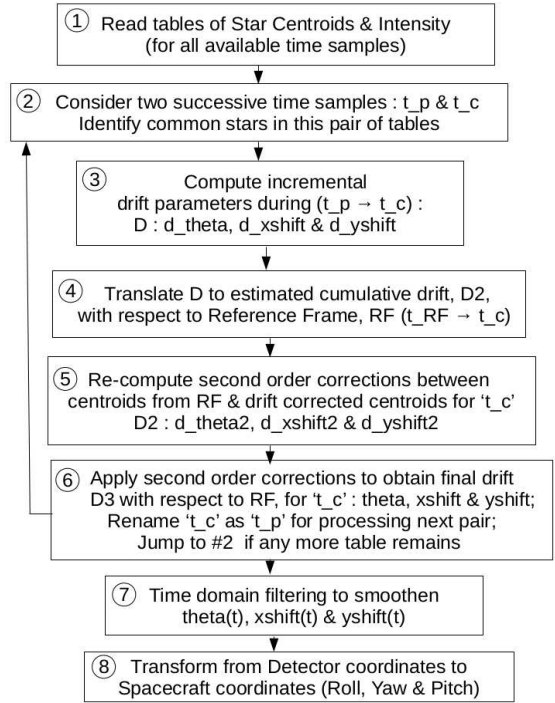}
\caption{
Flow chart for extraction of drift (uvtComputeDrift module).
Using centroids of detected stars in successive frames (sampled every $\sim$ 1 sec for VIS tracking)
the incremental relative drift local in time is estimated first. 
 The quantified drift includes 2 shifts along the detector axes and a possible rotation.
 Subsequently they are
translated to cumulative total drift as a function of time, with respect to the Reference Frame.  
Time domain filtering is used for smoothen the 3 drift variables. Finally they are
transformed to spacecraft coordinate system (for use by any of the bands). 
}
\end{figure*}

In order to limit the error propagation through addition of multiple incremental corrections over
multiple time steps, the above scheme is repeated again between the star centroids in the Reference
Frame (time\_RF) and the drift corrected centroids for any time sample time\_c obtained above. The
additional corrections obtained now (say, $\delta\theta$2, $\delta$x2\_shift \& $\delta$y2\_shift) are used to compute the final
drift values valid for time\_c, as :

{\small
 {$\theta$}\_RF\_fin(time\_c) = {$\theta$}\_RF\_est(time\_c) + {$\delta$}{$\theta$}2  \\
 x\_shift\_RF\_fin(time\_c) = x\_shift\_RF\_est(time\_c) + {$\delta$}x2\_shift  \\
 y\_shift\_RF\_fin(time\_c) = y\_shift\_RF\_est(time\_c) + {$\delta$}y2\_shift  \\
}

The complete process described above is repeated for the entire time range by advancing the time by
steps of one in a loop. The resulting time series of $\theta$\_RF\_fin,
 x\_shift\_RF\_fin \& y\_shift\_RF\_fin are
the Relative Aspect Series in Detector coordinate system. Other choices of algorithm for extraction of
drift parameters include : (a) dropping the assumption about `$\theta$' to be small; \& (b) assigning intensity
based weights to individual stars (weight proportional to intensity). In all the 3 algorithms, user has a
choice to drop the parameter  `$\theta$' from drift extraction (i.e. consider only shifts and no rotation).

The time domain low pass filtering is carried out to minimise errors in the drift parameters . A
sliding time window ($\Delta$T) selects the samples over (t- $\Delta$T/2) to (t + $\Delta$T/2) duration. They are then
fitted with a polynomial of degree `n\_poly' using a least square method.
 The value for fittedpolynomial at mid-point of the time window is assigned
 as the filtered result for time instance `t'. This
process is repeated through the complete time series. While $\Delta$T and n\_poly are selectable by the user
the most commonly used values for these have been 4 seconds \& 1 respectively.

Finally the drift parameters are transformed from the Detector coordinate system to the spacecraft
coordinate system using available transformation matrices. The convention followed for mapping them
are : $\theta {\Rightarrow}$ Roll, X $\Rightarrow$ Pitch \& Y $\Rightarrow$ Yaw. Details of the two way transformations between the Detector
and spacecraft coordinate systems are available in Appendix-3. The final Relative Aspect Series,
RAS, are made available in both the detector as well as spacecraft coordinate systems, which are
required by both versions of the Sky Imaging chains (L2\_PC \& L2\_INT). The standard bundle of the
pipeline products includes the two RAS tables for every individual Episode, one each from RA\_INT
and RA\_PC chains. The tables show the drifts at regular intervals of time.

\medskip
{\bf A1.15 uvtCentroidCorr}  \\

This Module was planned for the PC mode version of the Sky Imaging chain.
It corrects for a subtle effect which introduces shift in the centroid of a photon event computed on
board. It can arise if estimate of the background signal, which is equated to the lowest value of the four
corner signals of 5X5 window around centre of the event, used for on-board calculations of centroid
differs significantly from the actual value due to a large noise. The mitigation scheme employed use of
median filtered background from measured dark frames in orbit. During early operations in orbit, the
noise in background was found to be insignificant. Accordingly, need for activating this Module has
not arisen. This correction must precede the correction for centroid bias (uvtCentroidBias, Sec. A1.16)
described next.

\medskip
{\bf A1.16 uvtCentroidBias  }  \\

This Module is called by both the Relative Aspect \& Sky Imaging chains for the PC mode.
It applies corrections for bias introduced on the centroids of photon events computed by on board
signal processing. This systematic bias is inherent to the scheme adopted for obtaining the centroids
from raw pixel values in the event foot print (3x3 or 5x5), which slightly modifies the fractional part of
the computed centroid value from the corresponding true value. Sometimes this effect is also referred to
as Fixed Pattern Noise. The effects of this bias have been quantified from the extensive laboratory
calibrations which have been translated into look up tables correlating the measured with the true values
for the fractional part of the centroids along both the axes (X, Y). The CAL\_DB provides these look up
tables. This uvtCentroidBias Module recovers the true value from the measured one by applying the
corresponding correction (for the appropriate band FUV / NUV / VIS \& selected foot print 3x3 / 5x5;
e.g. BIAS\_CORR/FUV/3x3) to all centroids in the master table of events.

\medskip
{\bf A1.17 uvtShiftRot module \& further updating of event table }  \\

This Module is called by both INT \& PC versions of the Sky Imaging chain.
The uvtShiftRot Module carries out the following two tasks for INT mode observations : (A) apply
the drift to individual images - single frame or `N\_acc' frames clubbed together as Acc\_frame in the
Module uvtAccEveryTsec, to align these with the Reference Frame; (B) create Exposure Arrays
corresponding to each individual image taking into account the active areas of the detector and 
aligning them to the Reference Frame by applying drift corrections. For the PC mode, only the first of these two
functionalities is incorporated by applying the drift to the centroids of photons. In the PC mode, instead
of Exposure Arrays for individual exposure an integrated Exposure Array is created by adding the
Exposure Arrays corresponding to individual images. This has been done as the number of individual
images in an Episode could be about a million storing these individually would be too expensive in
memory space.

 {\it A1.17.1 Task-A :}

The drift parameters for every image are calculated by a linear interpolation of the entries in the table of
drifts for the band, provided by A1.14 uvtComputeDrift, at the relevant time.

 {\it A1.17.1a PC mode -}

For PC mode data, these interpolated drift parameters are used to apply identical corrections to
centroids of all photon events of the specific frame. \\
 {\small
 c\_x\_snr= x\_shift + (c\_x -x0)*cos$\theta$ - (c\_y -y0)*sin$\theta$ + x0  \\
 c\_y\_snr= y\_shift + (c\_x -x0)*sin$\theta$ - (c\_y -y0)*cos$\theta$ + y0  \\
}

where x0 = x\_size/2 \& y0 = y\_size/2;
where the `snr' subscript refers to the shift and rotation corrected centroids \& the (c\_x, c\_y) are the most
recently updated centroids from the master photon event list from the module ``uvtOpticAssDistCorr"
(Sec A1.12).
These drift corrected centroid values (c\_x\_snr, c\_y\_snr) for each photon event replace the earlier
entries for centroid values in the master table of events and remain available for subsequent processing
blocks till the very end of execution of the chain. In case the RAS does not cover the entire time range
for the Episode, only the part with time overlap is retained and the remaining events are discarded since
the drift corrections can not be applied to them. This updated table is preserved in the sub-directory
labeled `uvtShiftRot\_6.3' within the product bundle for the corresponding Episode, in a file named with
the suffix `\_snr' (see Appendix-9).

  {\it Further updating by a later processing (uvtFullFrameAst Module) :}

The photon event table finally updated by this uvtShiftRot Module (described above) gets further
refined by a later processing Module carrying out astrometric corrections by matching detected stars
with those in an existing standard catalogue (module uvtFullFrameAst; Sec. A1.20). A new event table
is generated which adds fresh information regarding coordinates and time while retaining all the past
details, located in the same sub-directory with the file name having the suffix `\_radec'.

Five additional columns are appended, which are named `RA', `DEC', `MJD\_L2', `RA-aaaaa' \&
`DEC-aaaaa'. The first pair contains the best estimate of the event coordinates in RA, Dec (ICRS;
J2000) system relative to the map centre, which are the final values obtained from the pipeline run for
the corresponding Episode including Astrometry. The last pair correspond to the RA, Dec obtained
through transformation of event centroids from Detector (X, Y) system to RA, Dec system using the
aspect information from the spacecraft. In the event of failure of the astrometry block, both sets of
coordinates would be identical.

The column named `TIME' was planned to provide the primary timing information for each photon
event. It is normally populated with SPS\_MJD, the absolute time provided by L1 through calibration of
the spacecraft’s on board clock, through selection of the `utcFlag' to be ON. Alternately, if the UTC
switch is selected to be OFF (when SPS\_MJD showed erratic values occasionally), this `TIME' column
is populated with the UVIT’s Master clock, and the `MJD\_L2' column in the table would still provide
estimated absolute time with about $\sim$ 1 sec accuracy, using local calibration of the UVIT’s Master clock
(carried out within the L2 pipeline, in DataIngest Module – Sec. A1.1). This enables astronomical
variability analysis at time scales shorter than typical duration of observation episodes which is $\sim$ 2000
seconds under all circumstances.

The standard bundle of the pipeline products disseminated by the ISSDC includes this final resulting
table (with suffix `\_radec'), as the event list table, corresponding to every data set from individual
Episodes.

\medskip
 {\it A1.17.1b INT mode -}

For INT mode data, individual image frames are first transformed by pixel sub-division from
original size 4800$\times$4800 to 9600$\times$9600. Then contents of each of these finer sub-pixels are moved to a
new location after application of the shift and rotation correction using relations similar to the above.
Finally, a 2x2 binning is carried out to bring the array dimensions back to original level. The process
ensures local `conservation' of signals and adequate handling of `Bad Pixels'. At the end, as many drift
corrected image frame products are generated as were available at the input. They are all in Detector
coordinate system (X, Y).

  {\it A1.17.2 Task-B :}

For the second task of generating Exposure Arrays, at first the relevant ``Template Exposure Array" is
accessed from CAL\_DB (EXPOSURE\_TEMPLATE / band / mode / window size). Primarily, entries in
this array are `1' for all sub-pixels corresponding to the good pixels within the selected window and `0'
for the rest. For every image qualified for inclusion in making sky image, a corresponding Exposure
Array is generated by applying drift corrections (on the template array) appropriate for that time instance
following the procedure described in Task-A (A1.17.1b INT mode).

\medskip
{\bf A1.18 uvtFrameIntegration  }  \\

This Module is called by both the Relative Aspect and Sky Imaging chains for PC mode data. It is
responsible for gridding the photon centroids from multiple (N\_combine) successive frames onto 2-D
array/(s). For the RA\_PC chain, a time series of images, Comb\_frames, are generated which are used for
identifying individual stars in a subsequent processing stage. For the L2\_PC chain this module is either
used to combine all the frames for the Episode into a single image or construct multiple images by sub-
dividing the full data into smaller time intervals corresponding to `pseudo-Episode'-s. The latter option
enables application of corrections for small mis-alignments due to any temporal effects, in a subsequent
processing step (module uvtRegAvg; Sec. A1.19).

The typical values of the time step `T' corresponding to the selectable parameter, N\_combine, are $\sim$ 1
s and $\sim$ 300 s for RA\_PC and L2\_PC chains respectively. Time sequence of images, Comb\_frames, are
referred to as 4800$\times$4800 ``Sig" arrays. The last image would be made with $<$ N\_combine frames in
case the Total number of frames is not integer multiple of N\_combine. There is a provision
 to skip N\_discard initial frames to discard the initial frames either due to any starting disturbance or during a
rapid drift of the pointing. Corresponding sole / sequence of ``Exposure"
 array/s and ``Uncertainty"
array/s are also generated in the L2\_PC chain. 
The ``Signal" array/s are obtained by pixel by pixel
division of the ``Sig" array/s by the corresponding ``Exposure" array/s and the statistical ``Uncertainty"
array/s are generated using the ``Signal" and ``Exposure" array/s
{
(based on square-root of estimated number of total EFFECTIVE photon events,
  corrected for the variation of responsivity on the detector,
   in each individual pixel and not the actual number of photon events).
}
The ``Exposure"
 products from this
module are stored in the directory `uvtExposureFrames\_6.3' and the rest (Sig, Signal \& Uncertainty) in
sub-directories under the directory `uvtFrameIntegration\_6.3'.

\medskip
{\bf A1.19 uvtRegAvg  }  \\

This Module is called by both INT \& PC versions of the Sky Imaging chain.
It may be noted that in an earlier processing block (PC : uvtFrameIntegration Module; Sec. A1.18;
INT : uvtFindWtdMean for L2\_INT (not described here), the dataset from a specific Episode has either
been treated as a whole (`Episode') or in smaller parts (`pseudo-Episode') as per user’s choice. Here, the
Module uvtRegAvg, combines the multiple Signal (``Sig") and Exposure (``Exposure") arrays in case
`pseudo-Episode' option was exercised in that earlier bloc. These multiple `Sig' \& `Exposure' arrays are
combined after determining relative Shifts and Rotation between them using an algorithm to align
detected brightest point sources from successive `Sig' arrays which were generated earlier on. The
alignment operations are identical for each pair of `Sig' \& `Exposure' arrays.

Additionally, for PC mode UV imaging, these combined `Sig' \& `Exposure' arrays are used to
generate the output `Signal' \& `Uncertainty' arrays in true ``count/second" unit, using the same scheme
as described earlier (uvtFrameIntegration Module; Section A1.18).
The same operation is carried out on the sole pair of `Sig' \& `Exposure' arrays in case the `Episode'
option was exercised in the preceding processing block to generate all 3 final products, viz., `Signal',
`Exposure' \& `Uncertainty'.

At the end of this uvtRegAvg process (irrespective of the choice between pseudo-Episode /
Episode), a single set of sky image, exposure \& uncertainty products are generated which utilize the
entire data set for the particular observation Episode (for a particular combination of Band, Filter \&
Window size in one orbit). These final array products Signal (counts/sec), Exposure (no of frames) \&
Uncertainty (counts/sec) are of 4800$\times$4800 size and in Detector coordinate (X, Y) system and stored in
the `uvtRegAvg\_6.3' directory (see Appendix-9).

Since the algorithms used in this uvtRegAvg Module for 3 different steps of processing are identical
to those described earlier for other Modules, only their references are mentioned here. The scheme to
find bright point sources in individual component images (from pseudo-Episodes) is similar to that
used for the Module uvtDetectStar (Sec. A1.10) \& to determine relative offset (shifts and rotation)
between them uses the scheme followed for the Module uvtComputeDrift (Sec. A1.14). Finally, the
scheme employed for aligning these individual images based on the locations of detected point
sources, is the same as followed for the Module uvtShiftRot (Sec. A1.17.1b; Task-A, for INT mode). It
may be noted that the alignment of pseudo-Episodes involves transformation of image sub-pixels
resulting in some loss of angular precision, unlike alignment among Episodes carried out in the Driver
Scheme (Sec. 3.2.4) in which individual photon centroids are transformed.

\medskip
{\bf A1.20 uvtFullFrameAst  }  \\

This Module is called by both INT \& PC versions of the Sky Imaging chain as well as the DriverScheme, which determines finer corrections to the sky coordinates of image products by correlating stars
detected in the UV image with standard astronomical catalogues of stars.

The Module uvtFullFrameAst attempts to improve the accuracy of the aspect of the UV sky image
products generated prior to invoking this Module. The input images have their sky coordinates
determined based on information of spacecraft attitude (Roll\_RA, Roll\_DEC, Roll\_ROT; ICRS J2000)
provided in the Level-1 Aux data (`*.att' file in `mL1' dataset). The typical accuracy of these coordinates
is $\sim$ 45 arc-sec RMS and up to 3 arc-min peak-to-peak. Hence, there is a large scope for improvement
by correlating stars detected in the image products with standard star catalogues to first determine the
offsets and then applying corresponding corrections.

The algorithm followed is described here. First identify brightest point like sources detected from the
UVIT image (NUV/FUV) and record their RA-Dec (ICRS) coordinates. They are then correlated with
brightest entries in the USNO A2 optical star catalogue within a selected search radius ($\sim$ 3 arc
minutes). The USNO catalogue was reformatted by – (a) retaining stars brighter than 16 magnitude \&
(b) embedding an indexing scheme to enable very fast search by the pipeline. `Good' or `accidental'
matches of UVIT-USNO pairs are classified based on a sequence of statistical tests carried out
iteratively. At first the brightest pairs are considered and progressively less bright pairs are also
included for this analysis. If successful in identifying confirmed matches, a least square method is
employed to extract the 3 parameters – shifts along RA \& Dec and a rotation around the centre of the
image (similar to the algorithm described for the Module uvtComputeDrift, Sec. A1.14). The next step
is to apply these corrections (shifts \& rotation) to the 3 standard products, viz., UV image (`Signal'),
Exposure and Uncertainty arrays. The scheme for application of these astrometric corrections to the
UV image differs between the case for individual Episode 
 (when sub-division into pseudo-Episodes is not selected),
from that for multi-Episode case. While the
astrometric corrections are applied to centroids of individual photons in the former case, the shift \&
rotational transformations are applied on the sub-divided array elements for the latter.

\medskip
 {\it In the Sky Imaging Chain L2\_PC -}

When this uvtFullFrameAst Module is called by the Sky Imaging chain L2\_PC for an individual
Episode, the corrections are applied to the centroids (RA, Dec) of each photon event and the master
table of events is updated. Two additional columns are introduced in this master table, which hold the
astrometry corrected coordinates (details given in Sec. A1.17 for Module uvtShiftRot, Task-A, PC
mode). The final UV image is generated by populating these photons in Signal array of
 4800$\times$4800 size.
This process ensures minimal loss of precision (angular resolution) while applying astrometric
corrections. Similar corrections for the Exposure and Uncertainty arrays are implemented by physical
movements of pixel contents as described for Module uvtShiftRot (Sec. A1.17, Task-A, INT mode; first
pixel sub-division from 4800$\times$4800 to 9600$\times$9600; re-gridding individual sub-pixels by applying
transformation; \& finally 2x2 binning).

\medskip
 {\it In the Driver Scheme -}

When this uvtFullFrameAst Module is called by the Driver Scheme to apply astrometric corrections
to the combined multi-Episode products, all 3 arrays viz., Signal, Exposure and Uncertainty are handled
identically. In this case, Signal array also undergoes the same scheme of corrections for the two shifts
and the rotation, as implemented for the Exposure and Uncertainty arrays in the case for individual
Episodes described in the preceding para.

In addition to using the optical catalogue USNO A2, bright NUV \& FUV source catalogues from
GALEX mission are also used to correlate bright point like sources from UVIT images. The good
matches of UVIT-GALEX pairs are also used to extract the offset corrections, but they are only
recorded in a log file. The details of star match `success' / `failure' with Optical \& UV catalogues are
recorded as HISTORY / COMMENT entries on image HEADERs.

All the image products Signal (counts/sec), Exposure (no of frames) \& Uncertainty (counts/sec) are of
4800$\times$4800 size and in RA-Dec (ICRS; J2000) coordinate system. They all are stored in the sub-
directory named `uvtFullFrameAst\_6.3' either below a directory labeled by the specific Episode (see
Appendix-9) or a directory identifying the ``group" (Band/ Filter/ Window) for multi-Episode case (see
Appendix-10).
In case of failure of this astrometry block, the set of resulting products contain
arrays identical to pre-astrometry images with adequate qualification 
in header keywords of the failure.

The standard bundle of pipeline products archived and disseminated by ISSDC includes the Signal
image product from this Module as final UV sky image in Astronomical coordinates, corresponding to
every Episode generated by the Sky Imaging chain (L2\_PC), as well as combined multi-Episode images
generated by the Driver Scheme for every ``group". The corresponding Exposure and Uncertainty
products are also included.

\medskip


{\bf Appendix 2 : Relative time shifts between the three Bands} \\
In spite of a common Master Clock used by all three Bands of UVIT, the timing details of detector
read out electronics lead to systematic shifts in the time stamps recorded on individual frames. These
shifts depend on the imaging parameters configured. The relative time shifts most relevant for the
pipeline are between the band used to obtain Relative Aspect Series and the Imaging UV bands. This
has been calibrated (in seconds unit) to be :

{\scriptsize

   t\_RAS - t\_img = t0\_RAS - (-T\_Img\_frame/ 2)
}
\medskip

where `t0\_RAS' depends on the band generating RAS, its imaging mode \& parameters; \&
`T\_Img\_frame' is the frame integration time of the imaging Band (NUV / FUV);
t0\_RAS =-0.427; for drift tracking by VIS band in INT mode; full 512$\times$512 field;
t0\_RAS = -(T\_RAS\_frame/2); for drift tracking by NUV band in PC mode; any window size;
where ‘T\_RAS\_frame’ is the frame integration time of the tracking Band (NUV).





\medskip
{\bf Appendix 3 : Transformations between the Detector and Spacecraft coordinate systems} \\

The Detector coordinate system for each band is defined by the X-Y axes of its electronic
 sensor (Star-250), and the axis normal to this plane corresponds to the optic axis of the telescope (either directly as
for the FUV \& VIS bands, or via one reflection with a plane surface; see Fig. 1). The rotation angle
about the optic axis is denoted by `$\theta$' in the detector system. The Spacecraft coordinate system comprise
of the axes Roll, Yaw \& Pitch which is defined by the structure of the spacecraft. These two systems are
interconnected through the details of mounting of sub-systems of UVIT on to the spacecraft’s
mechanical interface. Nominally, the optic axis is aligned parallel to the Roll axis and the (X, Y) and
(Yaw, Pitch) are connected through a single angle of rotation about an axis normal to all these 4 axes.
Similarly, one rotation angle connects any pair of detectors for two bands. All these angles were initially
measured during optical alignment tests in the laboratory. These were subsequently validated andrefined by in orbit measurements during the Performance Verification phase.

The transformations between detector (X, Y, $\theta$) and spacecraft (Roll, Yaw \& Pitch) are implemented
as matrices for band, BND, as RPYTOXYTHETA\_BND, \{BND : FUV / NUV / VIS\}, populated with
numerical values, along with a corresponding plate scale. The reverse transformations,
XYTHETATORPY\_BND, are generated internally by matrix inversion.

{\tiny

{\noindent}RPYTOXYTHETA\_FUV : -0.00249798, 0.29636841, 0.29636841, 0.00249798\\
RPYTOXYTHETA\_NUV : 0.15423, 0.25153, -0.25153, 0.15423\\
RPYTOXYTHETA\_VIS : 0.1627, 0.2400, 0.2400, -0.1627\\
}


The above are consistent with the angles between +Y and the -Yaw (measured CCW) to be +0.483,
+31.515 and +34.134 degrees for the FUV, NUV \& VIS bands respectively. They are also consistent with
the following directly observable relations which connect image coordinates of a star in the NUV \& VIS
detectors (accounting for slightly different plate scale for VIS; dX \& dY are referenced with respect to
the respective array centres) :

{\tiny

{\noindent}dX\_NUV = 1.01652 * dX\_VIS - 0.04649 * dY\_VIS\\
dY\_NUV = - 0.04649 * dX\_VIS - 1.01652 * dY\_VIS\\
}

The corresponding relations connecting image coordinates of a star in the FUV \& VIS detectors are :
{\tiny

{\noindent}dX\_FUV = -0.85093 * dX\_VIS + 0.56645 * dY\_VIS\\
dY\_FUV = 0.56645 * dX\_VIS - 0.85093 * dY\_VIS\\
}





{\bf Appendix-4 : Directory structure of merged Level-1 data set (mL1)} \\

%

{\tiny

{\noindent}-AS1G06\_157T01\_9000000772uvt\_level1\_mcap.xml\\
-AS1G06\_157T01\_9000000772uvt\_level1.mkf\\

---aux\\
-AS1G06\_157T01\_9000000772uvt\_level1.tct\\
---aux1\\
-AS1G06\_157T01\_9000000772uvt\_level1.att\\
-AS1G06\_157T01\_9000000772uvt\_level1.gyr\\
-AS1G06\_157T01\_9000000772uvt\_level1.orb\\
---aux2\\
-AS1G06\_157T01\_9000000772uvt\_level1.lbt\\
---aux3\\
---AS1G06\_157T01\_9000000772uvt\_level1.orb\\

-hk\\
-AS1G06\_157T01\_9000000772uvt\_level1.hk\\

---uvtF\\
------DarkF\\
------uvtF.01\\
-AS1G06\_157T01\_9000000772uvtFIIPC00F2\_level1.bti\\
-AS1G06\_157T01\_9000000772uvtFIIPC00F2\_level1.fits\\
-AS1G06\_157T01\_9000000772uvtFIIPC00F2\_level1.gti\\
------uvtF.02\\
-AS1G06\_157T01\_9000000772uvtFIIPC00F0\_level1.bti\\
-AS1G06\_157T01\_9000000772uvtFIIPC00F0\_level1.fits\\
-AS1G06\_157T01\_9000000772uvtFIIPC00F0\_level1.gti\\
------uvtF.03\\

...\\

---uvtN\\
------DarkN\\
------uvtN.01\\
-AS1G06\_157T01\_9000000772uvtNIIPC00F5\_level1.bti\\
-AS1G06\_157T01\_9000000772uvtNIIPC00F5\_level1.fits\\
-AS1G06\_157T01\_9000000772uvtNIIPC00F5\_level1.gti\\
------uvtN.02\\
-AS1G06\_157T01\_9000000772uvtNIIPC00F0\_level1.bti\\
-AS1G06\_157T01\_9000000772uvtNIIPC00F0\_level1.fits\\
-AS1G06\_157T01\_9000000772uvtNIIPC00F0\_level1.gti\\
------uvtN.03\\

    ...\\

---uvtV\\
------DarkV\\
------uvtV.01\\
-AS1G06\_157T01\_9000000772uvtVIIIM00F2\_level1.bti\\
-AS1G06\_157T01\_9000000772uvtVIIIM00F2\_level1.fits\\
-AS1G06\_157T01\_9000000772uvtVIIIM00F2\_level1.gti\\
------uvtV.02\\
-AS1G06\_157T01\_9000000772uvtVIIIM00F2\_level1.bti\\
-AS1G06\_157T01\_9000000772uvtVIIIM00F2\_level1.fits\\
-AS1G06\_157T01\_9000000772uvtVIIIM00F2\_level1.gti\\
------uvtV.03\\

...\\
}




%

{\bf Appendix-5 : Directory structure of UVIT Calibration Database (CAL\_DB)} \\

{\tiny

{\noindent}-change.log\\
-readme.txt\\

-AV\_PH\_ENERGY (scalar)\\
--FUV\\
-calibfile.fits\\
--NUV\\
-calibfile.fits\\
--VIS\\
-calibfile.fits\\

-BAD\_PIXELS (2-D array 512 x 512)\\
--FUV\\
---IM\\
----100X100\\
-calibfile.fits\\

...\\
----512X512\\
-calibfile.fits\\
---PC\\
----100X100\\
-calibfile.fits\\

...\\
----512X512\\
-calibfile.fits\\

{\noindent}--NUV\\
---IM\\
----100X100\\
-calibfile.fits\\

...\\
----512X512\\
-calibfile.fits\\
---PC\\
----100X100\\
-calibfile.fits\\

...\\
----512X512\\
-calibfile.fits\\

{\noindent}--VIS\\
---IM\\
----100X100\\
-calibfile.fits\\

...\\
----512X512\\
-calibfile.fits\\
---PC\\
----100X100\\
-calibfile.fits\\

...\\
----512X512\\
-calibfile.fits\\

-BIAS\_CORR (table : 3 col x 4097 row)\\
--FUV\\
---3X3\\
-calibfile.fits\\
---5X5\\
-calibfile.fits\\

{\noindent}--NUV\\
---3X3\\
-calibfile.fits\\
---5X5\\
-calibfile.fits\\

{\noindent}--VIS\\
---3X3\\
-calibfile.fits\\
---5X5\\
-calibfile.fits\\

-DARK (2-D array 512 x 512)\\
-darkend.fits\\
-darkstart.fits\\

-DISTORTION (Two 2-D arrays 512 x 512;\\
........ first for X axis \& second for Y axis)\\
--DETECTOR\\
---FUV\\
-calibfile.fits\\

{\noindent}---NUV\\
-calibfile.fits\\

{\noindent}---VIS\\
-calibfile.fits\\

{\noindent}--OPTICS\\
---FUV\\
-----F0\\
-calibfile.fits\\

...\\
-----F7\\
-calibfile.fits\\

{\noindent}---NUV\\
-----F0\\
-calibfile.fits\\

...\\
-----F7\\
-calibfile.fits\\

{\noindent}---VIS\\
-----F0\\
-calibfile.fits\\

...\\
-----F5\\
-calibfile.fits\\

-EXPOSURE\_TEMPLATE (2-D array 1200 x 1200)\\
--FUV\\
---IM\\
----100X100\\
-calibfile.fits\\

...\\
----512X512\\
-calibfile.fits\\
---PC\\
----100X100\\
-calibfile.fits\\

...\\
----512X512\\
-calibfile.fits\\

{\noindent}--NUV\\
---IM\\
----100X100\\
-calibfile.fits\\

...\\
----512X512\\
-calibfile.fits\\
---PC\\
----100X100\\
-calibfile.fits\\

...\\
----512X512\\
-calibfile.fits\\

{\noindent}--VIS\\
---IM\\
----100X100\\
-calibfile.fits\\

...\\
----512X512\\
-calibfile.fits\\
---PC\\
----100X100\\
-calibfile.fits\\

...\\
----512X512\\
-calibfile.fits\\


-FLAT\_FIELDS\_FILTER (2-D array 512 x 512)\\
--FUV\\
---IM\\
----F0\\
-calibfile.fits\\

...\\
{\noindent}----F7\\
-calibfile.fits\\

{\noindent}---PC\\
----F0\\
-calibfile.fits\\

...\\
{\noindent}----F7\\
-calibfile.fits\\

{\noindent}--NUV\\
---IM\\
----F0\\
-calibfile.fits\\

...\\
{\noindent}----F7\\
-calibfile.fits\\

{\noindent}---PC\\
----F0\\
-calibfile.fits\\

...\\
{\noindent}----F7\\
-calibfile.fits\\

{\noindent}--VIS\\
---IM\\
----F0\\
-calibfile.fits\\

...\\
{\noindent}----F5\\
-calibfile.fits\\

{\noindent}---PC\\
----F0\\
-calibfile.fits\\

...\\
{\noindent}----F5\\
-calibfile.fits\\

-QE\_TEMP (table : 9 col x 20 row)\\
--FUV\\
---IM\\
-calibfile.fits\\
---PC\\
-calibfile.fits\\

{\noindent}--NUV\\
---IM\\
-calibfile.fits\\
---PC\\
-calibfile.fits\\

{\noindent}--VIS\\
---IM\\
-calibfile.fits\\
---PC\\
-calibfile.fits\\

}




%
 {\bf           Appendix-6 Directory structure for DataIngest output
                        (Integration mode; INT) } \\


{\tiny

{\noindent}-AS1A03\_103T01\_9000001132uvtVIIIM00F4\_l2\_di.crcFailedList\\
-AS1A03\_103T01\_9000001132uvtVIIIM00F4\_l2\_di.dataIngestOut\\
-AS1A03\_103T01\_9000001132uvtVIIIM00F4\_l2\_di.info\\
-AS1A03\_103T01\_9000001132uvtVIIIM00F4\_l2\_di.missingFrameList\\
-AS1A03\_103T01\_9000001132uvtVIIIM00F4\_l2\_di.timeCorrection\\

-Dark\\
--AS1A03\_103T01\_9000001132uvtVIIIM00F4\_l2\_darkbegin.fits\\
--AS1A03\_103T01\_9000001132uvtVIIIM00F4\_l2\_darkend.fits\\

-SignalFrames\\
--AS1A03\_103T01\_9000001132uvtVIIIM00F4\_l2\_t2734052.8140\_f16\_di.fits\\
--AS1A03\_103T01\_9000001132uvtVIIIM00F4\_l2\_t2734053.7900\_f32\_di.fits\\
--AS1A03\_103T01\_9000001132uvtVIIIM00F4\_l2\_t2734054.7670\_f48\_di.fits\\

...\\

...\\

}
%




%
{\bf    Appendix-7 Directory structure for DataIngest output
                 (Photon Counting mode; PC) } \\


{\tiny

{\noindent}-AS1G06\_157T01\_9000000772uvtFIIPC00F6\_l2\_di.crcFailedList\\
-AS1G06\_157T01\_9000000772uvtFIIPC00F6\_l2\_di.dataIngestOut\\
-AS1G06\_157T01\_9000000772uvtFIIPC00F6\_l2\_di.events\\
-AS1G06\_157T01\_9000000772uvtFIIPC00F6\_l2\_di.image\\
-AS1G06\_157T01\_9000000772uvtFIIPC00F6\_l2\_di.info\\
-AS1G06\_157T01\_9000000772uvtFIIPC00F6\_l2\_di.missingFrameList\\
-AS1G06\_157T01\_9000000772uvtFIIPC00F6\_l2\_di.timeCorrection\\

--Dark\\
-AS1G06\_157T01\_9000000772uvtFIIPC00F6\_l2\_darkbegin.fits\\
-AS1G06\_157T01\_9000000772uvtFIIPC00F6\_l2\_darkend.fits\\
 
}
%




{\bf  Appendix 8 Directory structure for the outputs from the
      Relativistic Aspect Chain for Integration Mode (RA\_INT)}\\

{\tiny

{\flushleft{DataIngest\_6.3}}\\
-AS1A03\_103T01\_9000001132uvtVIIIM00F4\_l2\_di.crcFailedList\\
-AS1A03\_103T01\_9000001132uvtVIIIM00F4\_l2\_di.crcFailedList\\
-AS1A03\_103T01\_9000001132uvtVIIIM00F4\_l2\_di.dataIngestOut\\
-AS1A03\_103T01\_9000001132uvtVIIIM00F4\_l2\_di.info\\
-AS1A03\_103T01\_9000001132uvtVIIIM00F4\_l2\_di.missingFrameList\\
-AS1A03\_103T01\_9000001132uvtVIIIM00F4\_l2\_di.timeCorrection\\

-Dark\\
-AS1A03\_103T01\_9000001132uvtVIIIM00F4\_l2\_darkbegin.fits\\
-AS1A03\_103T01\_9000001132uvtVIIIM00F4\_l2\_darkend.fits\\

-SignalFrames\\
-AS1A03\_103T01\_9000001132uvtVIIIM00F4\_l2\_t2734052.8140\_f16\_di.fits\\
-AS1A03\_103T01\_9000001132uvtVIIIM00F4\_l2\_t2734053.7900\_f32\_di.fits\\

...\\

...\\

-uvtComputeDrift\_6.3\\
-AS1A03\_103T01\_9000001132uvtVIIIM00F4\_l2\_dr.fits\\
-AS1A03\_103T01\_9000001132uvtVIIIM00F4\_l2\_dr.info\\
-observation.txt\\

-uvtRefFrameCal\_6.3\\
-AS1A03\_103T01\_9000001132uvtVIIIM00F4\_l2\_rfc.info\\

-Centroid\\           
-AS1A03\_103T01\_9000001132uvtVIIIM00F4\_l2\_t2734053.790000\_f1\_rf\_centroid.fits\\
-AS1A03\_103T01\_9000001132uvtVIIIM00F4\_l2\_t2734054.767000\_f2\_rf\_centroid.fits\\

...\\

...\\

}





%
{\bf Appendix 9 : Directory structure for the outputs from the Level-2
         Sky Imaging Chain for Photon Counting Mode (L2\_PC)} \\

{\tiny

{\flushleft{DataIngest\_6.3}}\\
-AS1A03\_103T01\_9000001132uvtNIIPC00F1\_l2\_di.crcFailedList\\
-AS1A03\_103T01\_9000001132uvtNIIPC00F1\_l2\_di.dataIngestOut\\
-AS1A03\_103T01\_9000001132uvtNIIPC00F1\_l2\_di.events\\
-AS1A03\_103T01\_9000001132uvtNIIPC00F1\_l2\_di.image\\
-AS1A03\_103T01\_9000001132uvtNIIPC00F1\_l2\_di.info\\
-AS1A03\_103T01\_9000001132uvtNIIPC00F1\_l2\_di.missingFrameList\\
-AS1A03\_103T01\_9000001132uvtNIIPC00F1\_l2\_di.timeCorrection\\

-Dark\\
-AS1A03\_103T01\_9000001132uvtNIIPC00F1\_l2\_darkbegin.fits\\
-AS1A03\_103T01\_9000001132uvtNIIPC00F1\_l2\_darkend.fits\\

-uvtExposureFrames\_6.3\\
-AS1A03\_103T01\_9000001132uvtNIIPC00F1\_l2\_t0.0000\_f1\_exp\_fi.fits\\

-uvtFlippedRegImage\_6.3\\
-AS1A03\_103T01\_9000001132uvtNIIPC00F1\_l2\_t0.0000\_f1\_10Per\_rg.fits\\
-AS1A03\_103T01\_9000001132uvtNIIPC00F1\_l2\_t0.0000\\
............................\_f1\_expFlipped\_rg.fits\\
-AS1A03\_103T01\_9000001132uvtNIIPC00F1\_l2\_t0.0000\\
............................\_f1\_NoiseMapFlipped\_rg.fits\\
-AS1A03\_103T01\_9000001132uvtNIIPC00F1\_l2\_t0.0000\\
............................\_f1\_sigFlipped\_rg.fits\\

-uvtFrameIntegration\_6.3\\
--NOISE\_MAP\\
-AS1A03\_103T01\_9000001132uvtNIIPC00F1\_l2\_t2734148.0000\\
...........................\_f1\_sig\_noise\_map\_sig.fits\\
--SignalFrames\\
-AS1A03\_103T01\_9000001132uvtNIIPC00F1\_l2\_t2734148.0000\\
...........................\_f1\_sig\_fi.fits\\
--SignalFrames\_DividedWithExposure\\
-AS1A03\_103T01\_9000001132uvtNIIPC00F1\_l2\_t2734148.0000\\
...........................\_f1\_sig\_Sig\_DividedWithExp.fits\\

-uvtFullFrameAst\_6.3\\
-AS1A03\_103T01\_9000001132uvtNIIPC00F1\_l2\_as\_Exp.fits\\
-AS1A03\_103T01\_9000001132uvtNIIPC00F1\_l2\_as\_NoiseMap.fits\\
-AS1A03\_103T01\_9000001132uvtNIIPC00F1\_l2\_as\_Sig.fits\\
-star\_raDec\_frmOptics\_catalogueWith\_10Stars.txt\\
-star\_raDec\_frmOptics\_catalogueWith\_5Stars.txt\\
-star\_radec.txt\\

-uvtRADECImage\_6.3\\
-AS1A03\_103T01\_9000001132uvtNIIPC00F1\_l2\_t0.0000\\
..............................\_f1\_noise\_ra-decnoise.fits\\
-AS1A03\_103T01\_9000001132uvtNIIPC00F1\_l2\_t0.0000\\
..............................\_f1\_sig\_ra-decexp.fits\\
-AS1A03\_103T01\_9000001132uvtNIIPC00F1\_l2\_t0.0000\_f1\_sig\_ra-dec.fits\\

-uvtRegAvg\_6.3\\
-AS1A03\_103T01\_9000001132uvtNIIPC00F1\_l2\_t0.0000\_f1\_exp\_regAvg.fits\\
-AS1A03\_103T01\_9000001132uvtNIIPC00F1\_l2\_t0.0000\\
..............................\_f1\_noiseMap\_regAvg.fits\\
-AS1A03\_103T01\_9000001132uvtNIIPC00F1\_l2\_t0.0000\_f1\_sig\_regAvg.fits\\

-uvtShiftRot\_6.3\\
-AS1A03\_103T01\_9000001132uvtNIIPC00F1\_l2\_radec.fits\\
-AS1A03\_103T01\_9000001132uvtNIIPC00F1\_l2\_snr.fits\\
-AS1A03\_103T01\_9000001132uvtNIIPC00F1\_l2\_t9999.0000\_f1\_img\_fi.fits\\

}




{\bf Appendix-10 : Top level directories for products from Driver Scheme
 (with some additional details for Combined multi-Episode products)} \\

{\tiny

-/FUV\_FullFrameAst\_F3\_W511\\
--uvtFullFrameAst\_6.3\\
-AS1A03\_103T01\_9000001132uvtFIIPC00F3\_l2\_as\_Exp.fits\\
-AS1A03\_103T01\_9000001132uvtFIIPC00F3\_l2\_as\_NoiseMap.fits\\
-AS1A03\_103T01\_9000001132uvtFIIPC00F3\_l2\_as\_Sig.fits\\
-star\_raDec\_frmOptics\_catalogueWith\_5Stars.txt\\
-star\_radec.txt\\

--/FUV\_Final\_F3\_W511\\
-F3\_W511\_FinalImage\_Exp.fits\\
-F3\_W511\_FinalImage\_NoiseMap.fits\\
-F3\_W511\_FinalImage\_Sig.fits\\

{\noindent}-/FUV\_FullFrameAst\_F2\_W511\\
-/FUV\_Final\_F2\_W511\\
-/NUV\_FullFrameAst\_F6\_W511\\
-/NUV\_Final\_F6\_W511\\
-/NUV\_FullFrameAst\_F3\_W511\\
-/NUV\_Final\_F3\_W511\\
-/NUV\_FullFrameAst\_F2\_W511\\
-/NUV\_Final\_F2\_W511\\
-/NUV\_FullFrameAst\_F1\_W511\\
-/NUV\_Final\_F1\_W511\\
-/\_RAPC\\
-/\_FUV\_10\\
-/\_NUV\_10\\
-/\_FUV\_9\\
-/\_NUV\_9\\
-/\_FUV\_8\\
-/\_NUV\_8\\
-/\_FUV\_7\\
-/\_NUV\_7\\
-/\_FUV\_6\\
-/\_NUV\_6\\
-/\_FUV\_5\\
-/\_NUV\_5\\
-/\_FUV\_4\\
-/\_NUV\_4\\
-/\_FUV\_3\\
-/\_NUV\_3\\
-/\_FUV\_2\\
-/\_NUV\_2\\
-/\_FUV\_1\\
-/\_NUV\_1\\
-/uvit\\

}




{\bf Appendix-11 : All user selectable parameters }

List of all user selectable switch settings and parameters for the pipeline (through
Parameter Interface Library, PIL), are presented in a tabular form (see Table 7).
The sequence of parameters are grouped by :
DRIVER\_SCHEME and RA\_INT, RA\_PC \& L2\_PC chains. Parameter names in bold font
indicate them to be tweaked frequently by users.

\begin{table*}[th]
\scriptsize
\caption{ 
List of all user selectable switch settings and parameters available through
Parameter Interface Library (PIL). The sequence of items are grouped in the 
following order :
(a) DRIVER\_SCHEME and (b) RA\_INT, RA\_PC \& L2\_PC chains. Parameters
frequently tweaked by the users are highlighted by bold font.
}
\begin{tabular}{|l|l|l|l|l|l|}
\hline
\textbf{Serial}&\textbf{Parameter name}&\textbf{Processing}&\textbf{Suggested}&\textbf{Description}&\textbf{Comments}\\
\textbf{No.}&{\it (bold indicates}&\textbf{Chain}&\textbf{Default}&\textbf{(switch logic:}& \\
 &{\it frequently}& &\textbf{value}&\textbf{`1'/`y' for ON;}& \\
 &{\it tweaked ones)}& & &\textbf{`0'/`n' for OFF)}& \\
\hline
& \textbf{Parameters for} & \textbf{Processing} & & & \\
& \textbf{Driver Scheme} & \textbf{Chain} & & & \\
\hline
1 & NUVonNUVflag & DRIVER\_SCHEME & n & Switch for forcing all drift generations &
  Turn ON only if \\
& & & & using NUV only (i.e. ignore all VIS) & VIS data NOT \\
& & & &  & reliable \\
\hline
2 & FUVonFUVflag & DRIVER\_SCHEME & n & Switch for forcing all drift generation & Turn ON only if \\
& & & & using FUV only (i.e. ignore all VIS \& & VIS \& NUV  \\
 & & & &  NUV data). Useful for generating& NOT available  \\
 & & & & FUV images only &   \\
\hline
3 & thresholdpc & DRIVER\_SCHEME & 50 & Parameter related to threshold for star & `AVG' \& \\
& & (\& L2PC-NUV) & & detection (starting value of `p' in : & `SIGMA' are  \\
 & & & & Threshold = AVG + p * SIGMA; `p' is & average and  \\
 & & & & gradually decreased iteratively till & standard  \\
 &&&& demanded no. of stars are achieved). & deviation of \\
 &&&& Used (1) for aligning UV images from & values (only non- \\
 &&&& different Episodes; \& (2) in Astrometry & zero entries \\
 &&&& module.             & considered) in \\
 &&&&                     & pixels of image \\
 &&&&                     & array. Only for \\
 &&&&                     & PC mode. \\
\hline
4 & minimumTargetedStars & DRIVER\_SCHEME & 6 & Targeted minimum no. of stars & (1) For PC mode \\
  & & (\& RA\_INT) & & (iterations proceed with lower `p' \& & \& (2) for INT \\ 
  & & & & detection threshold, till target is & mode. \\
  &&&&  achieved) : (1) to align images from & \\
  &&&&  different Episodes; \& (2) detection of & \\
  &&&&  stars in `uvtDetectStar' module (INT & \\
  &&&&  mode) & \\
\hline
5 & minimum\_targetedstarspc & DRIVER\_SCHEME & 5 & Success Criterion for Astrometry stage : &\\
& & (\& L2PC-NUV) && minimum no. of star matches with &\\
&&&& Catalogue required; &\\
\hline
& \textbf{Parameters } & \textbf{Processing} & & & \\
& \textbf{common to} & \textbf{Chains} & & & \\
& \textbf{multiple chains} & & & & \\
\hline
6 & \textbf{level1indir} & RA\_INT, RA\_PC, & - & Path to the directory holding Input & \\
 & & L2\_PC, L2\_INT & & Level-1 data & \\
\hline
7 & caldbdir & RA\_INT, RA\_PC, & - & Path to the directory holding Calibration & \\
 & & L2\_PC, L2\_INT & & Database (CAL\_DB) & \\
\hline
8 & \textbf{level2outdir} & RA\_INT, RA\_PC, & - & Path to the directory holding Calibration &\\
 & & L2\_PC, L2\_INT & & from the Driver Scheme (top level & \\
 & & & & location, below which products appear & \\
  & & & & in a structured manner) & \\
\hline
9 & ZipFlag & RA\_INT, RA\_PC, & y & Switch to indicate type of L1 data set : & ISSDC \\
 && L2\_PC, L2\_INT &&  `tar' (select `n') or `zip' (select `y') & disseminates in \\
 & & & & & `zip' format \\
\hline
10 & utcFlag & RA\_INT, RA\_PC, & n & Switch to select clock for time reference & Must be `n' \\
 && L2\_PC, L2\_INT && - `y' : use calibrated Universal Time &  \\
 && & & Clock (UTC); `n' : use UVIT's Master & \\
 &&&& clock & \\
\hline
%
\end{tabular}
\centerline {}\\
{\mbox{}\hfill  ... Continued} \\
\end{table*}
\addtocounter{table}{-1}
\begin{table*}[th]
\scriptsize
\caption{ {\it (Continued) :}
List of all user selectable switch settings and parameters available through
Parameter Interface Library, PIL. 
}
\begin{tabular}{|l|l|l|l|l|l|}
\hline
\textbf{Serial}&\textbf{Parameter name}&\textbf{Processing}&\textbf{Suggested}&\textbf{Description}&\textbf{Comments}\\
\textbf{No.}&&\textbf{Chain}&\textbf{Default}&& \\
 && &\textbf{value}&& \\
\hline
& \textbf{Parameters for } & \textbf{RA\_INT} & & \textbf{For `drift' extraction }  & \\
& \textbf{Relative Aspect} & & & \textbf{from VIS images }   & \\
& \textbf{Series Chain (INT)}  &&& \textbf{(in Integration Mode)} & \\
\hline
%
11 & history & RA\_INT & y & Switch for history to be written (`y') or & Records the \\
 & & & & not (`n'). History contains all the settings & settings : \#11 to \\
 & & & & of the selectable switches \& input & \#39 as listed in \\
 & & & & parameters with which the Chain has & this table) \\
 & & & &  ran. & \\
\hline
12 & clobber & RA\_INT & y & Action in case the output directory & Default to erase \\
   & & & & chosen by the user is already existing. If & past contents (if \\
   & & & & clobber is YES (‘y’) then this directory & any) \\
  & & & & will be removed and created afresh by & \\
  & & & & pipeline. If clobber is NO (`n') then & \\
  & & & & pipeline will exit with an appropriate & \\
  & & & & error message  & \\
\hline
13 & GTI\_FLAG & RA\_INT & 0 & GTI filtering to be done(1) or not(0). & Must be '0' (i.e. \\
 & & & & & not using GTI \\
 & & & & & filter) \\
\hline
14 & crcflag & RA\_INT & n & Switch to select if CRC check to be & CRC check is \\
 & & & & carried out (`y'), or not (`n'). If `y' is & ignored. Bit error \\ 
 & & & & selected, another switch (`dropframe', & rate in general is \\
 & & & & see \#20) selects the action : to drop the & very low. \\ 
 & & & & affected packet alone or all packets for & \\
 & & & & that image frame. & \\
\hline
15 & ManualMode & RA\_INT & n & Switch for Manual Mode operation to & Useful only when \\
 & & & & select stars - On(`y') or Off(`n'); & either star field \\
 & & & & [used in “uvtComputeDrift” block]; & very complex or \\
 & & & & enters interactive mode for this block & severe detector \\
 & & & & alone, and the rest of the processes run & artifacts (\& \\
 & & & & automatically]. & normal run with \\
 & & & & & `n' has FAILED) \\
\hline
16 & channel & RA\_INT & VIS & Selection of Band for which to run & Normally only \\
 & & & & & VIS band is used \\
 & & & & & in Integration \\
 & & & & & (INT) Mode \\
\hline
17 & junkfileflag & RA\_INT & 1 & Switch for handling artifacts of VIS & VIS artifact \\
 & & & & detector (`horizontal stripes' etc. in VIS & handler must be \\
 & & & & images) - `1' : ON; `0' : OFF & ON \\
\hline
18 & thresholdforjunkFrame & RA\_INT & 5000 & Threshold (pixel ADU) value for & Pixel threshold \\
 & & & & identifying `artifacts' in VIS images & value of VIS \\
 & & & & (Threshold for artifact detection) & artifacts from \\
 & & & & & experience \\
\hline
19 & darkframeFlag & RA\_INT & 1 & Switch for Dark Subtraction: `1' & Dark subtraction \\
 & & & & =substract Dark; `0' = no action & must be ON (`1') \\
\hline
20 & dropframe & RA\_INT-DataIngest & 0 & Switch to select action if CRC check & Relevant only if \\
 & & & & fails : `1' = entire image frame to be & switch for CRC \\
 & & & & discarded; or `0' = only the individual & test is turned ON \\
 & & & & failed packets (2048 bytes) to be & (`crcflag' = `y'; \\
 & & & & discarded. & see \#14 above) \\
\hline
21 & ThresholdValue & RA\_INT & 400000 & Threshold value to identify pixel with & Cosmic Ray \\
 & & & & Cosmic Ray hit (for INT mode) & detection is \\
 & & & & & effectively OFF \\
\hline
22 & flatfieldFlag & RA\_INT & 1 & Flat Field correction Switch : `1'= apply & Flat Field \\
 & & & & Flat Field correction; `0'= no action & correction must \\
 & & & & & be ON (`1') \\
\hline
23 & Nacc & RA\_INT & 1 & Effectively selects the size of the Time & Allowed values \\
 & & & & bin for drift extraction; `Nacc' is the no. & include any \\
 & & & & of successive raw INT mode frames & integer : 1 / 2 / 3 / \\
 & & & & accumulated to generate a single image & 4 ... etc. The \\
 & & & & (from which stars are identified) \& & value of 1 is \\
 & & & & processed further & optimal for \\
 & & & & & typical spacecraft \\
 & & & & & drift. \\
\hline
24 & refineWindow & RA\_INT & 15 & Size (pixel unit) of the square window & Selection MUST \\
 & & & & considered (centered on each star & be `15' \\
 & & & & candidate in the First-cut list) to further & \\
 & & & & shortlist identifying local maxima & \\
 & & & & (leading to generate Refined list); & \\
\hline
%
%
\end{tabular}
\centerline {}\\
{\mbox{}\hfill  ... Continued} \\
\end{table*}
\addtocounter{table}{-1}
\begin{table*}[th]
\scriptsize
\caption{ {\it (Continued) :}
List of all user selectable switch settings and parameters available through
Parameter Interface Library, PIL.
}
\begin{tabular}{|l|l|l|l|l|l|}
\hline
\textbf{Serial}&\textbf{Parameter name}&\textbf{Processing}&\textbf{Suggested}&\textbf{Description}&\textbf{Comments}\\
\textbf{No.}&&\textbf{Chain}&\textbf{Default}&& \\
 && &\textbf{value}&& \\
\hline
25 & centroidWindow & RA\_INT & 3 & Size (pixel unit) of the square window & Selection MUST \\
 & & & & used to compute Centroid values (along & be `3' \\
 & & & & X \& Y axes) for each entry in the & \\
 & & & & Refined list; & \\
\hline
26 & \textbf{threshold} & RA\_INT & 10 & Parameter related to threshold for star & `AVG' \& \\
 & & & & detection - starting value of `p' in : & `SIGMA' are \\
 & & & & Threshold = AVG + p * SIGMA; `p' is & average and \\
 & & & & gradually decreased iteratively till & standard \\
 & & & & required no. of stars are found. & deviation of \\
 & & & & & values in pixels \\
 & & & & & (with outliers \\
 & & & & & removed) in dark \\
 & & & & & substracted \\
 & & & & & image array (for \\
 & & & & & INT mode only). \\
\hline
27 & framesToBeDiscard & RA\_INT & 1 & Number of initial images (each obtained & Valid range of \\
 & & & & by accumulating `Nacc' raw images) to & values : 0/ 1/ 2/ \\
 & & & & be ignored (skipped) prior to identifying & ... less than the \\
 & & & & the Reference Frame (or its beginning & total no. of \\
 & & & & sample) & images. \\
 & & & & & Can be used to \\
 & & & & & skip data patch \\
 & & & & & with unusually \\
 & & & & & large drift. \\
\hline
28 & averageFactor & RA\_INT & 1 & A second level of Time bin selection : & Valid range of \\
 & & & & centroids of detected stars from selected & values : 1/ 2/ 3/... \\
 & & & & number of successive `accumulated' & etc. \\
 & & & & frames (each comprising of `Nacc' raw & Optimal value for \\
 & & & & frames) are averaged to define the & typical drift is \\
 & & & & Reference Frame \& similarly averaged & `1'. \\
 & & & & centroid tables for future time bins; & \\
\hline
29 & freqDomainFilterFlag & RA\_INT & 2 & Switch to select kind of filtering on the & Default value is \\
 & & & & Drift Series (low pass filtering) : `0'= & conservative (by- \\
 & & & & Time domain; `1'= Frequency domain; & passing the \\
 & & & & \& `2'= no Filtering; & filter). Under \\
 & & & & & special \\
 & & & & & circumstances, \\
 & & & & & the filter can be \\
 & & & & &  useful. \\
\hline
30 & freq\_value & RA\_INT & 0.2 & Cut-off frequency (Hz); & Default is max. \\
 & & & & Relevant only if Frequency domain & allowed, \\
 & & & & filtering is selected; (i.e. & corresponding to \\
 & & & & `freqDomainFilterFlag'=`1', see \#29 & the bandwidth for \\
 & & & & above) & drift extraction \\
 & & & & & from VIS. Lower \\
 & & & & & values are valid. \\
\hline
31 & orderPitch & RA\_INT & 1 & Polynomial order for filtering time series & Choice is \\
 & & & & of PITCH component of drift & appropriate for \\
 & & & & & default Time \\
 & & & & & Window (4 sec). \\
\hline
32 & orderYaw  & RA\_INT & 1 & Polynomial order for filtering time series & See \#31 \\
 & & & & of YAW component of drift & \\
\hline
33 & orderRoll  & RA\_INT & 1 & Polynomial order for filtering time series & See \#31 \\
 & & & & of ROLL component of drift & \\
\hline
34 & typeFiltering & RA\_INT & 2 & Switch for selection among 3 options in & Global option fits \\
 & & & & Time domain filtering : `0'= global; & the entire time \\
 & & & & 1=polynomial; 2= local (sliding & series of drift at \\
 & & & & polynomial) ; & once. Local \\
 & & & & & option fits many \\
 & & & & & times, using a \\
 & & & & & sliding Time \\
 & & & & & Window \\
 & & & & & (`deltaTime' = 4 \\
 & & & & & sec, see \#35 \\
 & & & & & below). \\
\hline
35 & deltaTime & RA\_INT & 4 & Selection of Time Window for the `local' & \\
 & & & (seconds) & option of Time domain filtering; & \\
 & & & & (Relevant only if `typeFiltering'=2, see & \\
 & & & & \#34 above) & \\
\hline
\end{tabular}
\centerline {}\\
{\mbox{}\hfill  ... Continued} \\
\end{table*}
\addtocounter{table}{-1}
\begin{table*}[th]
\scriptsize
\caption{ {\it (Continued) :}
List of all user selectable switch settings and parameters available through
Parameter Interface Library, PIL.
}
\begin{tabular}{|l|l|l|l|l|l|}
\hline
\textbf{Serial}&\textbf{Parameter name}&\textbf{Processing}&\textbf{Suggested}&\textbf{Description}&\textbf{Comments}\\
\textbf{No.}&&\textbf{Chain}&\textbf{Default}&& \\
 && &\textbf{value}&& \\
\hline
36 & diffDist & RA\_INT & 1 & Parameter for drift computation. & Search distance \\
 & & & & Selection of initial search distance (pixel & is gradually \\
 & & & & unit) for identifying star matches among & increased if no \\
 & & & & images from successive time bins; & match is found. \\
 & & & & & (up to a fixed \\
 & & & & & max. value). \\
\hline
37 & GenMatchStarsFile\_flag & RA\_INT & 0 & Switch regarding writing record of star & \\
 & & & & matched star-pair lists from two & \\
 & & & & successive time bins; `1'= write \& `0'= & \\
 & & & & do not write; & \\
\hline
38 & shiftRotDetAlgoFlag & RA\_INT & 1 & Switch to select (among 3 available & Equal weights \\
 & & & & options) the scheme for computation of & to each star \\
 & & & & drift. The 3 choices are : (1) Equal & provides uniform \\
 & & & & weight to each star \& rotation angle & representation to \\
 & & & & \textit{d\_theta}, between successive time  & the full sky field, \\
 & & & & bins is small; (2) Intensity based & improving \\
 & & & & weight; (3) \textit{d\_theta} can be large. & precision of the \\
 & & & & & extracted drift. \\
\hline
39 & flag\_thetaComp & RA\_INT & 0 & Switch to include or exclude `rotation' & \\
 & & & & (\textit{d\_theta}), in drift computation; `1' =  & \\
 & & & & \textit{d\_theta} parameter is included along with & \\
 & & & & `Xshift' \& `Yshift'; \& `0' = only `Xshift' & \\
 & & & & \& `Yshift' to be extracted; & \\
\hline
& \textbf{Parameters for } & \textbf{RA\_PC} & & \textbf{For `drift' extraction from }  & \\
& \textbf{Relative Aspect} & & & \textbf{NUV /FUV photon event list}   & \\
& \textbf{Series Chain (PC)}  &&& \textbf{(in Photon Counting Mode)} & \\
\hline
40 & historyrapc & RA\_PC & y & Switch to retain history or not; (similar & Records settings \\
   & & & & to `history', see \#11 above); & of all switches \& \\
   & & & & & parameters for \\
   & & & & & the RA\_PC \\
   & & & & & Chain (\#40 to \\
   & & & & & \#66 as listed in \\
   & & & & & this table) \\
\hline
41 & clobberrapc & RA\_PC & y & Action in case the output directory & \\
 & & & & chosen by the user is already existing; & \\
 & & & & (similar to `clobber', see \#12 above); & \\
\hline
42 & GTI\_FLAGrapc & RA\_PC-DataIngest & 0 & GTI filtering to be done(1) or not(0) & Not using GTI \\
 & & & & & filter \\
\hline
43 & crcflagrapc & RA\_PC & n & Switch to select if CRC check to be & CRC check is \\
   & & & & carried out (`y'), or not (`n'). If `y' is & ignored. Bit error \\
   & & & & selected, another switch & rate in general is \\
   & & & & (`dropframerapc', see \#44) selects the & very low. \\
   & & & & action : to drop the affected packet alone & \\
   & & & & or all packets for that image frame. & \\
\hline
44 & dropframerapc & RA\_PC-DataIngest & 0 & Switch to select action if CRC check & Relevant only if \\
   & & & & fails : `1' = entire image frame to be & switch for CRC \\
   & & & & discarded; or `0' = only the individual & test is turned ON \\
   & & & & failed packets (2048 bytes) to be & (`crcflagrapc' = \\
   & & & & discarded & `y'; see \#43 \\
   & & & & & above). \\
\hline
45 & parityFlagrapc & RA\_PC & 2 & Switch for qualifying criteria of photon & Default option \\
   & & & & events based on their parity tests; `1'= & tests the 2 crucial \\
   & & & & all three words for the event must have & words. The 3-rd \\
   & & & & correct parity, `2'= the two words & word holds event \\
   & & & & holding X \& Y centroids must have & asymmetry. \\
   & & & & correct parity. & Statistics of \\
   & & & & The option `0' by-passes parity tests & rejected events \\
   & & & & completely. & based correction \\
   & & & & & is applied to \\
   & & & & & retain \\
   & & & & &  photometric \\
   & & & & & accuracy. \\
\hline
\end{tabular}
\centerline {}\\
{\mbox{}\hfill  ... Continued} \\
\end{table*}
\addtocounter{table}{-1}
\begin{table*}[th]
\scriptsize
\caption{ {\it (Continued) :}
List of all user selectable switch settings and parameters available through
Parameter Interface Library, PIL.
}
\begin{tabular}{|l|l|l|l|l|l|}
\hline
\textbf{Serial}&\textbf{Parameter name}&\textbf{Processing}&\textbf{Suggested}&\textbf{Description}&\textbf{Comments}\\
\textbf{No.}&&\textbf{Chain}&\textbf{Default}&& \\
 && &\textbf{value}&& \\
\hline

46 & thresholdMultphrapc & RA\_PC & 9999 & Threshold for identifying multiple & Test to identify \\
   & & & & photon events (from asymmetry in the & multi-photon \\
   & & & & event’s 5x5 pixel footprint); & events is turned \\
   & & & & Event with `Max-Min' among 4 corners & OFF. \\
   & & & & above selected `thresholdMultphrapc' & Recommended \\
   & & & & are flagged as multi-photon event. & value to turn the \\
   & & & & & test ON is `100'. \\
\hline
47 & flatfieldFlagrapc & RA\_PC & 0 & Switch for applying Flat field correction & \\
   & & & & : `1' = apply; `0' = no action; & \\
\hline
48 & thresholdrapc & RA\_PC & 10 & Parameter related to threshold for star & Similar to \\
   & & & & detection (first-cut peaks); Starting value & `thresholdpc', see \\
   & & & & of `p' in : Threshold = AVG + p * & \#3 above. \\
   & & & & SIGMA; `p' is gradually decreased & \\
   & & & & iteratively till required no. of stars are & \\
   & & & & found. & \\ 
\hline
49 & GenMatchStarsFile\_flagrapc & RA\_PC & 0 & Switch regarding writing record of star & Similar to \\ & & & & matched star-pair lists; `1'= write \& & `GenMatchStarsF \\
 & & & & `0'= do not write; & ile\_flag', see \#37 \\
 & & & & & above. \\
\hline
50 & refinedWinSizerapc & RA\_PC & 15 & Neighbourhood criteria for identifying & Selection MUST \\
 & & & & stars; (pixel unit) & be `15'. Similar to \\
 & & & & & `refineWindow', \\
 & & & & & see \#24 above. \\
\hline
51 & centroidWinSizerapc & RA\_PC & 3 &  Size of square box to compute Centroids & Selection MUST \\
 & & & & for detected stars (pixel unit); & be `3'. Similar to \\
 & & & & & `centroidWindow', \\
 & & & & & see \#25 above. \\
\hline
52 & framesDiscardrapc & RA\_PC & 2 & Number of initial frames to be discarded & Default choice of \\
 & & & & before initiating accumulation of frames; & `2' is optimal, \\
 & & & & Valid values : 0 / 1 / 2/ ... less than total & since only the \\
 & & & & frames. & very first frame is \\
 & & & & & found to be \\
 & & & & & `disturbed'. \\
\hline
53 & \textbf{framesComputerapc} & RA\_PC & 90 ( 3 & Number of successive frames to be & Selection \\
& & & sec per & combined (accumulated) for generating & depends on NUV \\
& & & image) & every image to be used for extracting & brightness of \\
& & & & drift; (effectively selects the Time bin) & field stars \\
\hline
54 & framesToBeDiscardrapc & RA\_PC & 1 & Number of initial `combined' & The default value \\
& & & & (accumulated) frames to be discarded & `1', ignores \\
& & & & prior to identifying the Reference Frame & initial  3 sec, \\
 & & & & (or its beginning sample). & which is \\
 & & & & Valid choices are 0 /1 / 2/ 3 ...etc., less & appropriate. \\
 & & & & than total data set. & Can be used to \\
 & & & & & skip data patch \\
 & & & & & with unusually \\
 & & & & & large drift. \\
\hline
55 & averageFactorrapc & RA\_PC & 1 & A second level of selection to widen the & Optimal value \\
 & & & & Time bin : number of successive & for typical drift is \\
 & & & & 'combined' (accumulated) frames, whose & `1'. Similar to \\
 & & & & star centroids need to be averaged to & `averageFactor', \\
 & & & & construct Reference Frame (\& also for & see \#28 above. \\
 & & & & subsequent future Time bins) & \\
 & & & & Valid range of numbers : 1/ 2/ 3/... etc. & \\
 & & & & Larger number implies cruder time  & \\
 & & & & sampling of drift.  & \\
\hline
56 & flag\_thetaComprapc & RA\_PC & 0 & Switch to include (`1') or exclude (`0') & Similar to \\
 & & & & `rotation' (\textit{d\_theta}) in drift & `flag\_thetaComp', \\
 & & & &  computation; & see \#39 above. \\
\hline
57 & diffDistrapc  & RA\_PC & 1 & Parameter for drift computation. & Search distance is \\
 & & & &  Selection of initial search distance (pixel & gradually \\
 & & & &  unit) for identifying star matches among & increased if no \\
 & & & &  centroids from successive time bins; & match is found. \\
 & & & & & (up to a fixed \\
 & & & & & max. value). \\
 & & & & & Similar to \\
 & & & & & `diffDist'; see \\
 & & & & & \#36 above. \\
\hline
\end{tabular}
\centerline {}\\
{\mbox{}\hfill  ... Continued} \\
\end{table*}
\addtocounter{table}{-1}
\begin{table*}[th]
\scriptsize
\caption{ {\it (Continued) :}
List of all user selectable switch settings and parameters available through
Parameter Interface Library, PIL.
}
\begin{tabular}{|l|l|l|l|l|l|}
\hline
\textbf{Serial}&\textbf{Parameter name}&\textbf{Processing}&\textbf{Suggested}&\textbf{Description}&\textbf{Comments}\\
\textbf{No.}&&\textbf{Chain}&\textbf{Default}&& \\
 && &\textbf{value}&& \\
\hline

\hline
58 & freqDomainFilterFlag & RA\_PC & 2 & Switch to select kind of low pass & Similar to \\
 &rapc & & & filtering on the Drift Series; & `freqDomainFilter \\
  & & & & & Flag', see \#29 above. \\
\hline
59 & freqValuerapc & RA\_PC & 0.2 & Cut-off frequency (Hz); (Only for & Similar to \\
 & & & & Frequency domain filtering selection) & `freq\_value', see \\
  & & & & & see \#30 above. \\
\hline
60 & orderPitchrapc & RA\_PC & 1 & Polynomial order for filtering PITCH & Similar to \\
 & & & & component of drift & `orderPitch', see \\
  & & & & & see \#31 above. \\
\hline
61 & orderYawrapc & RA\_PC & 1 & Polynomial order for filtering YAW & See \#60 \\
 & & & & component of drift & \\
\hline
62 & orderRollrapc & RA\_PC & 1 & Polynomial order for filtering ROLL & See \#60 \\
 & & & & component of drift & \\
\hline
63 & typeFilteringrapc & RA\_PC & 2 & Switch for selection among 3 options in & Similar to \\
 & & & & Time domain filtering; & `typeFiltering', \\
  & & & & & see \#34 above. \\
\hline
64 & deltaTimerapc & RA\_PC & 4 & Selection of Time Window for the `local' & \\
 & & & (seconds) & option of Time domain filtering; & \\
 & & & & (Relevant only if `typeFilteringrapc'=2, & \\
 & & & & see \#63 above) & \\
\hline
65 & shiftRotDetAlgoFlag & RA\_PC & 3 & Switch to select (among 3 available & The 3 algorithms \\
 &rapc & & & options) the scheme for computation of & are described for \\
 & & & & drift & `shiftRotDetAlgo \\
  & & & & & Flag', see \#38 above \\
\hline
66 & minimumTargetedStars & RA\_PC & 4 & Minimum no. of stars demanded in the & Similar to \\
 &rapc & & & module for star detection & `minimumTargeted \\
 & & & & (uvtDetectStar; PC mode) & Stars', see \#4 above; \\
\hline
& \textbf{Parameters for } & \textbf{L2\_PC (for} & & \textbf{For generating NUV /FUV sky}  & \\
& \textbf{Sky Imaging chain} & \textbf{NUV /FUV)} & & \textbf{images from photon event list}   & \\
& \textbf{(Level2PC) :}  &&& \textbf{(in Photon Counting Mode)} & \\
\hline
67 & historypc & L2\_PC-NUV & y & Switch to write history or not; & Records settings \\
/68 & /historypcfuv & /L2\_PC-FUV &  &(similar to `history' in \#11 above); & of all switches \& \\
  & & & & & parameters for \\
  & & & & & the L2\_PC Chain \\
  & & & & & for NUV/ FUV; \\
  & & & & & (\#67/68 to \\
  & & & & & \#119/120 as \\
  & & & & & listed in this \\
  & & & & & table) \\
\hline
69 & clobberpc & L2\_PC-NUV & y & Action in case the output directory &  \\
/70 & /clobberpcfuv & /L2\_PC-FUV &  & chosen by the user is already existing; & \\
  & & & & (similar to ‘clobber’ in \#12 above); &  \\
\hline
71 & GTI\_FLAGpc & L2\_PC-NUV & 0 & GTI filtering to be done(1) or not(0) & Not using GTI \\
/72 & /GTI\_FLAGpcfuv & DataIngest & & & filter \\
 & & /L2\_PC-FUV  & & & \\
 & & DataIngest  & & & \\
\hline
73 & crcflagpc & L2\_PC-NUV & n & Switch to select if CRC check to be & CRC check is \\
/74 & /crcflagpcfuv & /L2\_PC-FUV & &carried out (`y'), or not (`n'). If `y' is & ignored. Bit error \\
&&&& selected, another switch (`dropframepc' & rate in general is \\
&&&& /`dropframepcfuv', see \#81/82 below) & very low. \\
&&&& selects the action : to drop the affected & \\
&&&& packet alone or all packets for that& \\
&&&& image frame. & \\
\hline
75 & pathToOutputTarpc & L2\_PC-NUV & - & path for storing the output products of &Not relevant \\
/76 & /pathToOutputTarpc & /L2\_PC-FUV && L2\_PC chain, when run as a stand alone &  when running full\\
 & fuv &&& chain; & Driver Scheme \\
\hline
77 & parityFlagpc & L2\_PC-NUV & 2 & Switch for qualifying criteria of photon & Details similar to \\
/78 & /parityFlagpcfuv & /L2\_PC-FUV &  &events based on their parity tests; & `parityFlagrapc', \\
& &&&& see \#45 above. \\
\hline
79 & thresholdMultphpc & L2\_PC-NUV & 9999 & Threshold for identifying multiple &Test to identify \\
/80 & /thresholdMultphpcfuv & /L2\_PC-FUV &  & photon events. & multi-photon \\
&&&& Similar to `thresholdMultphrapc', & events is turned OFF\\
&&&&  in \#46 above; & Recommended value \\
&&&&& to turn it ON is `100'. \\
\hline
\end{tabular}
\centerline {}\\
{\mbox{}\hfill  ... Continued} \\
\end{table*}
\addtocounter{table}{-1}
\begin{table*}[th]
\scriptsize
\caption{ {\it (Continued) :}
List of all user selectable switch settings and parameters available through
Parameter Interface Library, PIL.
}
\begin{tabular}{|l|l|l|l|l|l|}
\hline
\textbf{Serial}&\textbf{Parameter name}&\textbf{Processing}&\textbf{Suggested}&\textbf{Description}&\textbf{Comments}\\
\textbf{No.}&&\textbf{Chain}&\textbf{Default}&& \\
 && &\textbf{value}&& \\
\hline
81 & dropframepc & L2\_PC-NUV & 0 & Switch to select action if CRC check & Relevant only if \\
/82 & /dropframepcfuv & DataIngest && fails : `1' = entire image frame to be & switch for CRC \\
& & /L2\_PC-FUV &  & discarded; or `0' = only the individual & test is turned ON \\
&&DataIngest && failed packets (2048 bytes) to be & (`crcflagpc' / \\
&&&& discarded & `crcflagpcfuv' = \\
&&&&& `y'; see \#73/ \#74 \\
&&&&& above). \\
\hline
83 & thr\_One\_crpc & L2\_PC-NUV & 9999 & Value of the parameter no. 1 for & Cosmic Ray \\
/84 & /thr\_One\_crpcfuv & /L2\_PC-FUV & & identifying Cosmic Ray affected frames, & affected frame \\
&&&& ``N", in equation : threshold = AVG + & identification is \\
&&&& N*sqrt (AVG) + ST/(sqrt(AVG)) ; & effectively turned \\
&&&& where AVG= average no of events per & OFF. \\
&&&& frame; & Recommended \\
&&&&& value to turn it \\
&&&&& ON is `3'. \\  
\hline
85 & thr\_Two\_crpc & L2\_PC-NUV & 9999 & Value of the parameter no. 2 for & Cosmic Ray \\
/86 & /thr\_Two\_crpcfuv &/L2\_PC-FUV & &identifying Cosmic Ray affected frames, & affected frame \\
&&&& ``ST"; (see `thr\_One\_crpc' / & detection is \\
&&&& `thr\_One\_crpcfuv', \#83/ \#84 above); & turned OFF. \\
&&&& & Recommended \\
&&&&& value to turn it \\
&&&&& ON is `10'. \\
\hline
87 & CentCorr\_tobedonepc & L2\_PC-NUV & 0 & Switch regarding application of & \\
/88 & /CentCorr\_tobedonepc & /L2\_PC-FUV &  & corrections to event centroids due to & \\
 & fuv &&& Dark : `1' = apply; `0' = no action & \\
\hline
89 & CentBias\_tobedonepc & L2\_PC-NUV & 0 & Switch regarding application of & \\
/90 & /CentBias\_tobedonepc & /L2\_PC-FUV &  & corrections to event centroids due to & \\
 & fuv &&& Bias (FPN) : `1' = apply; & \\
&&&& `0' = no action & \\
\hline
91 & DetectDist\_tobedone  & L2\_PC-NUV & 1 & Switch for Detector Distortion & \\
/92 & pc & /L2\_PC-FUV &  & correction : `1' = apply; & \\
 & /DetectDist\_tobedone & & & `0' = no action & \\
 & pcfuv &&&& \\
\hline
93 & OpticDist\_tobedone  & L2\_PC-NUV & 1 & Switch for Detector Distortion & \\
/94 & pc & /L2\_PC-FUV &  & correction : `1' = apply; & \\
 & /OpticDist\_tobedone & & & `0' = no action & \\
 & pcfuv &&&& \\
\hline
95 & frameIntFlagpc & L2\_PC-NUV & 0 & Switch selection for 'Frame Integration' & Default is \\
/96 & /frameIntFlagpcfuv & /L2\_PC-FUV && block : to divide the full data from the & `Episode' case, to \\
&&&& `Episode' into smaller parts (`pseudo- & use the full data \\
&&&& Episode'-s) or not; `1' = divide into & together (i.e. no \\
&&&& multiple parts (`pseudo-Episode' case); & division into \\
&&&& `0' = all data used together as a & `pseudo- \\
&&&& single set (`Episode' case); & Episode'-s). \\
\hline
97 & framesDiscardpc & L2\_PC-NUV & 2 & Number of initial frames to be discarded & Valid range of \\
/98 & /framesDiscardpcfuv & /L2\_PC-FUV && (prior to gridding photon events into & values : 0/ 1/ 2/ \\
& &&& 2-D array/(s) in `Frame Integration' block) & ... less than the \\
&&&&& total no. of \\
&&&&& images. \\
&&&&& Choice optimal, \\
&&&&& since only the \\
&&&&& very first frame is \\
&&&&& found to be \\
&&&&& `disturbed'. \\
\hline
99 & framesComputepc & L2\_PC-NUV & 9400 & Number of successive frames to be & Relevant only if \\
/100 & /framesComputepcfuv & /L2\_PC-FUV & (say) & combined together to form a `pseudo- & `pseudo-Episode' \\
&&&& Episode', in the `Frame Integration' & case is selected \\
&&&& block. & for dividing the \\
&&&& Should not exceed the total no. of & full data from the \\
&&&& frames in the full Episode. & `Episode' into \\
&&&& In general, the last `pseudo-Episode' & multiple parts \\
&&&& will consist of fewer no. of frames, when & (i.e.`frameIntFlagpc' \\
&&&& total no. is not an integral multiple of & /`frameIntFlagpcfuv' \\
&&&& this selection. & =`1'; see \#95 /\#96) \\
\hline

\end{tabular}
\centerline {}\\
{\mbox{}\hfill  ... Continued} \\
\end{table*}
\addtocounter{table}{-1}
\begin{table*}[th]
\scriptsize
\caption{ {\it (Continued) :}
List of all user selectable switch settings and parameters available through
Parameter Interface Library, PIL.
}
\begin{tabular}{|l|l|l|l|l|l|}
\hline
\textbf{Serial}&\textbf{Parameter name}&\textbf{Processing}&\textbf{Suggested}&\textbf{Description}&\textbf{Comments}\\
\textbf{No.}&&\textbf{Chain}&\textbf{Default}&& \\
 && &\textbf{value}&& \\
\hline
101 & refinedWinSizepc & L2\_PC-NUV & 15 & Neighbourhood criteria for identifying & Selection MUST \\
/102 & /refinedWinSize & /L2\_PC-FUV & & stars (pixel unit); & be `15' (similar to \\
 & pcfuv  &&&& `refineWindow', \\
&&&&&  see \#24 above); \\
\hline
103 & centroidWinSizepc & L2\_PC-NUV & 3 & Size of square box to compute Centroids & Selection MUST \\
/104 & /centroidWinSize & /L2\_PC-FUV & & for detected stars (pixel unit); & be `3' (similar to \\
 & pcfuv &&&& `centroidWindow', \\
&&&&&  see \#25 above); \\
\hline
105 & diffDistpc & L2\_PC-NUV & 1 & Selection of initial search distance (pixel & Search distance is \\
/106 & /diffDistpcfuv & /L2\_PC-FUV & & unit) for identifying star matches & gradually \\
&&&& between a pair of images from `pseudo- & increased if no \\
&&&& Episode'-s; (used for `pseudo-Episode' & match is found \\
&&&& case in Frame Integration block) & (up to a fixed \\
&&&&& max. value). \\
\hline
107 & shiftRotDetAlgo & L2\_PC-NUV & 1 & Switch to select (among 3 available & The 3 algorithms \\
/108 & Flagpc & /L2\_PC-FUV & & options) the algorithm for determining & are similar to \\
& /shiftRotDetAlgo & & & offsets (shift \& rotation) between images & those described for\\
& Flagpcfuv & && from a pair of `pseudo-Episode'-s; & `shiftRotDetAlgoFlag',\\
&&&& (used for `pseudo-Episode' case in & see \#38 above \\
&&&& Frame Integration block) & \\
\hline
109 & flag\_thetaComppc & L2\_PC-NUV & 0 & Switch to include (`1') or exclude (`0') & Similar to \\
/110 & /flag\_thetaComp & /L2\_PC-FUV & & `rotation' (\textit{d\_theta}) in computation & `flag\_thetaComp', \\
& pcfuv & & &of offset; (used for `pseudo-Episode' & see \#39 above; \\
&&&& case in Frame Integration block) & \\ 
\hline
111 & thresholdpc & L2\_PC-NUV (\& & 50 & Threshold for star detection (first-cut & See details in \#3 \\
/112 & /thresholdpcfuv & DRIVER\_SCHEME) & /10 & peaks) - starting multiplier, `p', of sigma & above.\\
& & /L2\_PC-FUV & & in : Threshold = AVG + p * SIGMA; & The parameter \\
&&&& [used by Astrometry block; and & `thresholdpc' is \\
&&&& used for `pseudo-Episode' case in Frame & common to \\
&&&& Integration block] & Driver Scheme \& \\
&&&&& L2\_PC Chain for\\
&&&&&  NUV.\\
\hline
113 & database\_namepc & L2\_PC-NUV & - & Path pointing to star catalogue (USNO & \\
/114 & /database\_name & /L2\_PC-FUV & & A2) database; used for matching stars in & \\
& pcfuv & & & Astrometry block. & \\
\hline
115 & search\_algo\_for  & L2\_PC-NUV & 2 & Switch to select type of Search area in & \\
/116 & FullFrameAstpc & /L2\_PC-FUV & & the star Catalog : `1' = Square box; & \\
 & /search\_algo\_for & & & `2' = Circular; & \\
& FullFrameAstpcfuv &&&& \\
\hline
117 & Radi\_searchpc & L2\_PC-NUV & 0.05 & Dimension of Search area (length of & Default choice is \\
/118 & Radi\_searchpcfuv & /L2\_PC-FUV & & square box or radius of circle) to find & based on \\
 & &&& matching stars from Catalogue; & accuracy of \\
 &&&& (in degrees) & spacecraft's \\
 &&&&& attitude. \\
\hline
119 & minimum\_targeted & L2\_PC-NUV & 5 & Success Criterion for Astrometry stage : & The parameter \\
/120 & starspc & /L2\_PC-FUV & & minimum no. of star matches with & `minimum\_ \\
 & /minimum\_targeted & & & Catalogue demanded; & targetedstarspc' is \\
 & starspcfuv &&&& common to \\
 &&&&& Driver Scheme \& \\
 &&&&& L2\_PC Chain for \\
 &&&&& NUV (see \#5 above). \\
\hline

\end{tabular}
\end{table*}



\section*{Acknowledgements}

We thank the anonymous referees for their useful suggestions which led to improvements.
The UVIT project is a result of collaboration between IIA, Bangalore, IUCAA, Pune, TIFR,
Mumbai, many centers of the Indian Space Research Organization (ISRO), and the Canadian Space
Agency. We thank these organizations for their general support. We thank D Bhattacharya
for his role in organizing the pipeline requirements. The Ground Segment software teams of ISRO at
ISAC/URSC, Banglaore, ISTRAC, Bangalore \& SAC, Ahmedabad have provided significant support
throughout the development of the pipeline. We gratefully thank all the members of these teams for
their support. We also thank members of the ASTROSAT Project \& the ASTROSAT Science Working
Group for their critical feedbacks. Finally, we thank numerous users of early versions of this pipeline
for their feedbacks.






\begin{theunbibliography}{}
\vspace{-1.5em}


\bibitem{latexcompanion}
Ghosh, S. K., Tandon, S. N., Joseph, P., et al., ~2021, JApA, 42, 29



\bibitem{latexcompanion}
Kumar, A., Ghosh, S. K., Hutchings, J., et al., ~2012, {SPIE}, {8443}, 84431N




\bibitem{latexcompanion}
Tandon, S. N., Ghosh, S. K., Hutchings, J. B., et al., ~2017a, CSci, 113, 583




\bibitem{latexcompanion}
Tandon, S. N., Hutchings, J. B., Ghosh, S. K., et al., ~2017b, JApA, 38, 28




\bibitem{latexcompanion}
Tandon, S. N., Postma, J., Joseph, P., et al., ~2020, AJ, 159, 158





\bibitem{latexcompanion}
Tandon, S. N., Subramaniam, A., Girish, V., et al., ~2017c, AJ, 154, 128







\end{theunbibliography}
\end{document}